\documentclass[aps,prd,reprint,superscriptaddress,nofootinbib,amsfonts,amssymb,amsmath]{revtex4-1}

\usepackage[utf8x]{inputenc}
\usepackage[dvipdf,dvips,dvipdfmx]{graphicx}
\usepackage{float}
\usepackage{wrapfig}
\usepackage{bm}
\usepackage{color}
\usepackage{comment}
\usepackage{xcolor}
\usepackage[normalem]{ulem}
\usepackage{hyperref}
\hypersetup{
colorlinks=true,
citecolor=blue,
citebordercolor=red,
linktoc=all,
linkcolor=blue,
urlcolor=blue
}

\catcode`\@=11
\def\lsim{\mathrel{\mathpalette\@versim<}}
\def\gsim{\mathrel{\mathpalette\@versim>}}
\def\@versim#1#2{\vcenter{\offinterlineskip
\ialign{$\m@th#1\hfil##\hfil$\crcr#2\crcr\sim\crcr } }}
\catcode`\@=12
\newcommand{\Slash}[1]{{\ooalign{\hfil/\hfil\crcr$#1$}}}

\newcommand{\p}{\partial}

\newcommand{\al}[1]{\begin{align}#1\end{align}}
\newcommand{\bp}{\begin{pmatrix}}
\newcommand{\ep}{\end{pmatrix}}
\newcommand{\nn}{\nonumber\\}

\newcommand{\df}{\text{d}}

\newcommand{\bs}[1]{\boldsymbol}

\newcommand{\Tr}{{\rm Tr}\,}

\newcommand{\pmat}[1]{\begin{pmatrix}#1\end{pmatrix}}

\newcommand{\fn}[1]{\!\left(#1\right)}

\graphicspath{{./figs/}}

\begin{document}

\title{
Variable Planck mass from the gauge invariant flow equation
}

\author{Christof \surname{Wetterich}}
\affiliation{Institut f\"ur Theoretische Physik, Universit\"at Heidelberg, Philosophenweg 16, 69120 Heidelberg, Germany}

\author{Masatoshi \surname{Yamada}}
\affiliation{Institut f\"ur Theoretische Physik, Universit\"at Heidelberg, Philosophenweg 16, 69120 Heidelberg, Germany}

\begin{abstract}
Using the gauge invariant flow equation for quantum gravity we compute how the strength of gravity depends on the length or energy scale.
The fixed point value of the scale-dependent Planck mass in units of the momentum scale has an important impact on the question, which parameters of the Higgs potential can be predicted in the asymptotic safety scenario for quantum gravity?
For the standard model and a large class of theories with additional particles the quartic Higgs coupling is an irrelevant parameter at the ultraviolet fixed point.
This makes the ratio between the Higgs boson and the top-quark mass predictable.
\end{abstract}

\maketitle

\section{Introduction}
The asymptotic safety scenario~\cite{Hawking:1979ig,Reuter:1996cp} realizes quantum gravity as a nonperturbatively renormalizable quantum field theory, as summarized in~\cite{Niedermaier:2006wt,Niedermaier:2006ns,Percacci:2007sz,Reuter:2012id,Codello:2008vh,Eichhorn:2017egq,Percacci:2017fkn,Eichhorn:2018yfc}.
If a particle physics model coupled to quantum gravity can be extended to an infinitely short distance, the free parameters of the model correspond to the relevant parameters at the ultraviolet (UV) fixed point.
On the other hand, one can predict every renormalizable coupling of the effective low energy theory of particle physics at length scales much larger than the Planck length that corresponds to an irrelevant parameter at the fixed point.
More precisely, for $n$ relevant parameters at the UV-fixed point and $m$ renormalizable couplings of the low energy theory below the Planck mass, there exist $m-n$ constraints relating the low energy couplings.

Within this general setting the mass of the Higgs boson has been predicted to be $126$\,GeV with a few giga-electron-volt uncertainty~\cite{Shaposhnikov:2009pv}.
This prediction is based on three assumptions: (i) The quartic scalar coupling $\lambda_H$ is an irrelevant coupling at the UV-fixed point. (ii) The fixed point value $\lambda_{H*}$ is close to zero. (iii) The flow equations for $\lambda_H$ for momenta below the Planck scale do not deviate much from the ones for the standard model.
The present paper addresses the first assumption (i).
We want to know for which type of particle physics models, specified by the number of (almost) massless scalars, fermions, and gauge bosons at the fixed point, the quartic Higgs coupling is an irrelevant parameter.

A previous detailed investigation~\cite{Pawlowski:2018ixd} of this question by the use of the gauge invariant flow equation for quantum gravity has been performed with the strength of gravity at the fixed point $w^{-1}$ taken as an unknown parameter.
The question of which parameters of the Higgs potential are relevant or irrelevant depends in an important way on the value of $w$ at the fixed point, possibly being influenced as well by a nonminimal coupling $\xi_H$ of the Higgs doublet to gravity.
By taking $w$ and $\xi_H$ as fixed parameters also the stability matrix for the flow away from the fixed point is reduced to the parameters in the scalar potential.
This approximation influences the critical exponents $\theta_i$, which are the eigenvalues of the stability matrix.
The critical exponents decide if a coupling is relevant ($\theta_i>0$) or irrelevant ($\theta_i<0$).
The present paper computes the flow equations for $w$ and $\xi_H$ and determines their values at the fixed point.
The stability matrix at the fixed point is extended to include these parameters.
Within the Einstein-Hilbert truncation for gravity we compute the critical exponents in dependence on the number of massless scalars, fermions, and gauge bosons at the fixed point.

In a quantum field theory for gravity the Planck mass $M_\text{p}(k)$ depends on the renormalization scale $k$, which is a typical inverse length scale at which the effective laws are investigated.
Fluctuations with wave length shorter than $k^{-1}$ are included in the scale-dependent effective action (or effective average action) $\Gamma_k$.
Lowering $k$ includes additional fluctuations and induces a scale dependent $M_\text{p}(k)$.
The flow equation for $M_\text{p}^2\fn{k}$ takes the general form 
\al{
k\p_k M_\text{p}^2\fn{k}=4c_Mk^2\,.
\label{general form of RG for M2}
}
The gravitational interactions are universal and all particles contribute to $c_M$.
In particular, the contribution of free massless scalars, fermions, or vector bosons yield constant contributions to $c_M$ since no mass scale is present besides $k$, and dimensionless coupling constants are absent.
The structure of the flow equation \eqref{general form of RG for M2} is very general and does not need the contribution from metric fluctuations.
While the metric fluctuations do not change the structure \eqref{general form of RG for M2}, they induce a quantitatively important part of $c_M$ that depends on the value of the (dimensionless) effective scalar potential or ``cosmological constant."

For the dimensionless ratio $w=M_\text{p}^2\fn{k}/(2k^2)$ the flow equation \eqref{general form of RG for M2} shows a fixed point for $w_*=c_M$.
In the presence of additional dimensionless couplings $c_M$ has to be evaluated for the fixed point values of these couplings. 
At the fixed point $M_\text{p}(k)$ scales according to its canonical dimension~\cite{Reuter:1996cp}
\al{
M_\text{p}^2(k)=2w_*k^2\,,
}
while for $k$ much smaller than the observed reduced Planck mass $\bar M_\text{p}=2.435\times 10^{18}$\,GeV the running of the Planck mass stops, 
\al{
M_\text{p}^2(k)=\bar M_\text{p}^2+2w_* k^2=2w(k) k^2.
}
Here $\bar M_\text{p}^2$ may depend on fields, being independent of $k$.
We are interested in the UV regime for which we want to compute the fixed point value $w_*$.
For this purpose we need the flow equation for the dependence of $w(k)$ on $k$.

Functional renormalization~\cite{Wetterich:1992yh,Reuter:1993kw,Reuter:1996cp} permits one to compute the flow equation for $w(k)$, both for pure quantum gravity~\cite{Reuter:1996cp} and for gravity coupled to matter~\cite{Dou:1997fg}.
We need $c_M$ in dependence on the number of massless real scalars $N_S$, massless Weyl or Majorana fermions $N_F$, and massless gauge bosons $N_V$.
There is already a rather substantial body of work for the computation of $c_M$ in various truncations for the effective action of gravity~\cite{
Reuter:1996cp,Souma:1999at,Percacci:2002ie,Percacci:2003jz,Benedetti:2009rx,Benedetti:2009gn,Narain:2009fy,Manrique:2010mq,Manrique:2010am,Harst:2011zx,Eichhorn:2011pc,Folkerts:2011jz,Donkin:2012ud,Eichhorn:2012va,Christiansen:2012rx,Dona:2013qba,Codello:2013fpa,Falls:2013bv,Christiansen:2014raa,Falls:2014tra,Demmel:2014hla,Percacci:2015wwa,Labus:2015ska,Oda:2015sma,Dona:2015tnf,Christiansen:2015rva,Meibohm:2015twa,Meibohm:2016mkp,Christiansen:2016sjn,Denz:2016qks,Gies:2016con,Eichhorn:2016vvy,Biemans:2016rvp,Christiansen:2017gtg,Christiansen:2017cxa,Christiansen:2017bsy,Hamada:2017rvn,Falls:2017lst,Eichhorn:2017als,Biemans:2017zca,Alkofer:2018fxj,deBrito:2018jxt, Alkofer:2018baq,Falls:2018ylp,Eichhorn:2018akn,Eichhorn:2018ydy,Eichhorn:2018nda,deBrito:2019epw}.
The existing quantitative results are, however, more widely scattered than needed for our purpose.

To increase the robustness of the result for $c_M(N_S,N_F,N_V)$, we employ here the gauge invariant flow equation for a single metric field~\cite{Wetterich:2016ewc}.
It has the important advantage that all physical information is contained in the gauge invariant or diffeomorphism symmetric effective action $\bar \Gamma_k$ which depends on a single macroscopic metric field $g_{\mu\nu}$.
Diffeomorphism invariance imposes an important restriction on the allowed couplings, reducing greatly the number of possible couplings as compared to an effective action without gauge invariance, or with gauge invariance only realized by simultaneous transformations of a background field $\bar g_{\mu\nu}$ and an independent macroscopic metric $g_{\mu\nu}$.
For example, keeping only up to two derivatives the gauge invariant effective action for scalars coupled to gravity reads
\al{
\bar \Gamma_k=\int_x\sqrt{g}\left\{ -\frac{F}{2}R +U+\sum_{i=1}^{N_S}\frac{K_i}{2}\p^\mu \varphi_i \p_\mu \varphi_i\right\}\,,
}
where we use the shorthand convention $\int_x=\int \df^4x$, and $F$, $U$, and $K$ depend on $k$ and are functions of the scalar fields $\varphi_i$.
Here $R$ is the curvature scalar and $g=\det g_{\mu\nu}$.
We will work in this truncation, setting further $K_i=1$.

We investigate the flow of $U(\rho)$, as embodied in the dimensionless quantity $u(\tilde\rho)=U/k^4$, with $\rho$ a suitable bilinear of scalar fields and $\tilde\rho=\rho/k^2$.
Similarly, we establish flow equations for $F(\rho)$ or $w(\tilde\rho)=F/2k^2$.
For the gauge invariant flow equation one finds a rather simple result:
\al{
k\p_k u&= 2{\tilde\rho}\,\p_{\tilde\rho} u-4u +\frac{1}{32\pi^2}\left( N_S-2N_F+2N_V-\frac{8}{3}\right)\nn
&\quad+\frac{5}{24\pi^2}\left(1-\frac{u}{w}\right)^{-1}\,,
}
and
\al{
k\p_k w&= 2{\tilde\rho}\,\p_{\tilde\rho} w-2w +\frac{1}{96\pi^2}\left( -N_S-N_F+4N_V +\frac{43}{6}\right)\nn
&\quad
+\frac{25}{64\pi^2}\left(1-\frac{u}{w}\right)^{-1} \,.
}
This result uses the Litim cutoff function~\cite{Litim:2001up} and makes a mild simplification in the sector of scalar fluctuations.
Constant scaling solutions for an UV-fixed point are found by setting $\p_k u=\p_{\tilde \rho}u=\p_k w=\p_{\tilde\rho}w=0$.

Within our truncation we find an acceptable UV-fixed point with stable gravity for a large region in ``theory space" $(N_S,N_F,N_V)$.
This region includes pure gravity, the standard model, and many grand unified models.
Our truncation becomes doubtful for large positive values of $N_S+N_F-4N_V$.
Discarding this doubtful extreme region the quartic scalar coupling is found to be an irrelevant parameter.
It can therefore be predicted, giving support to the prediction of the Higgs boson mass~\cite{Shaposhnikov:2009pv}.
The validity for a large positive value of $N_{S}+N_{F}-4N_{V}$ may be enlarged by extending the truncation for the gravity system.
The limitation of the truncation will be discussed in Sec.\,\ref{sect: Limitation of truncation}.

In Sec.\,\ref{description on gauge invariant flow} we briefly recapitulate the gauge invariant flow equation for the effective average action for quantum gravity.
Subsequently, we discuss separately the different contributions to the flow of the effective scalar potential $U\fn{\rho}$ and the effective squared Planck mass $F\fn{\rho}$.
We start in Sec.\,\ref{scalar contributions} with the fluctuations of massless scalars and continue in Sec.\,\ref{gauge bosons section} with massless gauge bosons. 
The gauge boson fluctuations alone are sufficient to generate an acceptable UV-fixed point for quantum gravity.
Section\,\ref{gauge and scalar contributions} addresses the coupled system of gauge bosons and scalars for nonvanishing gauge couplings, and Sec.\,\ref{fermionic contribution} includes fermionic fluctuations.
Matter fluctuations alone generate an UV-fixed point with stable gravity provided $N_S+N_F<4N_V$.

In Sec.\,\ref{Metric fluctuation section} we discuss the flow contributions from fluctuations of the metric.
The gauge invariant flow equation offers the advantage that the contributions from physical fluctuations are independent of the ones from gauge modes and the regularized Faddeev--Popov determinant.
Section\,\ref{physical and gauge modes section} describes the dominant graviton contribution from the traceless transverse metric fluctuations.
In Sec.\,\ref{evaluation of graviton contributions} we discuss the combined ``measure contribution" from gauge modes and the Faddeev-Popov determinant.
Section\,\ref{evaluation of measure contribution} presents a simplified version of the subleading contribution from the physical scalar metric fluctuation.
The full contribution is displayed in the Appendixes~\ref{formulations} and \ref{evaluation of flow equations}.

In Sec.\,\ref{UV fixed point analysis} we discuss in detail the UV-fixed point solution for $\rho$-independent $U$ and $F$.
An approximate treatment of the subleading physical scalar metric fluctuations and their mixing with other scalar fluctuations allows us to discuss many aspects analytically.
The contributions from matter fluctuations, as well as the measure contribution and the contribution from the physical scalar metric fluctuations can be combined into two effective parameters $\tilde{\mathcal N}_U=N_S-2N_F+2N_V-8/3$ and $\tilde{\mathcal N}_M=-N_S-N_F+4N_V+43/6$.
For all these contributions the propagator is the one for massless fields.
Only for the graviton contribution does the value of $U$, corresponding to a cosmological constant, influence the propagator.

Section\,\ref{Higgs and quantum gravity} addresses the consequences of our investigation for the predictability of the parameters of the Higgs potential.
This issue depends on the particle physics model coupled to quantum gravity.
The precise number of massless scalars, fermions, and vector bosons for the UV completion of the standard model influences $\tilde{\mathcal N}_U$ and $\tilde{\mathcal N}_M$ and therefore the precise location and properties of the fixed point.

Many of the particles may acquire a mass proportional to the Planck mass $\bar M_\text{p}$ as the flow of couplings moves away from the fixed point.
This is typically the case for grand unified theories (GUTs).
The effective low energy theory below the Planck mass may be only the standard model.
Nevertheless, predictions for the fixed point and critical exponents for small deviations from it depend on the complete microphysical particle model.
If the microscopic model remains the standard model, the quartic Higgs coupling is an irrelevant parameter and can be predicted.
In contrast, the mass term is a relevant parameter such that the gauge hierarchy is a free parameter that cannot be predicted.
This is similar for a minimal GUT based on SU(5).
For microscopic GUTs with a large number of scalar $N_S$ the gravity induced anomalous dimension for the scalar mass term and quartic coupling increases due to the graviton propagator moving close to the onset of instability.
This is the region where our truncation becomes doubtful.
Unfortunately, at the present stage no robust statement is possible on the question if the mass term for the Higgs scalar becomes irrelevant (self-induced criticality) in GUTs with large $N_S$ as SO(10).
In Sec.\,\ref{conclusions} we present our conclusions.

\section{Gauge invariant flow equation}
\label{description on gauge invariant flow}
The gauge invariant flow equation for the gauge invariant effective average action $\bar\Gamma_k$ takes the form~\cite{Wetterich:2016ewc,Wetterich:2017aoy}
\al{
k\p_k \bar\Gamma_k=\p_t \bar\Gamma_k=\pi_k -\delta_k\,,
\label{general gauge inv. flow equation}
}
with $\pi_k$ the contribution of physical fluctuations that depends on $\bar \Gamma_k$, and $\delta_k$ a universal measure contribution that is independent of $\bar \Gamma_k$.
The contribution from physical fluctuations takes the one-loop form
\al{
\pi_k=\frac{1}{2}\text{Str}\left\{\p_t R_P\, G_P\right\},
\label{flow equation for GP}
}
where Str denotes a momentum integration and summation over internal indices, with an additional minus sign for fermions arising from their Grassmann nature.
The full propagator $G_P$ for the physical modes is a functional of arbitrary macroscopic fields, such that Eq.\,\eqref{general gauge inv. flow equation} is a functional differential equation.
With $P$ the projector on the physical fluctuations, the physical propagator obeys $PG_P=G_PP^T=G_P$.
The infrared cutoff function $R_P$ acts on the physical fluctuations.

The relation between $\bar \Gamma_k$ and $G_P$ involves the projector on the physical fluctuations,
\al{
\left(\bar\Gamma_k^{(2)}+R_P\right)G_P=P^T\,.
\label{propagator and Gamma2}
}
In the presence of a local gauge symmetry the second functional derivative $\bar\Gamma_k^{(2)}$ of a gauge invariant effective action has zero modes corresponding to the gauge degrees of freedom.
It is therefore not invertible.
It is invertible, however, on the projected subspace of physical fluctuations.
This underlies the relation \eqref{propagator and Gamma2}, which remains meaningful even in the limit $k\to 0$ where $R_P$ vanishes.
In short, the physical propagator is the inverse of the second functional derivative of $\bar \Gamma_k$ on the projected subspace.
For the flow equation, it is the inverse in the presence of the IR-cutoff $R_P$.
Insertion of Eq.\,\eqref{propagator and Gamma2} into Eq.\,\eqref{flow equation for GP} closes the flow equation, which becomes a functional differential equation for $\bar \Gamma_k$.

Projection operators on physical fluctuations are necessarily nonlocal objects.
An example is the projection on a transverse photon, $P_\mu{}^{\nu}=\delta_\mu^\nu-q_\mu q^\nu /q^2$.
At first sight the gauge invariant flow equations \eqref{flow equation for GP} and \eqref{propagator and Gamma2} seem therefore to be plagued by severe nonlocalities.
The explicit use of projectors can be circumvented, however, by a simple procedure.
One adds to $\bar \Gamma_k^{(2)}$ the second functional derivative of a physical gauge fixing term $\tilde \Gamma_\text{gf}^{(2)}/\alpha$, $\Gamma^{(2)}_k=\bar \Gamma_k^{(2)}+\tilde \Gamma_\text{gf}^{(2)}/\alpha$, which renders $\Gamma^{(2)}_k$ invertible.
A physical gauge fixing acts only on the gauge fluctuations, obeying 
\al{
P^T \tilde\Gamma_\text{gf}^{(2)}=\tilde \Gamma_\text{gf}^{(2)}P=0\,.
\label{condition for projection}
}
Adding also an IR cutoff for the gauge modes $R_\text{gf}/\alpha$, the propagator in the presence of gauge fixing and IR cutoffs is given by 
\al{
G=\left[ \bar\Gamma_k^{(2)}+R_P+\frac{1}{\alpha}\left( \tilde\Gamma_\text{gf}^{(2)}+R_\text{gf}\right) \right]^{-1}.
}
No projectors are needed any longer for the inversion.

For $\alpha\to0$ one finds a block diagonal form of the propagator matrix $G$, with a physical block $G_P$ and a block for the gauge modes that vanishes $\sim \alpha$.
As a consequence, one has
\al{
&\frac{1}{2}\text{tr}\left\{ \p_t \left(R_P+R_\text{gf}/\alpha \right)\left( \Gamma_k^{(2)}+R_P+R_\text{gf}/\alpha \right)^{-1}\right\}\nn
&\qquad=\pi_k+\delta_k\,.
\label{flow equation structure}
}
The part $\delta_k$ arises from the block in $G$ for the gauge modes.
This part of $G$ is proportional to $\alpha$, such that multiplication with $\p_t R_\text{gf}/\alpha$ yields a result that remains finite for $\alpha\to0$.
It is given by a simple determinant in the projected space of gauge modes
\al{
\delta_k=\frac{1}{2}\p_t \ln\det\left( \tilde \Gamma_\text{gf}^{(2)}+R_\text{gf}\right),
\label{measure contribution general}
}
with $\tilde \Gamma_\text{gf}^{(2)}$ a fixed differential operator, such that the $k$ dependence arises only from $R_\text{gf}$.
With given $\delta_k$ we can compute the flow contribution of the physical fluctuations $\pi_k$ from Eq.\,\eqref{flow equation structure} without any explicit use of projections.

Despite the close resemblance to the method of gauge fixing the gauge invariant flow equation \eqref{flow equation for GP} does not use any gauge fixing.
The addition of $\tilde\Gamma_\text{gf}^{(2)}/\alpha$ should be seen as a purely technical device for an effective computation of $G_P$, as defined by Eq.\,\eqref{propagator and Gamma2}. 
The relation \eqref{condition for projection} and the limit $\alpha\to0$ are mandatory, and there is no freedom for the choice of a gauge fixing.

The measure factor in Eq.\,\eqref{general gauge inv. flow equation} amounts to $-\delta_k$, with $\delta_k$ a universal expression given by Eq.\,\eqref{measure contribution general}~\cite{Wetterich:2016ewc,Wetterich:2017aoy}.
It expresses the presence of nonlinear constraints for the physical fluctuations.
Omitting the measure term $-\delta_k$ in the gauge invariant flow equation \eqref{general gauge inv. flow equation} would erroneously treat the physical fluctuations as unconstrained fields.
In the present approach the measure contribution is universal since the presence of constraints does not involve the form of the effective action $\bar\Gamma_k$.
It is based \cite{Wetterich:2016ewc} on the direct regularization of the Faddeev-Popov determinant, and no ghosts are introduced.
It is not known if this type of IR regularization is sufficient for all purposes
For the present level of truncation we establish explicitly in Appendix\,\ref{evaluation of flow equations} the equivalence with a regularization of the ghost propagator.

Despite the conceptional difference, Eq.\,\eqref{flow equation structure} can also be viewed as the flow generator for a gauge fixed theory with a truncation of the form
\al{
&\Gamma_k=\bar \Gamma_k +\Gamma_\text{gf}\,,&
&\Gamma_\text{gf}^{(2)}=\frac{1}{\alpha}\tilde \Gamma_\text{gf}^{(2)}\,,
\label{set of effective action}
}
up to a part from ghost fluctuations.
This holds provided one uses for $\Gamma_\text{gf}$ a physical gauge fixing that acts only on the gauge modes.
The ghost contribution amounts to $-2\delta_k$, having the same structure as the contribution from gauge fluctuations, but with opposite sign and a factor of 2.
We therefore observe a rather close relation between the gauge invariant flow equation and the background field method with a particular physical gauge fixing and a particular truncation.
This relation is discussed in detail in Appendixes~\ref{formulations} and \ref{evaluation of flow equations}.

The quantity $\tilde \Gamma_\text{gf}^{(2)}$ appearing in the measure term \eqref{measure contribution general} follows from 
\al{
\Gamma_\text{gf}=\frac{1}{\alpha} \tilde\Gamma_\text{gf}=\frac{1}{2\alpha}\int_x\sqrt{g}\,D^\mu a_{\mu\nu}D^\rho a_\rho{}^{\nu}
}
by second variation with respect to the gauge fluctuation $a_{\mu\nu}$.
It corresponds to a physical gauge fixing condition $D^\mu h_{\mu\nu}=0$, with $h_{\mu\nu}=f_{\mu\nu}+a_{\mu\nu}$ the fluctuation of the metric around the macroscopic metric $g_{\mu\nu}$, and $f_{\mu\nu}$ the physical fluctuation.
Gauge transformations act only on $a_{\mu\nu}$, leaving $f_{\mu\nu}$ invariant.
By construction, $\tilde\Gamma_\text{gf}^{(2)}$ obeys the projection condition \eqref{condition for projection}.
The precise form of $\tilde\Gamma_\text{gf}^{(2)}$ will be discussed in Sec.\,\ref{Metric fluctuation section}.

The infrared cutoff functions $R_P$ and $R_\text{gf}$ involve covariant derivatives formed with the macroscopic metric.
This dependence on the macroscopic metric is a crucial feature for guaranteeing gauge invariance in a formulation with a single macroscopic metric and no separate ``background metric."
As a consequence of the formulation in terms of a single metric, the derivatives $\p_t=k\p_k$ and $\p/\p g_{\mu\nu}$ commute.
For example, the flow equation for the graviton propagator follows directly from the second functional derivative of $\pi_k-\delta_k$ with respect to the metric.
This feature is an important difference as compared to the background field formalism, even if one uses for the latter the truncation \eqref{set of effective action}.
Derivatives of the flow generator with respect to the metric contain parts that involve field derivatives of $R_k$.
This results in additional diagrams for the flow equation for propagators or vertices.
These additional diagrams involve external lines ``ending" in the $R_k$ insertion.
They are not present in the background field formalism.
This is true only for the average fluctuations lines~\cite{Codello:2013wxa}.
It is a crucial advantage of the gauge invariant formulation with a single metric that physical propagators and vertices can be extracted directly from functional derivatives of the gauge invariant effective action $\bar\Gamma_k$ for $k\to0$.
The gravitational field equations imply that source terms always involve a covariantly conserved energy momentum tensor. 
This property does not hold automatically in the background field formalism.
We here comment on the modification of local symmetries in the background field formalism.
In the standard background field formalism for the gravity--Yang-Mills system, the effective action, especially the ghost action, loses the SU($N$) gauge invariance due to the non-commutative feature between diffeomorphisms and the SU($N$) gauge transformations.
For this issue one would define modified diffeomorphisms~\cite{Percacci:2008zf,Daum:2009dn} such that the effective action is invariant under both the modified diffeomorphisms and the SU($N$) gauge transformations.
On the other hand, in the present gauge invariant formalism such a modification is not required since the projection operators $P$ for local transformations, which define the propagators of the physical modes as Eq.\,\eqref{propagator and Gamma2}, are commutative and then the gauge invariant theory space is automatically projected out.

We conclude that the gauge invariant flow equation has many attractive properties.
What is not settled at the present stage is the question whether this equation is exact, or whether it is only an approximation to a more complicated functional differential equation for $\bar\Gamma_k$.
If the macroscopic metric is identified with the expectation value of the microscopic metric, and $\bar\Gamma_k$ is defined by the standard implicit functional integral over fluctuations (functional differential equation or ``background field identity"), the flow equations \eqref{flow equation for GP} and \eqref{propagator and Gamma2} are only an approximation~\cite{Wetterich:2017aoy}.
In this case the exact gauge invariant flow equation for $\bar\Gamma_k$ involves a gauge invariant correction term.
It has been argued that Eqs.\,\eqref{flow equation for GP} and \eqref{propagator and Gamma2} can become exact if one chooses a different macroscopic metric and modifies the definition of $\bar\Gamma_k$.
This requires that the differential equation relating an optimized macroscopic metric to the expectation value of the microscopic metric admits a solution \cite{Wetterich:2016ewc}.
Only the existence of a solution is needed, but a proof or disproof of existence is not available so far.
We note here that there exists an exact formula if one chooses $g_{\mu\nu}=\langle g_{\mu\nu}\rangle$.
In this version there are correction terms that may be absorbed by a different definition of the macroscopic metric $g_{\mu\nu}$.
One could estimate the relative importance of the correction term.
At the present level we think, however, that the truncation error is dominant.
As a check of the validity of different truncations, the stability of critical exponents could still be an indicator for the approximation to the flow equation.

\section{Matter induced flowing Planck mass}
In this section we compute the contribution to the flow equations for the effective scalar potential $U$ and the coefficient of the curvature scalar $F$ from fluctuations of scalars, gauge bosons, and fermions.
We partly recover results of earlier work for a subclass of employed methods and choices of cutoff functions, and we trace the origin of the differences to other results.
Since no metric fluctuations are involved at this stage, the issue of gauge fixing for diffeomorphism symmetry does not matter at this stage.
What is important for the differences between earlier results in the background formalism or flow equations violating gauge symmetry is the treatment of terms in $\Gamma_k$ involving the differences between macroscopic fields and background fields, and the choices of infrared cutoffs.
Our approach of gauge invariant flow equations, combined with requirements of locality for the choice of cutoff functions, eliminates many earlier ambiguities in the computation of $c_M$.

We find that matter fluctuations alone induce a fixed point for the flowing dimensionless Planck mass, provided that the number of gauge bosons $N_V$ exceeds the value $(N_S+N_F)/4$.
This can serve as a demonstration for the solidity of the concept of a nonperturbative fixed point for quantum gravity.
Particle physics models with $4N_V-N_S-N_F>0$ and $N_V\to\infty$ are easily constructed.
In this limit the strength of gravity at the fixed point $w^{-1}$ tends to zero $\sim N_V^{-1}$.
Thus metric fluctuations play a subdominant role and may be neglected, eliminating thereby many associated conceptional issues. 
The case of matter domination for $N_V\to \infty$ realizes in a certain sense old ideas of ``induced gravity"~\cite{Adler:1982ri}.
In contrast to the divergent expressions in a simple loop expansion, which involve often a problem of interpretation and preservation of symmetries, our flow equation is UV finite and gauge invariant. 

\subsection{Flow contribution from scalar field}
\label{scalar contributions}
Basic properties can be understood from the contribution of a scalar field with effective action
\al{
\Gamma_{k}^{(S)}=\int_x \sqrt{g}\left\{ \frac{1}{2} \p^\mu \varphi\, \p_\mu \varphi +U\fn{\varphi} -\frac{1}{2}F\fn{\varphi}R \right\} \,.
\label{effective action for varphi}
}
The second functional derivative with respect to $\varphi$ reads
\al{
\Gamma_{k}^{(S,2)}=\sqrt{g}\left(- D^2 +m^2\fn{\varphi} -\frac{1}{2}\tilde \xi\fn{\varphi}R \right)\,,
}
where $D^2=D^\mu D_\mu$, $D_\mu$ is a covariant derivative, and 
\al{
&m^2\fn{\varphi}=\frac{\p^2 U}{\p \varphi^2}\,,&
&\tilde \xi^2\fn{\varphi}=\frac{\p^2 F}{\p \varphi^2}\,.&
\label{mass and non-minimal}
}

Adding an appropriate IR-cutoff term $\sqrt{g}R_k\fn{-D^2}$ modifies 
\al{
\sqrt{g}(-D^2)\to \sqrt{g}P_k\fn{-D^2}=\sqrt{g}\left( -D^2+R_k\fn{-D^2} \right).
}
The scalar contribution to the flow equation reads
\al{
&\p_t \Gamma_{k}\nn
&= \frac{1}{2}\text{tr}_{(0)}\left\{ \p_t R_k\fn{-D^2}\left( P_k\fn{-D^2} +m^2 -\frac{\tilde \xi R}{2} \right)^{-1} \right\} .
}
We note that the factor $\sqrt{g}$ drops out, multiplying both $-D^2$ and $R_k\fn{-D^2}$.
We can write 
\al{
\p_t \Gamma_{k} = \frac{1}{2} \text{tr}_{(0)}\,\tilde \p_t \ln \fn{ P_k \fn{-D^2} +m^2-\frac{\tilde\xi R}{2}}\,,
\label{flow generator of scalar}
} 
where $\tilde \p_t$ acts only on the $k$ dependence of the IR cutoff $R_k=P_k+D^2$, e.g., not on $m^2$, on $\tilde \xi$, or on parameters or fields appearing in $D^2$.
The derivative $\tilde \p_t$ makes the trace finite. 
This is a central difference as compared to one-loop perturbation theory.

We can write 
\al{
&\p_t \Gamma_{k} =\frac{1}{2} \text{tr}\,W\fn{\Delta_S}, &
&\Delta_S=-D^2,&
}
with
\al{
W\fn{z}=\p_t R_k\fn{z} \left(z+R_k\fn{z}+m^2 -\frac{\tilde\xi R}{2}\right)^{-1}.
}
The trace can be evaluated by the heat kernel expansion, see Appendix\,\ref{heat kernel methods},
\al{
\text{tr}\, W\fn{\Delta_S}= \frac{1}{16\pi^2} \sum^\infty_{n=0}Q_{2-n}\fn{\tilde w} \int_x\sqrt{g}\,c_{2n}^{S}\fn{\Delta_S}\,.
}
The heat kernel coefficients for the operator $\Delta_S=-D^2$ are well known,
\al{
&c_0^{S}=b_0^S=1\,,&
&c_2^{S}=b_2^SR=\frac{R}{6}\,.&
}
The functions $Q_n$ depend on the field $\varphi$ via
\al{
\tilde w = \frac{m^2\fn{\varphi}}{k^2}-\frac{\tilde\xi R}{2k^2}\,,
}
with
\al{
Q_2&= \int^\infty_0 \df z\,z W\fn{z},&
Q_1&= \int^\infty_0\df z\, W\fn{z},&\nn[2ex]
Q_0&=W\fn{z=0},&
W\fn{z}&=\frac{\p_t R_k\fn{z}}{P_k\fn{z}+\tilde w k^2}\,.&
}
The first two terms in the expansion yield
\al{
\p_t \Gamma_{k} &= \frac{1}{32\pi^2} \int _x\sqrt{g}\int^\infty_0 \df z\frac{\p_t R_k \fn{z}}{P_k\fn{z}+\tilde wk^2} \left( z+ \frac{R}{6}\right).
\label{Scalar contribution to F and U}
}

The functions $Q_n$ are directly related to the threshold functions $\ell_0^d$ that have been investigated in functional renormalization for many different cutoffs \cite{Wetterich:1991be,Berges:2000ew,Litim:2000ci,Wetterich:2001kra,Pawlowski:2015mlf},
\al{
Q_2\fn{\tilde w}&= 2\ell_0^4 \fn{\tilde w}k^4 = \frac{k^4}{1+\tilde w}\, ,\nn[2ex]
Q_1\fn{\tilde w}&=2\ell_0^2\fn{\tilde w}k^2 =\frac{2k^2}{1+\tilde w}\, ,
\label{threshold functions}
}
where the second identity uses the specific Litim cutoff \cite{Litim:2001up}.
One infers
\al{
&\p_t \int_x \sqrt{g}\left( U -\frac{F}{2}R\right)=\pi_k^{(S)}\nn
&\qquad=
\frac{1}{16\pi^2} \int_x\sqrt{g} \left( k^4 \ell_0^4 \fn{\tilde w} +\frac{1}{6} k^2 \ell _0^2\fn{\tilde w}R\right)\,,
\label{pi_S}
}
For $R=0$ this yields the flow equation for the effective potential
\al{
\p_t U=\frac{k^4}{16\pi^2}\ell_0^4\fn{\tilde m^2} =\frac{k^4}{32\pi^2(1+\tilde m^2)}\, ,
\label{flow of U}
}
with 
\al{
\tilde m^2=\frac{m^2}{k^2}\,.
}
One recovers the standard flow for a scalar model~\cite{Wetterich:1992yh,Wetterich:1991be,Wetterich:2001kra}.

For the flow of the coefficient of the curvature scalar $F$ we expand in linear order in $R$
\al{
\ell_0^4\fn{\tilde w}=\ell_0^4\fn{\tilde m^2}+\frac{1}{2}\tilde\xi \ell_1^4\fn{\tilde m^2}\frac{R}{k^2}\,,
}
with
\al{
&\ell_1^d\fn{\tilde w}=-\frac{\p \ell_0^d\fn{\tilde w}}{\p\tilde w}\,,&
&\ell_1^4\fn{\tilde m^2}=\frac{1}{2(1+\tilde m^2)^2}\,,&
}
where the last identity applies for the Litim cutoff.
The flow equation for $F$ therefore receives an additional contribution $\sim \tilde\xi$,
\al{
\p_t F&=-\frac{k^2}{48\pi^2}\ell_0^2\fn{\tilde m^2}-\frac{k^2\tilde\xi}{16\pi^2}\ell_1^4\fn{\tilde m^2}\nn
&=-\frac{k^2}{48\pi^2 (1+\tilde m^2)}-\frac{k^2\tilde\xi}{32\pi^2(1+\tilde m^2)^2}\,.
\label{flow of F}
}
These results agree with a computation for a fixed background geometry~\cite{Merzlikin:2017zan}.
For $\tilde\xi=0$ the result agrees with Refs.\,\cite{Dou:1997fg,Codello:2008vh,Narain:2009fy,Dona:2013qba,Percacci:2015wwa,Labus:2015ska}.

For the dimensionless functions and field variables
\al{
u&= \frac{U}{k^4}\,,&
w&=\frac{F}{2k^2}\,,&
\tilde \rho&= \frac{\rho}{k^2}=\frac{\varphi^2}{2k^2}\,,&
\label{dimensionless functions}
}
we obtain
\al{
\p_t u\fn{\tilde \rho}&=-4u +2\tilde \rho\, \p_{\tilde\rho} u +4c_V\,, \nn[2ex]
\p_t w\fn{\tilde \rho}&=-2w +2\tilde \rho \, \p_{\tilde\rho} w +2c_M\,,
\label{flow equations}
}
with scalar contributions to $c_V$ and $c_M$
\al{
c^{(S)}_V&= \frac{1}{128\pi^2(1+\tilde m^2)}\,,\nn
c^{(S)}_M&=-\frac{1}{192\pi^2(1+\tilde m^2)}-\frac{\tilde \xi}{128\pi^2(1+\tilde m^2)^2}\,.
\label{cV and cM with xi}
}
Here we have switched from fixed $\rho$ for Eqs.\,\eqref{flow of U} and \eqref{flow of F} to fixed $\tilde \rho$, and we assume a discrete symmetry $\varphi\to -\varphi$ such that $U$ and $F$ depend only on $\rho$, with 
\al{
\tilde m^2=\p_{\tilde \rho} u +2\tilde \rho\, \p_{\tilde \rho}^2 u=u'+2\tilde \rho u''\, .
}

The flow equations \eqref{flow equations} have a fixed point or scaling solution with $\tilde \rho$-independent $u\fn{\tilde \rho}=u_*$
 and $w\fn{\tilde \rho}=w_*$,
\al{
u_*&= c_V\fn{0},&
w_*&=c_M\fn{0},&
}
where $c_V$ and $c_M$ are evaluated for $\tilde m^2=0$, $\tilde\xi=0$. 
Indeed, for constant $u$ and $w$ one has $\tilde w=0$.
This fixed point occurs for negative $w_*$ or a negative squared running Planck mass $M_\text{p}^2\fn{k}=2w_*k^2$.
It does not correspond to stable gravity and is therefore not acceptable for the definition of quantum gravity.

One may alternatively extract the flow of the Planck mass and other aspects of the gravitational effective action from the flow of the graviton propagator.
This is discussed in Appendix\,\ref{Flow of the graviton propagator appendix}.
For the gauge invariant flow equation one finds the same results as for the heat kernel expansion.
This differs from computations for which the IR-cutoff $R_k$ does not involve the macroscopic metric through covariant derivatives, but rather only involves momenta in flat space, or covariant derivatives with a fixed background metric.
A comparison of our gauge invariant approach with the background formalism can be found in Appendix\,\ref{Flow of the graviton propagator appendix}.

\subsection{Gauge bosons}
\label{gauge bosons section}
We next consider the flow of $F$ and $U$ induced by the fluctuations of gauge bosons.
We employ the gauge invariant flow equation~\cite{Wetterich:2016ewc,Wetterich:2017aoy}.
The connection to a gauge fixed version in the background field formulation for the Landau gauge is similar to the case of gravity discussed in Sec.\,\ref{description on gauge invariant flow}.
The contribution of the gauge boson fluctuations to the flow of $U$ and $F$ involves again a physical part $\pi_k^{(V)}$ and a universal measure part $-\delta_k^{(V)}$, which is a fixed functional of the metric and gauge fields,
\al{
\p_t \Gamma_k =\zeta_k^{(V)}=\pi_k^{(V)}-\delta_k^{(V)}\,.
\label{flow equation for Vector}
} 
The measure factor arises from the nonlinearity of the constraint for the physical gauge boson fluctuations.
The physical part $\pi_k^{(V)}$ depends on the gauge invariant effective action $\bar\Gamma_k^{(V)}$ for the gauge bosons.

We concentrate on a single gauge boson of an Abelian U(1)-gauge theory ---a photon coupled to gravity.
This is sufficient for the flow of $U$ and $F$.
Both are evaluated for zero macroscopic gauge fields.
We first assume a fixed point with a vanishing gauge coupling.
Nonvanishing gauge couplings are discussed in Sec.\,\ref{gauge and scalar contributions}.
For a vanishing gauge coupling any Abelian or non-Abelian gauge theory with a total of $N_V$ gauge bosons gives the same contribution as $N_V$ photons.
The measure term $-\delta_k^{(V)}$ only appears due to the metric dependence of the covariant derivatives in the projection on physical gauge boson fluctuations.

For the truncation of the gauge invariant effective action $\bar\Gamma_k^{(V)}$ we consider here the minimal kinetic term
\al{
\bar \Gamma_{k}^{(V)}= \frac{1}{4e^2}\int_x\sqrt{g}\,F_{\mu\nu}F^{\mu\nu}\,,
\label{gauge inv gauge field action}
}
with field strength
\al{
F_{\mu\nu}=D_\mu A_\nu -D_\nu A_\mu =\p_\mu A_\nu-\p_\nu A_\mu\, ,
}
and gauge coupling $e$.
Covariant derivatives involve the macroscopic metric, while they are independent of gauge fields in the case of an Abelian gauge symmetry.

For the photon field the physical degrees of freedom are the transverse field, where the longitudinal field is a gauge mode.
We introduce projection operators
\al{
&\left(P_\text{L} \right)_\mu{}^{\nu}=D_\mu D^{-2} D^\nu \,,&
&\left(P_\text{T}\right)_\mu{}^{\nu}=\delta_\mu^\nu-(P_\text{L})_\mu{}^{\nu}\,,&
}
which obey [$\left(P_\text{L}^2 \right)_\mu{}^{\nu}=\left(P_\text{L} \right)_\mu{}^{\rho}\left(P_\text{L} \right)_\rho{}^{\nu}$, etc.]
\al{
&P_\text{L}^2=P_\text{L},&
&P_\text{T}^2=P_\text{T},&
&P_\text{L}P_\text{T}=P_\text{T}P_\text{L}=0.&
}
The corresponding transversal and longitudinal gauge fluctuations are
\al{
&A_\mu^\text{T}=\left(P_\text{T}\right)_\mu{}^{\nu} A_\nu,&
&A_\mu^\text{L}=\left(P_\text{L}\right)_\mu{}^{\nu} A_\nu,&
&A_\mu= A_\mu^\text{T} +A_\mu^\text{L}.&
}
Since we evaluate the flow of the effective action for vanishing macroscopic gauge fields, we make here no difference between the gauge fields and the fluctuations of gauge fields around a macroscopic field.
For Yang-Mills theories the projection of infinitesimal physical fluctuations remains well defined for arbitrary macroscopic gauge fields, while no global definition of a physical field exists~\cite{Wetterich:2016ewc,Wetterich:2017aoy}.

The longitudinal gauge field can be written as the derivative of a scalar
\al{
&A_\mu^\text{L}=\p_\mu {a} =D_\mu {a},&
&\left(P_\text{T}\right)_\mu{}^{\nu}D_\nu {a}=0 .&
\label{longitudinal mode written as a scalar field}
}
Under an infinitesimal gauge transformation $A_\mu \to A_\mu +\delta A_\mu $, $\delta A_\mu =\p_\mu {b}$ the transversal gauge field is invariant, while the longitudinal gauge field transforms as 
\al{
&\delta A_\mu^\text{L}=\p_\mu {b},&
&\delta {a}={b},&
&\delta A_\mu ^\text{T}=0.&
}
We can therefore consider $A_\mu ^\text{L}$ as a gauge degree of freedom, while $A_\mu^\text{T}$ is a gauge invariant ``physical degree of freedom.''

By partial integration we can write 
\al{
\bar \Gamma_{k}^{(V)} = \frac{1}{2e^2} \int_x\sqrt{g}\,A_\mu (\mathcal D^{\mu\nu}+D^\mu D^\nu) A_\nu
}
with
\al{
\mathcal D^{\mu\nu}=-D^2 g^{\mu\nu} +R^{\mu\nu}\,.
}
The part involving the transversal gauge fields therefore reads
\al{
\bar\Gamma_{k}^{(V)}=\frac{1}{2e^2}\int_x \sqrt{g}\,A_\mu^\text{T} \mathcal D_T ^{\mu\nu}A_\nu^\text{T}\, ,
}
Indeed, we can project ${\mathcal D}_\mu{}^{\nu}={\mathcal D}$ on the subspace of transversal and longitudinal fluctuations,  
\al{
(\mathcal D_\text{T})_\mu{}^{\nu}&=\left( P_\text{T} \mathcal D P_\text{T}\right)_\mu{}^{\nu}=-D^2 \delta_\mu^\nu +R_\mu{}^{\nu}+D_\mu D^\nu\, ,\nn[2ex]
(\mathcal D_\text{L})_\mu{}^{\nu}&=\left( P_\text{L} \mathcal D P_\text{L}\right)_\mu{}^{\nu}=-D_\mu D^\nu\, .
}
Even though the projectors $P_\text{T}$ and $P_\text{L}$ are nonlocal, the projected operators $\mathcal D_\text{T}$
 and $\mathcal D_\text{L}$ involve only two derivatives.
One has 
\al{
&\mathcal D=\mathcal D_\text{T}+\mathcal D_\text{L}\,,\qquad
P_\text{T}\mathcal D P_\text{L}= P_\text{L}\mathcal  D P_\text{T}=0\,,
\label{projection properties of derivatives} \\[2ex]
&\qquad\qquad
\mathcal D_\text{L}\mathcal D_\text{T}=\mathcal D_\text{T}\mathcal D_\text{L}=0\,.\nonumber
}

The computation of $\pi_k^{(V)}$ from a gauge invariant flow equation involves the projector $P_T$ similar to Eq.\,\eqref{propagator and Gamma2}.
We employ again the trick of first computing $\pi_k^{(V)}+\delta_k^{(V)}$ by adding to the action a formal gauge fixing term
\al{
\Gamma^{(V)}_\text{gf}=\frac{1}{2\alpha}\int_x \sqrt{g}\,(D_\mu A^\mu)^2\,.
\label{gauge fixing for Yang-Mills field}
}
We need to take the limit $\alpha\to0$, such that Eq.\,\eqref{gauge fixing for Yang-Mills field} corresponds to Landau gauge fixing.
By partial integration the gauge fixing part reads
\al{
\Gamma^{(V)}_\text{gf}=-\frac{1}{2\alpha} \int_x\sqrt{g}\,A_\mu D^\mu D^\nu A_\nu \,,
}
or
\al{
\Gamma^{(V)}_\text{gf}=\frac{1}{2{\alpha}}\int_x \sqrt{g}\,A_\mu^\text{L} \mathcal D_L ^{\mu\nu}A_\nu^\text{L}\, . 
}
As it should be for a physical gauge fixing, $\Gamma_\text{gf}$ involves only the gauge degree of freedom $A_\mu^\text{L}$.
In contrast, for our ansatz \eqref{gauge inv gauge field action} $\bar \Gamma$ depends only on the gauge invariant field $A_\mu ^\text{T}$. 

If we define $\Gamma_k^{(V)}=\bar\Gamma_k^{(V)}+\Gamma_\text{gf}^{(V)}$, the second functional derivative is block diagonal in the transverse and longitudinal fields,
\al{
\Gamma_{k}^{(V,2)}&=\frac{\sqrt{g}}{e^2}P_\text{T}\mathcal D P_\text{T}+\frac{\sqrt{g}}{\alpha}P_\text{L}\mathcal D P_\text{L}\nn
&=\sqrt{g}\left( \frac{1}{e^2}\mathcal D_\text{T} +\frac{1}{\alpha} \mathcal D_\text{L}\right)\,.
}
In consequence, also its inverse or the propagator is block diagonal for $\alpha\to0$, with a block for the gauge modes $\sim\alpha$.
For the particular case of a maximally symmetric space, 
\al{
R_{\mu\nu}=\frac{1}{4}R g_{\mu\nu}\,,
}
one has
\al{
\mathcal D^{\mu\nu}=\left( -D^2 +\frac{R}{4}\right) g^{\mu\nu}\,.
\label{covariant derivative for gauge field}
}

We introduce the infrared cutoff function $R_k$ such that
\al{
\Gamma_{k}^{(V,2)}+{\mathcal R}_k =\sqrt{g} \left( \frac{1}{e^2}P_k\fn{\mathcal D_\text{T}}+\frac{1}{\alpha} P_k\fn{\mathcal D_\text{L}}\right) \, ,
}
with
\al{
P_k\fn{z}=z+R_k\fn{z}.
}
One infers for the sum
\al{
\pi_k^{(V)}+\delta_k^{(V)}=\frac{1}{2}\text{tr}\left\{ \left( \Gamma_k^{(V,2)}+{\mathcal R}_k \right)^{-1}\p_t {\mathcal R}_k\right\}\,.
\label{flow of gauge thories}
} 
The part $\delta_k^{(V)}$ is connected to the gauge fluctuations, and the separate contributions are given by
\al{
&\pi_k^{(V)}=\frac{1}{2}\text{tr}\,W\fn{\mathcal D_\text{T}}\, ,&
&\delta_k^{(V)}=\frac{1}{2}\text{tr}\, W\fn{\mathcal D_\text{L}}\,,&\nonumber
}
with 
\al{
W\fn{z}=\p_t R_k\fn{z} \left( z +R_k\fn{z}\right)^{-1}\,.
}
The quantity $\delta_k$ also determines the measure contribution in Eq.\,\eqref{flow equation for Vector}.

Let us first evaluate the measure contribution $\delta_k^{(V)}$.
Writing the longitudinal gauge field as the derivative of a scalar \eqref{longitudinal mode written as a scalar field}, one has 
\al{
(\mathcal D_\text{L})_\mu{}^{\nu}A_\nu^\text{L}=(\mathcal D_\text{L})_\mu{}^{\nu}D_\nu a=D_\mu (-D^2)a\,.
}
The eigenvalues of the (negative) scalar Laplacian $\Delta_S=-D^2$ acting on $a$ are the eigenvalues of $\mathcal D_\text{L}$ acting on $A_\mu^\text{L}$, implying 
\al{
\delta_k^{(V)}=\frac{1}{2}\text{tr}_{(0)}\,W\fn{\Delta_S},
\label{measure contribution of gauge mode in Yang-Mills theory}
}
with $\text{tr}_{(0)}$ the trace over scalar fields.
Therefore the flow contribution $\delta_k^{(V)}$ is the same as the one for a massless scalar
\al{
\delta_k^{(V)}=\frac{1}{16\pi^2}\int_x\sqrt{g}\left( \ell_0^4k^4 +\frac{1}{6}\ell_0^2 k^2 R\right).
\label{deltaV}
}

We next turn to the contribution $\pi_k^{(V)}$ from the physical gauge boson fluctuations,
\al{
&\pi_k^{(V)}=\frac{1}{2}\text{tr}_{(1)}\,W\fn{\mathcal D}\,,&
&(\mathcal D)_\mu{}^{\nu}=\Delta_V\delta^\nu_\mu+R_\mu^\nu\,.
}
Here $\text{tr}_{(1)}$ is the trace over transverse vector fields (spin one) and $\Delta_V=-D^2$ the (negative) Laplacian acting on vector fields.
We will evaluate the trace \eqref{measure contribution of gauge mode in Yang-Mills theory} again by the heat kernel method; see Appendix\,\ref{heat kernel methods}.

For the evaluation of $\text{tr}\,\exp\fn{-s\mathcal D_1}$ we employ the fact that the result is a series of integrals over local terms that are invariant under diffeomorphism transformations.
Dimensional analysis implies $c_2=b_2 R$.
For the computation of $b_2$ we can choose the geometry of a maximally symmetric space with constant $R$, for which Eq.\,\eqref{covariant derivative for gauge field} implies
\al{
\text{tr}_{(1)}\,e^{-s\mathcal D}&=\text{tr}_{(1)}\,e^{-s(\Delta_V+R/4)}
=\left( \text{tr}_\text{(1)}\,e^{-s\Delta_V}\right)\left( e^{-sR/4}\right)\nn
&=\left( \text{tr}_{(1)}\, e^{-s\Delta_{V}}\right) \left( 1-\frac{sR}{4}+\cdots \right)\, .
}
From
\al{
\text{tr}_{(1)}\,e^{-s\Delta _{V}}=\frac{1}{16\pi^2}\int_x\sqrt{g}\left( b_0^{V} s^{-2} +b_2^{V} R s^{-1}+\cdots \right),
}
one infers
\al{
&\text{tr}_{(1)}\, e^{-s\mathcal D_1}\nn
&\quad=\frac{1}{16\pi^2} \int_x\sqrt{g}
 \left[ b_0^{V} s^{-2}+\left( b_2^{V}-\frac{b_0^{V}}{4}\right)R s^{-1}+\cdots \right].
}
With
\al{
b_0^{V}&=3\,,&
b_2^{V}&=\frac{1}{4}\,,&
b_2^{V}-\frac{b_0^{V}}{4}&=-\frac{1}{2}\,,&
}
one obtains a negative contribution of the term $\sim R$.

Employing again the relation between the function $Q_n$ and the threshold functions, the flow contribution from the transverse vector fields is
\al{
\pi_k^{(V)}=\frac{1}{16\pi^2}\int_x\sqrt{g}\left(3\ell_0^4k^4-\frac{1}{2}\ell_0^2k^2 R\right),
}
with threshold functions evaluated for $\tilde w=0$.
We will see in Sec.\,\ref{gauge and scalar contributions} that in the case of a nonvanishing gauge coupling $\tilde w$ will depend on $\tilde\rho$, i.e., $\tilde w=e^2\tilde\rho$.
Then also $\pi_k^{(V)}$ will depend on $\tilde\rho$, while for $\delta_k^{(V)}$ the threshold functions remain evaluated at $\tilde w=0$.

Taking terms together the flow contribution of a massless gauge field is 
\al{
&\p_t \int_x \sqrt{g}\left( U-\frac{F}{2}R\right) =\zeta_k^{(V)} \nn
&\qquad =\frac{1}{16\pi^2} \int\sqrt{g}\left( 2k^4 \ell_0^4-\frac{2k^2}{3}\ell_0^2 R\right)\,,
}
or
\al{
&\p_t U=\frac{k^4 \ell_0^4}{8\pi^2}\,,&
&\p_t F=\frac{k^2 \ell_0^2}{12\pi^2}\,.&
\label{beta functions for U and F}
}
For $N_V$ gauge fields one has in Eq.\,\eqref{flow equations}, using the Litim cutoff \eqref{threshold functions},
\al{
&c^{(V)}_V=\frac{N_V \ell^4_0}{32\pi^2}=\frac{N_V}{64\pi^2}\,,&
&c^{(V)}_M=\frac{N_V \ell^2_0}{48\pi^2}=\frac{N_V}{48\pi^2}\,.&
\label{vector matter contributions}
}
Our final result is very simple.
It agrees with Refs.\,\cite{Dou:1997fg,Codello:2008vh,Dona:2013qba}.

The flow equations \eqref{beta functions for U and F}, translated to the dimensionless quantities $u$ and $w$ in Eq.\,\eqref{dimensionless functions}, admit a simple fixed point with $\tilde \rho$-independent $u$ and $w$
\al{
&u_*=c_V\,,&
&w_*=c_M\,.&
}
This fixed point occurs now for a positive running Planck mass $M_\text{p}^2\fn{k}=2c_M k^2$.
In contrast to scalar fluctuations, a fixed point induced by $N_V$ massless vector boson fluctuations leads to stable gravity.
A UV-fixed point of this type can be used to define quantum gravity as an asymptotically safe quantum field theory.
A measure for the strength of gravity is $w^{-1}$.
In the limit $N_V\to \infty$ one finds that $w^{-1}$ tends to zero such that gravity is weak.
This may allow for a valid perturbative expansion in $w^{-1}$.

The positive sign of $c_M$ is a crucial ingredient.
This property should not depend on the precise implementation of the IR cutoff.
The threshold function $\ell_0^2$ is positive for every cutoff function with positive $\p_tR_k$.
We may also consider a different implementation with $R_k$ depending only on $-D^2$ instead of $\mathcal D_\text{T}$ and $\mathcal D_\text{L}$.
This is investigated in Appendix\,\ref{IR cutoff scheme}.
It leads to a similar result, with threshold function $\ell_0^2$ replaced by a different threshold function $\ell_1^4$.
When the so-called type I is employed, one should replace $N_V\to 7N_V/16\approx 0.438N_V$ for $c_M^{(V)}$ given in Eq.\,\eqref{vector matter contributions}.

\subsection{Gauge bosons coupled to scalars}
\label{gauge and scalar contributions}
We next investigate the role of a possible nonzero gauge coupling for the properties of the fixed point.
For this purpose we consider scalar quantum electrodynamics with a complex scalar coupled to a gauge boson.
Because of the U(1)-gauge symmetry $U$ and $F$ can depend only on $\rho=\varphi^*\varphi$.
We normalize the gauge coupling $e$ such that the gauge boson mass is given by $m_g^2=e^2\rho$.
The contribution \eqref{cV and cM with xi} to the flow from the two real scalar fields is
\al{
c_V^{(S)}&=\frac{1}{128\pi^2}\left( \frac{1}{1+u'+2\tilde \rho u''}+\frac{1}{1+u'}\right),\nn[2ex]
c_M^{(S)}&=-\frac{1}{192\pi^2}\left( \frac{1}{1+u'+2\tilde \rho u''}+\frac{1}{1+u'}\right),
\label{cV and cM}
}
The first term in these equations arises from the ``radial mode'' with $\tilde m^2=u'+2\tilde\rho u''$, while the second corresponds to the ``Goldstone mode'' with $\tilde m^2=u'$.
In Eq.\,\eqref{cV and cM} we have omitted a nonminimal coupling $\tilde\xi$ to gravity, which would give an additional contribution to $c_M^{(S)}$.
The contributions from scalar fluctuations are not modified in the presence of a nonzero gauge coupling $e$.

The contribution from the physical gauge boson fluctuations $\pi_k^{(V)}$ is sensitive to $e^2\neq0$ through a different argument $\tilde w$ of the threshold functions. 
Indeed, a nonzero $\rho$ induces a mass term for the transverse gauge boson fluctuations, resulting in $\tilde w=e^2\tilde \rho$.
As a consequence, the threshold functions $\ell_0^4$ and $\ell_0^2$ in Eq.\,\eqref{beta functions for U and F} are now functions of $\tilde w=e^2\tilde \rho$, instead of being evaluated at $\tilde w=0$.

For the gauge invariant flow equation this is the only change.
Because of the projection on the physical fluctuations there are no mixing effects.
The transverse gauge bosons are vectors that cannot mix with scalars for any rotation invariant geometry.

A similar behavior is found for a gauge fixed background field approach.
In this case the physical gauge fixing (Landau gauge) is crucial for the absence of mixed terms.
The gauge invariant scalar kinetic term can be written in terms of a scalar fluctuation $\chi$ around the macroscopic field $\varphi$,
\al{
\Gamma_\text{kin}=\int_x\sqrt{g}\left[ \left( D^\mu -iA^\mu \right) (\varphi+\chi)\right]^*\left[ \left( D_\mu -iA_\mu \right) (\varphi+\chi)\right].
}
For $\Gamma^{(2)}$ we need the terms quadratic in $A_\mu$ and $\chi$,
\al{
\Gamma_\text{kin,2}&=\int_x\sqrt{g}\left[ ( D^\mu \chi)^*(D_\mu \chi) +\chi^*\chi A^\mu A_\mu \right] +\Gamma_\text{mix}\,,\nn[2ex]
\Gamma_\text{mix}&=-i\int_x\sqrt{g}\,A^\mu \left[ (D_\mu \chi)^*\varphi -\varphi^* D_\mu \chi \right]\,.
}
The mixed term involves only the longitudinal gauge fixed $A_\mu^\text{L}$.
We can take $\varphi$ real, and $\chi=(\chi_\text{R}+i\chi_\text{I})$, such that
\al{
\Gamma_\text{mix}=-2\varphi \int_x\sqrt{g}\,A^\mu D_\mu\chi_\text{I}
=2\varphi\int_x\sqrt{g}\,\chi_\text{I}  D_\mu A^\mu\,.
}
Because of the divergence of $\Gamma_k^{(2)}+\mathcal R_k$ in the longitudinal sector $\sim \alpha^{-1}$, the effect of the mixed term vanishes in the propagator $\left(\Gamma^{(2)}_k+\mathcal R_k\right)^{-1}$ proportional to $\alpha$.
It does not contribute to the flow equation.

In consequence of the absence of mixing and the simple change of argument in the photon threshold function, one obtains rather simple flow equations.
They read for a Litim cutoff  
\al{
\p_t u&= -4u+2\tilde \rho\, \p_{\tilde \rho} u+4c_V\, ,\nn[2ex]
\p_t w&= -2w+2\tilde \rho\, \p_{\tilde \rho} w+2c_M\,,
}
with
\al{
c_V&=\frac{1}{128\pi^2}\left( \frac{3}{1+e^2\tilde \rho}-1 +\frac{1}{1+u'}+\frac{1}{1+u'+2\tilde \rho u''}\right),\nn[2ex]
c_M&=\frac{1}{192\pi^2}\left( \frac{3}{1+e^2\tilde \rho}+1-\frac{1}{1+u'}-\frac{1}{1+u'+2\tilde \rho u''}\right).
\label{c_M with scalar with gauge boson}
}
This generalizes to the case of several vector bosons and more complicated scalar sectors.
The dependence on the gauge coupling arises only from the physical gauge boson fluctuations.
For each gauge boson one needs the field-dependent squared mass or the corresponding eigenvalue of the mass matrix.

The qualitative effects of nonzero $e^2$ can already be seen for the Abelian case of a single photon with $\tilde w=e^2\tilde \rho$.
We concentrate on the effects for the properties of the scaling solution.
For the scaling solution $\p_t u$ and $\p_t w$ vanish.
The scaling solution therefore obeys the differential equations
\al{
&\tilde \rho\,\p_{\tilde \rho} u=2(u-c_V),&
&\tilde \rho\,\p_{\tilde \rho} w=w-c_M,&
\label{scaling solution equations}
}
For $e^2\neq 0$ both $c_V$ and $c_M$ depend explicitly on $\tilde \rho$.
This has an immediate important consequence: a $\tilde \rho$-independent $u$ and $w$ no longer solve Eq.\,\eqref{scaling solution equations}.

For $e^2\neq0$ a possible UV-fixed point is characterized by $\tilde\rho$-dependent scaling solutions of Eq.\,\eqref{scaling solution equations}.
We briefly discuss here the limits $\tilde\rho\to0$ and $\tilde\rho\to\infty$.
For $\tilde \rho=0$ the solutions of the flow equation have to obey
\al{
u_*\fn{0}&=c_V\fn{0}=\frac{1}{64\pi^2}\left( 1+\frac{1}{1+\tilde m_0^2}\right)\,,\nn[2ex]
w_*\fn{0}&=c_M\fn{0}= \frac{1}{96\pi^2}\left( 2-\frac{1}{1+\tilde m_0^2}\right)\,,
} 
with
\al{
\tilde m_0^2=u_*'\fn{0}.
}
One observes positive $w_*\fn{0}$ and $u_*\fn{0}$.
These values do not depend on $e^2$.

On the other hand, for $\tilde \rho\to \infty$ the gauge boson contribution vanishes if $e^2\neq 0$.
In this limit $c_M$ becomes negative.
A possible asymptotic solution for $\tilde \rho\to \infty$ is
\al{
&w\to \xi_\infty \tilde \rho\,,&
&u\to \frac{1}{2}\lambda_\infty \tilde \rho^2\,. 
}
With $u'\fn{\tilde \rho\to \infty}=\lambda_\infty \tilde \rho$ both $c_V$ and $c_M$ vanish in this limit if $\lambda_\infty>0$.

We may investigate directly the flow of the $\tilde\rho$ dependence of the functions $u(\tilde\rho)$ and $w(\tilde\rho)$.
For this purpose we define
\al{
\tilde m^2\fn{\tilde \rho}&=u'\fn{\tilde\rho}\,, &
\xi\fn{\tilde \rho}&=\frac{1}{2} w'\fn{\tilde \rho}\,,&
\lambda\fn{\tilde \rho}&=u''\fn{\tilde \rho}\,.&
}
(The factor $1/2$ in the definition of $\xi$ is chosen such that for constant $\xi$ one has $F=2w_0k^2+\xi \rho$.)
The flow equation for the dimensionless scalar mass term $\tilde m^2\fn{\tilde \rho}$ reads
\al{
\p_t \tilde m^2=-2\tilde m^2+2\tilde \rho \, \p_{\tilde \rho}\tilde m^2+4\p_{\tilde \rho}c_V\,,
}
where
\al{
\p_{\tilde \rho}c_V&=-\frac{1}{128\pi^2}\bigg( \frac{3e^2}{(1+e^2\tilde\rho)^2}+\frac{\lambda}{(1+\tilde m^2)^2}\nn[1ex]
&\qquad\qquad\qquad+\frac{3\lambda +2\tilde \rho u'''}{(1+\tilde m^2 +2\lambda \tilde \rho)^2}\bigg).
}
The fixed point value for $\tilde m_0^2=\tilde m^2\fn{\tilde \rho=0}$ is negative
\al{
\tilde m_0^2=-\frac{1}{64\pi^2}\left(3e^2 +\frac{4\lambda_0}{(1+\tilde m_0^2)^2}\right).
}
If $u(\tilde\rho)$ increases for $\tilde\rho\to\infty$ or goes to a constant, one expects a minimum of $u$ at some $\tilde\rho_0>0$, indicating spontaneous symmetry breaking for the scaling solution.

For the flow equation for $\xi$ one obtains
\al{
\p_t \xi=2\tilde \rho\, \p_{\tilde \rho} \xi+2\p_{\tilde \rho}c_M\,,
}
with
\al{
\p_{\tilde \rho}c_M&=-\frac{1}{192\pi^2}\bigg( \frac{3e^2}{(1+e^2\tilde\rho)^2}-\frac{\lambda}{(1+\tilde m^2)^2}\nn[1ex]
&\qquad\qquad\qquad-\frac{3\lambda +2\tilde \rho u'''}{(1+\tilde m^2 +2\lambda \tilde \rho)^2}\bigg).
}
For the scaling solution $\xi\fn{\tilde \rho\to 0}$ diverges logarithmically
\al{
\xi\fn{\tilde \rho}&=c_1+c_2\ln\, \tilde \rho\,,&
c_2&=-\p_{\tilde \rho}c_M\fn{\tilde \rho=0}\,.&
}
The effective Planck mass remains finite for $\tilde \rho\to 0$
\al{
w=w_0+2(c_1-c_2)\tilde \rho+2c_2\tilde \rho\, \ln\tilde \rho\,.
}
We conclude that a nonvanishing gauge coupling at the fixed point has important consequences for the behavior of the scaling solution. 

In the present paper we will concentrate on a UV-fixed point with $e_*^2=0$.
The constant scaling solutions for the fixed point exist in this case.
We note, however, that $e^2$ has to be a relevant parameter in this case since the extrapolation of the observed gauge couplings to the transition scale $k_t$ where the metric fluctuations decouple indicate nonzero $e^2\fn{k_t}$.
With $e^2$ increasing slowly in the vicinity of the fixed point, our discussion of constant $e^2>0$ may still be relevant for the flow away from the fixed point.
Instead of a fixed point one has in this case an approximate partial fixed point.

\subsection{Fermions}
\label{fermionic contribution}
The gravitational interactions of fermions cannot be described by a direct coupling to the metric.
In the presence of fermions the basic gravitational degree of freedom is the vierbein $e_\mu{}^{m}$.
The metric is subsequently associated with a bilinear of the vierbein, 
\al{
g_{\mu\nu}=e_\mu{}^{m}e_\nu{}^{n}\eta_{mn}\,,
\label{metric by vierbein}
}
with $\eta_{mn}=\text{diag}(-1,1,1,1)$ for the Minkowski signature and $\eta_{mn}=\delta_{mn}$ for the Euclidean signature.
As an alternative, one may always use $\eta_{mn}=\delta_{mn}$ and admit complex values for the vierbein.
The Minkowski signature for flat space is then realized for $e_\mu{}^{m}=\text{diag}(i,1,1,1)$.
Analytic continuation is achieved \cite{Wetterich:2010ni} by varying $e_\mu{}^{m}=(e^{i\phi},1,1,1)$  with $\phi=0$ for the Euclidean signature and $\phi=\pi/2$ for the Minkowski signature.
With Eq.\,\eqref{metric by vierbein} one has $\sqrt{g}=e=\text{det}(e_\mu{}^{m})$, which yields the factor $i$ multiplying the action for Minkowski space.
Our Euclidean notation \cite{Wetterich:2010ni} is adapted to the complex vierbein for which analytic continuation is straightforward even for curved space. 

The second central object is the spin connection $\omega_{\mu mn}$, which can be expressed in the simplest version of gravity by the vierbein and its first derivative 
\al{
\omega_{\mu mn}\fn{e}&=\frac{1}{2}\Big\{ e_m{}^{\rho} e_n{}^{\tau} e_\mu{}^{p}\p_\tau e_{\rho p} -  e_m{}^{\rho}\p_\rho  e_{\mu n}+ e_m{}^{\rho}\p_\mu e_{\rho n} \nn
&\qquad-(m \leftrightarrow n) \Big\}.
\label{spin connection by vierbein}
}
Here $e_m{}^{\mu}$ is the inverse vierbein 
\al{
 e_m{}^{\mu} e_\mu{}^{n}=\delta_m^n\,,
}
where we suppose that $e_\mu{}^{m}$ are elements of a regular matrix.
One can construct a curvature tensor in terms of the vierbein and the spin connection 
\al{
R_{\mu\nu mn}\fn{e,\omega}=\p_\mu \omega_{\nu mn}+\omega_{\mu m p}\omega_{\nu}{}^{p}{}_n-(\mu \leftrightarrow \nu).
}
For $\omega=\omega\fn{e}$ according to Eq.\,\eqref{spin connection by vierbein} this is a function of the vierbein involving two derivatives.
Contraction with the vierbein,
\al{
R_{\mu\nu\rho\tau}=e_\rho{}^{m}e_\tau{}^{n}R_{\mu \nu mn}\,,
}
yields with Eqs.\,\eqref{metric by vierbein} and \eqref{spin connection by vierbein} the curvature tensor of Riemannian geometry formed from $g_{\mu\nu}$.
For a computation of the fermionic contribution to the flow of $U$ and $F$ it is sufficient to investigate the fermion fluctuations in a geometric background.
We first perform this in a background given by the vierbein, and subsequently translate to the metric.

For the effective action for fermions we choose the truncation
\al{
\Gamma_{k}^{(F)}=\int _xe\Big\{ i\bar\psi \gamma^m e_{m}{}^{\mu}D_\mu \psi +y\bar\psi \gamma^5 \psi \varphi \Big\}
}
with $y$ a Yukawa coupling to a complex scalar field $\varphi$.
In our conventions for fermions, $\varphi$ is a scalar (not a pseudoscalar), and a fermion mass term involves $\gamma^5$, ${\mathcal L}_m \sim em\bar\psi \gamma^5 \psi$.
We consider here a Dirac fermion for which $\bar\psi$ and $\psi$ are independent Grassmann variables.
The covariant derivative involves the spin connection
\al{
D_\mu \psi=\p_\mu \psi +\frac{1}{8}\omega_{\mu mn}[\gamma^m,\gamma^n] \psi\,,
}
with Dirac matrices obeying
\al{
&\{ \gamma^m,\gamma^n\} =2\eta^{mn},&
&\{ \gamma^5,\gamma^m\}=0.&
}
For real values of $\varphi$ the second functional derivative reads in the $\bar\psi$-$\psi$ block
\al{
\Gamma^{(F,2)}_{k}=e\left( i\Slash{\mathcal D}+y\varphi \gamma^5\right).
}
It involves the covariant Dirac operator
\al{
\Slash{\mathcal D}=\gamma^me_m{}^{\mu}D_\mu\,.
}

The fermion contribution to the flow equation reads
\al{
\p_t \Gamma_k=-\text{tr}\left\{ \p_tR_k \left( i\Slash{\mathcal D}+ R_k^{(F)}+y\varphi\gamma^5 \right)^{-1} \right\}\,,
\label{flow equation of dirac operator}
}
where the minus sign reflects that $\psi$ and $\bar\psi$ are Grassmann variables.
For a Majorana fermion $\bar\psi$ is related to $\psi$ by charge conjugation.
For Majorana spinors the right-hand side of Eq.\,\eqref{flow equation of dirac operator} is multiplied by $1/2$.
(If convenient,  one may choose a normalization which multiplies $ \Gamma_{k}^{(F)}$ by a factor $1/2$.)
We will perform the computation for a Dirac spinor.
For $N_F$ Majorana or Weyl spinors the result will be multiplied by $N_F/2$.

We can decompose $\psi$ and $\bar\psi$ into Weyl spinors that are eigenstates of $\gamma^5$,
\al{
&\psi_\pm =\frac{1\pm \gamma^5}{2}\psi,&
&\bar\psi_\pm=\bar\psi \frac{1\mp \gamma^5}{2}.&
}
(One often uses left- and right-handed $\psi_+\equiv \psi_\text{L}$ and $\psi_-\equiv \psi_\text{R}$.)
The kinetic term does not mix different Weyl spinors,
\al{
\mathcal L_\text{kin}=ie\bar\psi \Slash{\mathcal D}\psi=-ie \left( \bar\psi_+\Slash{\mathcal D}\psi_+ +\bar\psi_-\Slash{\mathcal D}\psi_-\right)\,,
}
while the Yukawa coupling does,
\al{
\mathcal L_Y=ey \varphi \bar\psi  \gamma^5 \psi=ey\varphi\left( \bar\psi_-\psi_+-\bar\psi_+\psi_-\right)\,.
}

Chiral symmetries transform $\psi_+$ and $\psi_-$ independently.
They are violated for $y\neq0$.
We want a cutoff function that is compatible with the chiral symmetry~\cite{Wetterich:1990an}.
Since $\Slash{\mathcal D}$ maps $\psi_+$ to $\psi_-$ and vice versa, the cutoff should be chosen in the form 
\al{
R_k^{(F)}=ies_k(-{\Slash{\mathcal D}^2})\Slash{\mathcal D}\,,
\label{cutoff for fermion}
} 
such that it has the same chiral properties as $ie\Slash{\mathcal D}$.
In consequence,
\al{
P_k^{(F)}=ie\Slash{\mathcal D}+R_k^{(F)}=ie\left( 1+s_k(-\Slash{\mathcal D}^2)\right) \Slash{\mathcal D}
}
maps again $\psi_+$ to $\psi_-$.
We emphasize that $s_k$ should be a function of the operator $\Slash{\mathcal D}$ since we want to cut off small eigenvalues of the Dirac operator.
This function should be even for compatibility with chiral symmetry.

Employing the properties of Dirac matrices one has
\al{
-\Slash{\mathcal D}^2=-D^2+\frac{R}{4}\,,
}
with 
\al{
D^2=e_m{}^{\mu}e^{m\nu}D_\mu D_\nu \,.
}
The covariant Laplacian for spinors depends on the vierbein and spin connection and involves Dirac matrices.
(No Laplacian involving the metric is defined for spinors.)
With 
\al{
\pi_k^{(F)}&=-\text{tr}\,\tilde \p_t\ln\fn{P_k^{(F)} +ey\varphi \gamma^5}\nn
&=-\frac{1}{2}\text{tr}\,\tilde \p_t\ln\fn{P_k^{(F)} +ey\varphi \gamma^5}^2\nn
&=-\frac{1}{2}\text{tr}\,\tilde \p_t\ln\fn{ \left(P_k^{(F)}\right)^2 +e^2y^2\varphi^2}\,,
\label{pi F contribution}
}
and defining 
\al{
P_k=\frac{1}{e^2}\left(P_k^{(F)}\right)^2
&=-\Slash{\mathcal D}^2\left( 1+s_k(-\Slash{\mathcal D}^2) \right)^2\nn
&=-\Slash{\mathcal D}^2\left( 1+r_k(-\Slash{\mathcal D}^2)\right)\,,
}
one has
\al{
r_k=2s_k+s_k^2\,.
\label{identity for fermion cutoff}
}

Using $\rho=\varphi^2$ the fermion contribution is given by the familiar form
\al{
\pi_k^{(F)}=-\frac{1}{2}\text{tr}\,\tilde \p_t \ln\fn{P_k+y^2 \rho},
}
where
\al{
P_k=\left( -D^2+\frac{R}{4}\right) \left( 1+r_k\fn{-D^2+\frac{R}{4}}\right).
}
For complex $\varphi$ this generalizes to $\rho=\varphi^\dagger \varphi$, and similarly for multicomponent scalar fields.
For $P_k$ we will choose a function similar to the one for scalars or gauge bosons.
This determines the infrared cutoff function \eqref{cutoff for fermion} via the identity \eqref{identity for fermion cutoff}.

In consequence, one obtains from the heat kernel expansion for a Dirac fermion
\al{
\pi_k^{(F)}=-\frac{1}{16\pi^2}\int_xe \left\{ b_0^D \ell_0^4\fn{\tilde w}k^4+\left( b_2^D-\frac{1}{4}b_0^D \right) \ell_0^2\fn{\tilde w}R\right\}
\label{physical fermionic fluctuation}
}
with $b_0^D=4$, $b_2^D=2/3$, and 
\al{
\tilde w=\frac{y^2 \rho}{k^2}.
}
For Majorana or Weyl spinors $\pi_F$ is divided by two.

With Eqs.\,\eqref{metric by vierbein} and \eqref{spin connection by vierbein}, we translate $e=\sqrt{g}$ and $R$ becomes the curvature scalar for the metric $g_{\mu\nu}$.
For $y=0$ Eq.\,\eqref{physical fermionic fluctuation} agrees with Ref.\,\cite{Larsen:1995ax} and with the result of a ``type-II cutoff"~\cite{Percacci:2005wu,Codello:2008vh,Dona:2012am}.
The same result is obtained by a spectral sum on a sphere~\cite{Dona:2012am}.
The sign differs, however, from a ``type-I cutoff"~\cite{Dou:1997fg}.
We emphasize that the formulation of the cutoff in terms of the Dirac operator involving the vierbein, together with the preservation of chiral symmetry, imposes important constraints on the properties of the infrared cutoff. 
They lead naturally to a type-II cutoff.

With $N_F$ Weyl spinors one finds, with $\tilde \rho=\rho/k^2$,
\al{
c_V^{(F)}&=-\frac{N_F}{32\pi^2}\ell_0^4\fn{y^2\tilde \rho}=-\frac{N_F}{64\pi^2(1+y^2\tilde \rho)}\,,\nn[2ex]
c_M^{(F)}&=-\frac{N_F}{192\pi^2}\ell_0^2\fn{y^2\tilde \rho}=-\frac{N_F}{192\pi^2(1+y^2\tilde \rho)}\, .
\label{cV  and cM from gauge scalar}
}
The second identity specifies to the Litim cutoff.
For the flow of $U$ fermions contribute with the opposite sign as compared to scalars and gauge bosons.
For the flow of $F$ their contribution has the same sign as the scalar contribution, opposite to the sign of vector contributions.

For the gauge invariant flow equation one obtains the fermion contribution to the flow of the graviton propagator by taking two derivatives of $\pi_k^{(F)}$ in Eq.\,\eqref{pi F contribution} with respect to the vierbein (or similarly to the metric).
The flow of the graviton propagator can be evaluated in flat space.
The result is expected to agree with Eq.\,\eqref{cV  and cM from gauge scalar}.

The different matter contributions can be summarized by
\al{
c_V&=\frac{\mathcal N_U}{128\pi^2},&
c_M&=\frac{\mathcal N_M}{192\pi^2}.&
\label{cV and cM general form}
}
For massless fields one has
\al{
&\mathcal N_U=N_S+2N_V-2N_F\,,\nn[1ex]
&\mathcal N_M=-N_S+4N_V-N_F-\frac{3\tilde \xi}{2}N_\xi\,.
\label{particle numbers}
}
Here $N_\xi$ is the number of scalars that have the nonminimal coupling $\tilde\xi$ to gravity.
This number may be smaller than $N_S$ since not all scalars may participate in this coupling.
For massive fields the numbers $N_S$, $N_V$ and $N_F$ become effective particle numbers, multiplied by corresponding threshold functions with $\tilde m^2\neq0$.
In this case, $\mathcal N_U$ and $\mathcal N_M$ depend on $\tilde \rho$.

Matter fluctuations alone can induce an UV-fixed point for quantum gravity, $u_*=c_V$, $w_*=c_M$. 
This does not need any metric fluctuations.
We observe, however, that without metric fluctuations an UV-fixed point with positive effective Planck mass ($F>0$, attractive gravity) is possible only for a sufficiently large number of vector bosons, $\mathcal N_M>0$. 
For $N_V>(N_F+N_S)/4$ asymptotic safety can be realized even if fluctuations of the metric are neglected. 
We will see below that metric fluctuations induce an acceptable UV-fixed point with $F>0$ even for $N_V<(N_F+N_S)/4$.

\section{Metric fluctuations}
\label{Metric fluctuation section}
This section addresses the contribution of the metric fluctuations to the flow equations for $U$ and $F$.
The dominant contribution $\pi_2$ arises from the graviton fluctuations, e.g., the traceless transverse tensor modes.
For spaces with rotation symmetry the graviton fluctuations do not mix with the other metric fluctuations or matter fluctuations.
Furthermore, the graviton fluctuations are physical fluctuations, being invariant under gauge transformations.
Once the graviton propagator is known or assumed in a given truncation, the graviton contribution to the flow equations is rather unambiguous.
The other physical metric fluctuation is a scalar.
It can mix with other scalar fields in the matter sector.
Still, its contribution $\pi_0$ to the flow equation remains somewhat involved.
The gauge invariant flow equation has the important advantage that physical scalar fluctuations are not affected by gauge modes.
The universal measure contribution $-\delta^{(g)}$ reflecting the nonlinearity of the projection on the space of physical fluctuations involves both a vector and a scalar part.
It depends only on the macroscopic metric.
The total contribution to the flow from the metric sector is given by
\al{
\zeta_k^{(g)}=\pi_2+\pi_0-\delta_k^{(g)}.
\label{total gravitation contributions}
}

\subsection{Physical metric fluctuations and gauge modes}
\label{physical and gauge modes section}
To compute the contribution of metric fluctuations to the flow equation for $U$ and $F$, we need the second functional derivative of the gauge invariant effective average action $\bar\Gamma_k^{(2)}$ for a general macroscopic metric $g_{\mu\nu}$.
For this purpose we consider $\bar\Gamma_k[g_{\mu\nu}+h_{\mu\nu}]$ and compute the term $\bar\Gamma_2$ quadratic in the metric fluctuations $h_{\mu\nu}$. 
We split the metric fluctuation $h_{\mu\nu}$ into ``physical fluctuations'' $f_{\mu\nu}$ and ``gauge fluctuations'' $a_{\mu\nu}$.
In linear order the physical fluctuations are transverse,
\al{
&h_{\mu\nu}=f_{\mu\nu}+a_{\mu\nu},\nn[1ex]
D^\mu f_{\mu\nu}=0,& \qquad
a_{\mu\nu}=D_\mu a_\nu +D_\nu a_\mu.
\label{first metric decomposition}
}
Infinitesimal diffeomorphism transformations act as a shift of $a_\mu$ and leave $f_{\mu\nu}$ invariant. 
We will need $\bar\Gamma_k$ in second order in $f_{\mu\nu}$.

Inserting the linear decomposition \eqref{first metric decomposition} into a gauge invariant effective action $\bar \Gamma_k$ one finds in linear order in $h_{\mu\nu}$ that $\bar \Gamma_1$ depends only on $f_{\mu\nu}$, reflecting the gauge invariance of $\bar \Gamma$.
A gauge invariant effective action can depend only on physical fluctuations.
The association of the physical fluctuations with $f_{\mu\nu}$ as defined by Eq.\,\eqref{first metric decomposition} holds only in linear order, however.
(An exception are Abelian gauge theories.)
Physical fluctuations vanish for $f_{\mu\nu}=0$, but can contain  higher order terms $\sim f_\mu{}^{\rho}a_{\nu\rho}$.
As a consequence, $\bar \Gamma_2$ contains terms quadratic in $f_{\mu\nu}$ and may also contain mixed terms linear in $f_{\mu\nu}$ and linear in $a_{\mu\nu}$.
(For an explicit discussion for the analogous case of non-Abelian gauge theories, see Ref.\,\cite{Wetterich:2017aoy}.)
For maximally symmetric spaces the mixed terms vanish due to the identity
\al{
\int_x\sqrt{g}\,f^{\mu\nu}a_{\mu\nu}=0.
}
For the purpose of the present paper this is sufficient.
For more general macroscopic metrics the mixed terms are eliminated by the projection on the physical subspace that is implicit in the flow equation \eqref{propagator and Gamma2}.

Similar to the treatment of gauge bosons we perform the projection by adding formally a term $\Gamma_\text{gf}^{(2)}=\frac{1}{\alpha}\tilde\Gamma_\text{gf}^{(2)}$ to the inverse propagator.
This allows the computation of $\pi_2+\pi_0-\delta^{(g)}$ without the explicit use of projectors.
The measure part $\delta^{(g)}$ is computed separately.
The implementation of this procedure introduces in the effective action a term
\al{
\Gamma_k=\bar \Gamma_k+\Gamma_\text{gf}\,,
}
which resembles a physical gauge fixing in the background field formalism.
The second variation $\Gamma_\text{gf}^{(2)}$ should act only on the gauge fluctuations, and we take
\al{
\Gamma_\text{gf}&=\frac{1}{2\alpha}\int_x\sqrt{g}\,D_\mu h^{\nu\mu} D_\rho h_{\nu}{}^{\rho}\\
&=\frac{1}{2\alpha}\int_x\sqrt{g}\, (D_\mu a^{\mu\nu})(D_\rho a^\rho{}_{\nu})\nn
&=\frac{1}{2\alpha}\int_x\sqrt{g}\,(D^2a^\nu +D^\nu D_\mu a^\mu+R^\nu{}_{\mu}a^\mu )\nn[-1ex]
&\qquad\qquad\qquad \times(D^2a_\nu +D_\nu D_\rho a^\rho+R_\nu{}^{\rho}a_\rho ),\nonumber
}
with $\alpha\to 0$. 
This ``physical gauge fixing" is the generalization of covariant Landau gauge fixing for Yang-Mills theories to the case of gravity.

The gauge fixing term induces for $\Gamma_k^{(2)}$ a term that diverges for $\alpha\to 0$
\al{
\left( \Gamma_\text{gf}^{(2)}\right)^{\mu\nu\rho\tau}=\frac{\sqrt{g}}{2\alpha}\left( \mathcal D_a\right)^{\mu\nu\rho\tau}\,,
}
with
\al{
\left(\mathcal D_a\right)_{\mu\nu}{}{}^{\rho\tau}=&-\frac{1}{2}\big( \delta_\mu^\rho D_\nu D^\tau+\delta_\nu^\rho D_\mu D^\tau \nn
&\qquad +\delta_\mu^\tau D_\nu D^\rho +\delta_\nu^\tau D_\mu D^\rho \big).
\label{covariant derivative D-a}
}
One observes
\al{
&\left(\mathcal D_a\right)_{\mu\nu}{}{}^{\rho\tau} f_{\rho\tau}=0,\nn[1ex]
&\left(\mathcal D_a\right)_{\mu\nu}{}{}^{\rho\tau}a_{\rho\tau}=D_\mu \hat{\mathcal D}_\nu{}^{\rho}a_\rho+D_\nu \hat{\mathcal D}_\mu{}^{\rho}a_\rho\,,
\label{covariant derivative projecting metric}
}
with
\al{
\hat{\mathcal D}_\mu{}^{\rho}=-(D^2\delta_{\mu}^\rho+D_\mu D^\rho +R_\mu{}^{\rho}).
\label{covariant derivative D hat}
}

We may formally introduce projectors $P_f$ and $P_a$  on the physical metric fluctuations and gauge modes
\al{
&\left( P_f\right)_{\mu\nu}{}{}^{\rho\tau}h_{\rho\tau}=f_{\mu\nu},&
&\left( P_a\right)_{\mu\nu}{}{}^{\rho\tau}h_{\rho\tau}=a_{\mu\nu},&
}
obeying 
\al{
&P_f+P_a=1,\quad
P_f^2=P_f,\quad
P_a^2=P_a,\nn
&P_fP_a=P_aP_f=0.
}
In terms of these projectors one has
\al{
&\mathcal D_aP_a=\mathcal D_a,&
&\mathcal D_aP_f=0,&\nn[2ex]
&P_a^T\sqrt{g}\mathcal D_a=\sqrt{g}\mathcal D_a,&
&P_f^T\sqrt{g}\mathcal D_a=0.&
\label{properties of projectors}
}
With 
\al{
\Gamma_k^{(2)}=\bar\Gamma_k^{(2)}+\frac{1}{2\alpha}P_a^T\sqrt{g}\mathcal D_a P_a
}
the inverse of $\Gamma_k^{(2)}$ is block diagonal for $\alpha\to0$, 
\al{
G=\left( \Gamma_k^{(2)}\right)^{-1}=G_P+G_a
}
where
\al{
&G_P=P_fGP_f^T,&
&G_a=P_a GP_a^T,&
}
and 
\al{
&\bar\Gamma_k^{(2)}G_P=P_f^T,&
&\Gamma_\text{gf}^{(2)}G_a=P_a^T.&
\label{projectors from inverse propagator}
}
Here $G_a$ is proportional to $\alpha$ and $\Gamma_\text{gf}^{(2)}$ proportional to $\alpha^{-1}$, such that the factors of $\alpha$ cancel in the second equation \eqref{projectors from inverse propagator}.
The part $G_P$ is the propagator for the physical metric fluctuations that appears in Eqs.\,\eqref{flow equation for GP} and \eqref{propagator and Gamma2}.

The block diagonal structure for $\alpha\to0$ has an important consequence.
For the computation of $G_P$ only the projection $P_f^T\bar \Gamma_k^{(2)}P_f$ enters, as appropriate for the propagator of the physical metric fluctuations.
Possible mixed terms in $\bar \Gamma_2$ that involve powers of $a_{\mu\nu}$ play no role---they are projected out by the physical gauge fixing for $\alpha\to0$.
It is therefore sufficient to evaluate $\bar \Gamma_2$ for the physical metric fluctuations $f_{\mu\nu}$, which simplifies the task since $D^\nu f_{\mu\nu}=0$ can be used.

For the truncation
\al{
\bar\Gamma_k=\int_x\sqrt{g}\left\{- \frac{1}{2}F\fn{\rho}R +U\fn{\rho}\right\},
}
one finds \cite{Wetterich:2016vxu} for geometries with a vanishing Weyl tensor
\al{
\bar \Gamma_2=\frac{1}{8}\int_x\sqrt{g}\left\{ -U(2f^{\mu\nu}f_{\mu\nu} -\sigma^2)+Ff^{\mu\nu}(\hat{\mathcal D}_f)_{\mu\nu}{}^{\rho\tau}f_{\rho\tau}\right\},
}
with
\al{
(\hat{\mathcal D}_f)_{\mu\nu}{}^{\rho\tau}&=\left( -D^2+\frac{2R}{3}\right) E_{\mu\nu}{}^{\rho\tau}+D^2 g_{\mu\nu}g^{\rho\tau}\,,\label{derivative hat D}
}
and unit matrix
\al{
E_{\mu\nu}{}^{\rho\tau}&=\frac{1}{2}(\delta_\mu^\rho \delta_\nu^\tau +\delta_\mu^\tau \delta_\nu^\rho).
}
The physical scalar metric fluctuation corresponds to the trace of $f_{\mu\nu}$, 
\al{
\sigma=g^{\mu\nu}f_{\mu\nu}.
}
Thus the physical metric fluctuations can be decomposed into a traceless transverse tensor $t_{\mu\nu}$ (graviton) and a scalar $\sigma$
\al{
&f_{\mu\nu}=t_{\mu\nu}+s_{\mu\nu},\nn[1ex]
D_\nu t_{\mu}{}^{\nu}&=0,\qquad
t_\mu^\mu=0,\qquad
s_{\mu\nu}=\hat S_{\mu\nu}\sigma,
}
with $\hat S_{\mu\nu}=\hat S_{\nu\mu}$ obeying
\al{
&\hat S_{\mu\nu}g^{\mu\nu}=1,&
&D^\mu \hat S_{\mu\nu}=0.&
}
 
For a computation of the flow of $F$ and $U$ we can restrict the discussion to maximally symmetric geometries with constant curvature scalar $R$.
In this case one has \cite{Wetterich:2016vxu}
\al{
\hat S_{\mu\nu}=\left( D^2 g_{\mu\nu} -D_\mu D_\nu +R_{\mu\nu} \right) (3D^2+R)^{-1}.
}
There is no mixing between $t_{\mu\nu}$ and $\sigma$ in $\bar\Gamma_2$, e.g.
\al{
\bar\Gamma_2=\Gamma_2^{(t)}+\Gamma_2^{(\sigma)},
}
where
\al{
\Gamma_2^{(t)}=\frac{1}{8}\int_x\sqrt{g}\,t^{\mu\nu}\left\{ F\left(-D^2+\frac{2}{3}R\right) -2U\right\}t_{\mu\nu}
}
and
\al{
\Gamma_2^{(\sigma)}&=\frac{1}{4}\int_x\sqrt{g}\,\sigma\bigg\{ \bigg[F\left( D^2+\frac{3}{8}R\right) D^2\nn[1ex]
&\qquad+\frac{U}{2}\left(D^2+\frac{1}{2}R\right)\bigg](3D^2+R)^{-1}\bigg\}\sigma\,.
}

The infrared cutoff $(R_k)_{\mu\nu}{}{}^{\rho\tau}$ should respect the split into physical fluctuations and gauge fluctuations,
\al{
{\mathcal R}_k=\sqrt{g}\left\{ \frac{F}{8}R_k^{(f)}\fn{\mathcal D_f} +\frac{1}{2\alpha}R_k^{(a)}\fn{\mathcal D_a}\right\}.
\label{cutoff function for physical metric}
}
Here the operator $\mathcal D_f$ should obey 
\al{
&\mathcal D_f P_f=\mathcal D_f,\quad
P_f^T\sqrt{g}{\mathcal D}_f=
\sqrt{g}{\mathcal D}_f,\nn[2ex]
&
\mathcal D_f P_a=0,\qquad
P_a^T\sqrt{g}\mathcal D_f=0,
}
similarly to Eq.\,\eqref{properties of projectors}.
We can define it by using  $\hat{\mathcal D}_f$ in Eq.\,\eqref{derivative hat D},
\al{
\mathcal D_f=\frac{1}{\sqrt{g}}P_f^T \sqrt{g}\hat{\mathcal D}_fP_f,
} 
such that $\sqrt{g}\mathcal D_f$ equals $\sqrt{g}\hat{\mathcal D}_f$ on the subspace of divergenceless physical fluctuations $f_{\mu\nu}$.
For a cutoff of the type \eqref{cutoff function for physical metric} the flow equation can be separated into different pieces similar to Eq.\,\eqref{flow of gauge thories}, with $\pi^{(g)}_k$ the contribution from physical metric fluctuations and $-\delta_k^{(g)}$ the measure contribution.
For maximally symmetric spaces the physical contribution $\pi^{(g)}_k$ further decays into a graviton contribution $\pi_2$ from the fluctuations of $t_{\mu\nu}$, and a scalar contribution $\pi_0$ from the $\sigma$ fluctuations, as given by Eq.\,\eqref{total gravitation contributions}.

\subsection{Graviton contribution}
\label{evaluation of graviton contributions}
The graviton contribution $\pi_2$ typically dominates.
It is given by
\al{
\pi_2=\frac{1}{2}\text{tr}_{(2)}\left\{ \p_t R_k \left( \frac{F}{4}\hat{\mathcal D}_f+R_k-\frac{U}{2}\right)^{-1}\right\}\,,
}
with $\text{tr}_{(2)}$ the trace in the projected space of $t_{\mu\nu}$ (spin two fields). 
For the physical metric fluctuations the cutoff $R_k(\hat{\mathcal D}_f)$ is chosen as a function of the operator $\hat{\mathcal D}_f$.
In the projected space for the graviton fluctuations $\hat{\mathcal D}_f$ reduces to $-D^2+2R/3$.
For the heat kernel expansion (cf.~Appendix\,\ref{heat kernel methods}), one needs
\al{
\text{tr}_{(2)}\,e^{-sD_t}&=\text{tr}_\text{(2)}\,e^{-s\Delta_T}e^{-2sR/3}\nn
&=\text{tr}_{(2)}\,e^{-s\Delta_T}\left( 1-\frac{2R}{3}s\right),
}
where $\Delta_T=-D^2$ involves the Laplacian acting on traceless transverse tensor fields and we have taken constant $R$.
With $b_0^T=5$ and $b_2^T=-5/6$ the heat kernel coefficients of $\Delta_T$ for traceless transverse tensors, the graviton contribution reads
\al{
c_0^T=b_0^{T}=5,\qquad
c_2^T=\left( b_2^T -\frac{2}{3}b_0^T\right)R=-\frac{25}{6}R.
}

One obtains the flow contribution of the graviton fluctuations
\al{
\pi_2=\frac{5}{32\pi^2}\int_x\sqrt{g}& \int^\infty_0\df z\,\left( z-\frac{5R}{6}\right) \nn
& \times \frac{\p_t R_k^{(f)}+(\p_t \ln F)R_k^{(f)}}{z+R_k^{(f)}-2U/F}.
}
With $\eta_g=2-\p_t \ln F=-\p_t \ln w$ and employing the Litim cutoff $R_k^{(f)}=(k^2-z)\theta\fn{k^2-z}$ this yields
\al{
\pi_2=\frac{5}{32\pi^2(1-v)}\int_x\sqrt{g}&\bigg[ \frac{4}{3}k^4\left( 1- \frac{\eta_g}{8} \right) \nn
&\quad- \frac{5}{2} k^2 \left(  1- \frac{\eta_g}{6}\right) R \bigg],
}
where
\al{
v=\frac{2U}{k^2F}=\frac{u}{w}.
}

Correspondingly, the graviton contribution to $c_V$ and $c_M$ depends on $\tilde \rho$ via $v(\tilde \rho)$ and $\eta_g(\tilde \rho)$,
\al{
c_V^{(T)}&=\frac{5}{96\pi^2(1-v)}\left (1 - \frac{\eta_g}{8}\right),\nn[1ex]
c_M^{(T)}&=\frac{25}{128\pi^2(1-v)}\left (1 - \frac{\eta_g}{6}\right).
}
For scaling solutions with constant $w(\tilde \rho)=w_*=c_M$ one has $\eta_g=0$.
For $c_V^{(T)}$ one recovers the graviton contribution to the flow of the effective potential~\cite{Pawlowski:2018ixd,Wetterich:2017ixo}.
This result for $c_M^{(T)}$ agrees with Refs.\,\cite{Narain:2009fy,Labus:2015ska,Percacci:2015wwa}.
Further details and a comparison with the background field formalism can be found in Appendixes\,\ref{formulations} and \ref{evaluation of flow equations}.

\subsection{Measure contribution}
\label{evaluation of measure contribution}
The gauge invariant flow equation involves a universal ``measure contribution''.
It is given by
\al{
-\delta^{(g)}_k=-\frac{1}{2}\p_t \ln \det \fn{\mathcal D_a+R_k^{(a)}\fn{\mathcal D_a}},
\label{universal measure contributions}
}
with $\mathcal D_a$ given by Eq.\,\eqref{covariant derivative D-a}.
We will evaluate this expression here.
In Appendix\,\ref{formulations}, we show that the truncated background field flow equation with physical gauge fixing yields the same result.
A different possible choice of the cutoff function, where $R_k^{(a)}\fn{\mathcal D_a}$ is replaced by $R_k^{(a)}(\tilde{\mathcal D}_a)$, with $\tilde{\mathcal D}_a$ obtained from $\mathcal D_a$ by omitting in $\hat{\mathcal D}$ in Eq.\,\eqref{covariant derivative D hat} the term $R_\mu{}^{\rho}$, is discussed in Appendix\,\ref{IR cutoff scheme}.

We can express $\delta^{(g)}_k$ in terms of the eigenvalues $\lambda_i$ of $\mathcal D_a$ as
\al{
\delta^{(g)}_k&=\frac{1}{2}\text{tr}\, \p_t \ln \fn{\mathcal D_a+R_k^{(a)}\fn{\mathcal D_a}}\nn[1ex]
&=\frac{1}{2}\sum_i  \p_t \ln \fn{\lambda_i +R_k^{(a)}\fn{\lambda_i}}.
}
The eigenvalues of $\mathcal D_a$ are the same as the eigenvalues of the operator $\hat{\mathcal D}$ in Eq.\,\eqref{covariant derivative D hat}.
Indeed, for 
\al{
\hat{\mathcal D}_\mu{}^{\rho}a_\rho=\lambda \, a_\mu
}
one concludes from Eq.\,\eqref{covariant derivative projecting metric}
\al{
\left( \mathcal D_a\right)_{\mu\nu}{}{}^{\rho\tau}a_{\rho\tau}=(D_\mu \lambda \, a_\nu+D_\nu \lambda \, a_\mu )=\lambda \, a_{\mu\nu}\,.
}
The constraints on the gauge fluctuations $a_{\mu\nu}$ are precisely that $a_{\mu\nu}$ can be expressed in term of $a_\mu$ by Eq.\,\eqref{first metric decomposition}, such that the set of eigenvalues of $\hat{\mathcal D}$ completely covers the set of  eigenvalues of $\mathcal D_a$ in the presence of the constraints on $a_{\mu\nu}$.
The vector fields $a_\mu$ are unconstrained, and we can write 
\al{
\delta_k^{(g)}=\frac{1}{2}\text{tr}_{(V)} \p_t \ln\fn{ \hat{\mathcal D}+R_k(\hat{\mathcal D})},
\label{deltak contribution}
}
with $\text{tr}_{(V)}$ the unconstrained trace over vector fields.
We emphasize that the measure contribution \eqref{deltak contribution} is universal.
For a given cutoff prescription it depends only on the form of the IR cutoff, not on the approximations used for the gauge invariant effective action $\bar \Gamma$.
It depends only on the metric, and not on fields describing other particles.
In particular, it contributes to the flow of $U$ and $F$ only a $\rho$-independent term.

For the evaluation of $\delta_k$ we split $a_\mu$ into a transverse part and a longitudinal part, similar to the discussion of the vector gauge field $A_\mu$ in Sec.\,\ref{gauge bosons section}.
We decompose
\al{
&a_\mu=\kappa_\mu+D_\mu u,&
D^\mu\kappa_\mu=0,&
}
with 
\al{
\hat{\mathcal D}_\mu{}^{\rho}\kappa_\rho&=-D^2\kappa_\mu -R_\mu{}^{\rho}\kappa_\rho,\nn[1ex]
\hat{\mathcal D}_\mu{}^{\rho}D_\rho u&=-2D^2 D_\mu u=-2 D_\mu D^2 u-2R_\mu{}^{\rho} D_\rho u.
}
For maximally symmetric geometries one finds
\al{
&\hat{\mathcal D}_\mu{}^{\rho}\kappa_\rho=\left( \Delta_V -\frac{R}{4}\right)\kappa_\mu =\mathcal D_1 \kappa_\mu ,\nn
&\hat{\mathcal D}_\mu{}^{\rho}D_\rho u =2D_\mu \left( \Delta_S -\frac{R}{4}\right) u= 2D_\mu {\mathcal D}_0u,
}
with $\Delta=-D^2$, and $\Delta_V$ and $\Delta_S$ denoting the action of the Laplacian on vector or scalar fields, respectively.
The eigenvalues of the operator $\Delta_S-R/4$ acting on $u$ are the eigenvalues of $\hat{\mathcal D}/2$ acting on the longitudinal field $D_\mu u$.
The factor $1/2$ drops out in $\tilde \p_t \ln\hat{\mathcal D}$.
The measure contribution becomes the sum of a transverse vector and a scalar piece,
\al{
\delta_k&=\frac{1}{2}\bigg[ \text{tr}_{(1)}\, \p_t \ln\fn{ \mathcal D_1+R_k^{(1)}\fn{\mathcal D_1} } \nn
&\qquad\qquad+\text{tr}_{(0)}\, \p_t \ln\fn{ \mathcal D_0+R_k^{(0)}\fn{\mathcal D_0} }\bigg],
}
with $\text{tr}_{(1)}$ the trace over transverse vector fields (spin one for geometries with rotation symmetry) and $\text{tr}_{(0)}$  the trace over scalars (spin zero).

For the heat kernel expansion we employ 
\al{
\text{tr}_{(1)}\,e^{-s\mathcal D_1}&=\text{tr}_{(1)}\,e^{-s \Delta_V}e^{sR/4}\nn
&=\text{tr}\,e^{-s\Delta_V}\left( 1+\frac{R}{4}s\right)\nn
&=\frac{1}{16\pi^2}\int_x\sqrt{g}\left( b_0^V s^{-2} +b_2^V Rs^{-1} \right) \left( 1+\frac{Rs}{4}\right)\nn
&=\frac{1}{16\pi^2}\int_x\sqrt{g}\left[ b_0^V s^{-2} +\left( b_2^V +\frac{1}{4}b_0^V \right) Rs^{-1}\right],
}
with $b_0^V=3$, $b_2^V=1/4$.
Similarly, one has
\al{
&\text{tr}_{(0)}\,e^{-s\mathcal D_0} \nn
&\quad=\frac{1}{16\pi^2}\int_x\sqrt{g}\left[ b_0^S s^{-2} +\left( b_2^S +\frac{1}{4}b_0^S \right) Rs^{-1}\right],
}
with $b_0^S=1$, $b_2^S=1/6$.
This yields
\al{
\delta^{(g)}_k=\frac{1}{32\pi^2}\int_x\sqrt{g}\int^\infty_0\df z\, W\fn{z}\left(4z +\frac{17}{12}R\right),
}
with $W\fn{z}=\p_t R_k\fn{z}/(z+R_k\fn{z})$.
The universal measure contribution \eqref{universal measure contributions} is found as
\al{
-\delta^{(g)}_k=-\int_x \sqrt{g}\left( \frac{k^4 \ell_0^4}{4\pi^2} +\frac{17k^2 \ell_0^2}{192\pi^2}R \right),
\label{measure contribution from metric}
}
with threshold functions $\ell_0^4$ and $\ell_0^2$ evaluated for $\tilde w=0$.

For the Litim cutoff the measure contributions to the flow of $U$ and $F$ are constants
\al{
c_V^{(m)}&=-\frac{\ell_0^4}{16\pi^2}=-\frac{1}{32\pi^2},\nn[1ex]
c_M^{(m)}&=\frac{17\ell_0^2}{384\pi^2}=\frac{17}{384\pi^2}.
\label{measure contributions to cV cM}
}
One may compare this simple result with the usual treatment of a ghost sector and the vector and scalar fluctuations of the metric.
For a general gauge fixing, including the most commonly employed gauge fixings, no such simple result exists.
For a physical gauge fixing, however, the leading order yields a contribution of the gauge fluctuations $\delta_k^{(g)}$ and a ghost contribution $\epsilon_k^{(g)}=-2\delta_k^{(g)}$, reproducing the total measure contribution \eqref{measure contribution from metric}.
Details can be found in Appendix\,\ref{formulations}.
Comparing the measure contribution \eqref{measure contributions to cV cM} with the graviton contribution, one finds for $\eta_g=0$
that the measure part of $c_V$ is a factor of $-3(1-v)/5$ of the graviton part, while for $c_M$ it is a factor of $17(1-v)/75$ of the graviton contribution.
The graviton contribution dominates, in particular, for $v$ close to one.

\subsection{Physical scalar metric fluctuations}
\label{discussion on scalar metric fluctuation}
For the physical scalar fluctuations the inverse propagator matrix mixes the scalar contained in the metric fluctuations with the other scalars in the model.
As a result, the scalar contribution to the flow becomes somewhat lengthy.
We display this mixing for our truncation in Appendix\,\ref{formulations}.
In particular, for a single additional scalar field $\phi$ the scalar contribution $\pi_0$ involves the mixing effect between $\sigma$ and $\phi$ as given in Eq.\,\eqref{physical spin 0 mode two point function}.
We will neglect here this mixing effect and concentrate on the contribution of the scalar metric fluctuation $\sigma$, while the contribution of $\phi$ is contained in $N_S$ as discussed in Sec.\,\ref{scalar contributions}.
The resulting expression for $\pi_0$ can be extracted from Appendix\,\ref{evaluation of flow equations}.
It still remains a somewhat lengthy expression.
Since the scalar part in the metric fluctuations is small as compared to the graviton part, we approximate it by setting $v=0$.
This error is modest, given that factors such as $(1-v/4)^{-1}$ appearing in the propagator are reasonably close to one for the range of interest $|v|\leq 1$.

Under the approximations made above one can evaluate $\pi_0$ as
\al{
\pi_0&= \frac{1}{32\pi^2}\int_x\sqrt{g}\int_0^\infty \df z \left( z +\frac{R}{6} \right)\frac{\p_t R_k +(\p_t \ln F)R_k }{z+R_k} \nn
&\quad
-\frac{1}{32\pi^2}\int_x\sqrt{g}\int_0^\infty \df z\,\left(z\frac{R}{24}\right) \frac{\p_t R_k +(\p_t \ln F)R_k}{(z+R_k)^2}\nn[2ex]
&=\frac{1}{32\pi^2}\int_x\sqrt{g}\left[\frac{4k^4}{3}\left( 1-\frac{\eta_g}{8}\right) +\frac{4k^2}{9}\left( 1-\frac{11\eta_g}{64}\right)R \right],
\label{approximated pi0}
}
where we employ the Litim cutoff in the last equality.
The contributions to the flow equations \eqref{flow equations} read
\al{
c_V^{(\sigma)}&=\frac{1}{96\pi^2}\left( 1-\frac{\eta_g}{8}\right),\nn[1ex]
c_M^{(\sigma)}&=-\frac{1}{144\pi^2}\left( 1-\frac{11\eta_g}{64}\right).
\label{cV and cM from scalar metric fluctuations}
}
More precise calculations for the contributions from the physical scalar fluctuations, including the mixing term and a finite $v$, are presented in Appendix\,\ref{evaluation of flow equations}.

For $\eta_g=0$, as appropriate for the scaling solution at $\tilde \rho=0$, the contribution of the scalar in the metric fluctuations effectively adds to $N_S$ a term $4/3$.
Here the factor $4/3$ is due to the $k$ dependence of the factor $F$ multiplying the IR cutoff for the metric fluctuations.
We observe that $\pi_0$ is suppressed as compared to the graviton fluctuations $\pi_2$ by a typical factor of $(1-v)/5$, with even stronger suppression for the derivatives relevant for anomalous dimensions.
This justifies the simplification in the scalar metric sector.
For positive $v<1$ the error of setting $v=0$ used for the analytic discussion in the scalar propagator $\sim (1-v/4)^{-1}$ amounts at most to a factor of $4/3$.
For large negative $v$ the graviton fluctuations are less dominant as compared to the scalar fluctuations.
However, all effects of metric fluctuations are suppressed by $v^{-1}$ in this case.
For all models considered here Eq.\,\eqref{cV and cM from scalar metric fluctuations} is a rather good approximation.

\section{UV-fixed point}
\label{UV fixed point analysis}
In our truncation we find an UV-field point with $\tilde\rho$-independent $u_*$, $w_*$ for a large region of particle physics models characterized by $N_S$, $N_F$ and $N_V$.
This region covers the standard model as well as GUT models based on SU(5) and SO(10).
The fixed point exists for an arbitrary number of scalar fields $N_S$ for these models.
In the region of strong gravity for $v$ close to one or $w$ close to zero our truncation is presumably not reliable. 
These regions are encountered for large $N_S$.
We also find a new fixed point, in addition to the Reuter fixed point, in a small region of models.
Based on the argument in Sec.\,\ref{discussion on New fixed point} we use the simplification \eqref{cV and cM from scalar metric fluctuations}, which permits an analytic discussion of fixed points and critical exponents.

\subsection{Dependence on particle content}

For an understanding of the UV-fixed point we take all particles to be massless. We also set $\eta_g=0$. As discussed above, we omit for the physical scalar in the metric fluctuations the mixing with other scalars and we set $v=0$ for this contribution.
This results in the flow equations
\al{
&\p_t u\fn{\tilde \rho}=2\tilde \rho\, \p_{\tilde\rho} u
 -4\left[ u- \frac{1}{128\pi^2}\left(\tilde{\mathcal N}_U + \frac{20}{3(1-v)}  \right)  \right]\,, \nn[3ex]
&\p_t w\fn{\tilde \rho}=2\tilde \rho \, \p_{\tilde\rho} w
-2\left[ w-\frac{1}{192\pi^2}\left(\tilde{\mathcal N}_M+\frac{75}{2(1-v)}  \right)\right]\,,
\label{total results of beta functions}
}
where
\al{
v=\frac{u}{w}.
}

With Eq.\,\eqref{particle numbers} we define the effective particle numbers
\al{
\tilde{\mathcal N}_U&={\mathcal N}_U-\frac{8}{3}=N_S+2N_V-2N_F-\frac{8}{3}\,,\nn[2ex]
\tilde{\mathcal N}_M&={\mathcal N}_M+\frac{43}{6}=-N_S+4N_V-N_F+\frac{43}{6}\,.
}
The second identities neglect $\tilde\xi$.
The dependence of the flow equations on the particle content of the model is summarized in two numbers $\mathcal N_U$ and $\mathcal N_M$ or, equivalently $\tilde{\mathcal N}_U$ and $\tilde{\mathcal N}_M$.
We will display our results in dependence of $\tilde{\mathcal N}_U$ and $\tilde{\mathcal N}_M$.
This has the advantage that an improved treatment of the physical scalar metric fluctuations, as well as different treatments or cutoffs for the measure contribution, can partially be absorbed by small shifts in $\tilde{\mathcal N}_U$ and $\tilde{\mathcal N}_M$.
This extends to the effect of small particle masses.

In a ``lowest order approximation" one sets $v$ in the graviton propagator to be zero.
This applies to weak gravity of small $w^{-1}$, since $v\sim w^{-1}$.
In this limit the contributions from matter and metric fluctuations for $u$ and $w$ give the factors ${\mathcal N}_U+4$ and ${\mathcal N}_M+134/3$, respectively.
The matter contributions (${\mathcal N}_U$ and ${\mathcal N}_M$) agree with Refs.\,\cite{Codello:2008vh,Dona:2013qba}, whereas the contributions from the metric fluctuations generally differ due to the dependences on choices of the gauge parameters and regulators.
Reference\,\cite{Dona:2013qba}, in which the gauge choice $\alpha=\beta=1$ is taken and the same cutoff functions are used, reports the factors of $2$ for $u$ and $46$ for $w$ as contributions from the metric fluctuations.
Since the main contribution comes from the traceless transverse (TT) graviton mode, there is no drastic difference arising from variations of the gauge choice and the cutoff functions.
The metric contribution in ${\mathcal N}_U+4$ has a simple explanation.
The contribution of the 6 physical degrees of freedom is multiplied by $4/3$ due to $F$ multiplying the IR cutoff, resulting in $6\times(4/3)-4=4$.

Beyond the lowest order the gauge invariant flow equation leads to considerable simplifications as compared to earlier work, e.g., in Ref.\,\cite{Dona:2013qba}.
In particular, the effect of the scalar metric fluctuations remains always small, such that the mixing with other scalars can be neglected and the simple approximation \eqref{total results of beta functions} can be used.
This has important consequences for the existence of the fixed point, being at the origin of differences in the allowed range of $N_S$, $N_V$, $N_F$ as compared to Ref.\,\cite{Dona:2013qba}.

\subsection{Limitation of truncation}
\label{sect: Limitation of truncation}
A key element for our truncation for the metric fluctuations is the specific form of the propagator that is derived from an Einstein-Hilbert form of the gravitational effective average action.
In terms of the dimensionless quantities $u$ and $w$ the inverse graviton propagator in the presence of the IR cutoff is proportional to 
\al{
G^{-1}_\text{grav}=\left( \bar \Gamma_k^{(2)}+R_k\right)_\text{grav} \sim w\, p_k(\hat{\mathcal D}_f/k^2)-u\,,
}
with 
\al{
p_k(z/k^2)=(z+R_k(z))/k^2\,.
}
Because of the IR cutoff, $p_k$ has a minimum $\bar p>0$.
For the Litim cutoff, as well as other suitably normalized cutoffs, one has $\bar p=1$, and we will take this value.
The minimal value of $wp_k-u$ is therefore given by
\al{
y=w-u=w(1-v).
}

In the vicinity of $y=0$ our truncation is expected to become insufficient.
This concerns $w$ close to zero or $v$ close to one.
For example, adding to $\bar\Gamma_k$ a term quadratic in the Weyl tensor with coefficient $D/2$ adds to $G^{-1}_\text{grav}$ a piece $D\,\hat{\mathcal D}_f^2/2$, replacing 
\al{
wp_k-u\to w p_k -u+D\,\hat{\mathcal D}_f^2/k^4\,.
}
For $y\to 0$ the term $\sim D$ will dominate near the minimum of $G^{-1}_\text{grav}$ and can no longer be neglected.
We therefore expect our approximation to break down for small values of $y$, typically $y\lsim D$.

Within our truncation ($D=0$) the graviton propagator diverges for $y\to 0$ even in the presence of the IR cutoff.
It becomes unstable for $y<0$.
Such an instability is avoided by the flow for a valid truncation.
In the present truncation this is not the case.
With
\al{
\p_t y=\beta_y&=2\tilde \rho\, \p_{\tilde \rho} y-4y+w\left( 2+\frac{35}{192\pi^2 y}\right)\nn[1ex]
&\qquad+\frac{1}{96\pi^2}\left( \tilde{\mathcal N}_M-3\tilde{\mathcal N}_U\right)\, ,
}
we observe that for $y\to 0$ the flow generator $\beta_y$ is positive for $w>0$.
As a result, a $\tilde \rho$-independent $y$ could run into the singularity for decreasing $k$.
Similarly, a scaling solution for a $\tilde\rho$-dependent $y$ [$\p_t y(\tilde \rho)=0$] could run into the singularity for increasing $\tilde \rho$.
Such a behavior is not acceptable, indicating the need for an extension of the truncation for $y\to 0$.
The same problem is visible in the flow equation for $v$,
\al{
\p_t v&=2\tilde \rho\, \p_{\tilde \rho}v-2v \nn
&\quad+\frac{1}{192\pi^2w}\left( 6\tilde{\mathcal N}_U-2v\tilde{\mathcal N}_M+75-\frac{35}{1-v}\right)\,,
}
for which $v$ could run into the onset of the instability at $v=1$.
This contrasts with the behavior of the flow of $u$ or $v$ for constant $w$ (or with the inclusion of a constant $D$), for which the instability is repulsive and avoided by the flow~\cite{Wetterich:2017ixo,Wetterich:2018poo}.
We conclude that for $v$ very close to one or very small $w$ our approximation can no longer be trusted.

\subsection{Constant scaling solution}
In this paper we concentrate on the constant scaling solution for which $u(\tilde \rho)$ and $w(\tilde \rho)$ have $\tilde\rho$-independent scaling solutions, such that the terms $2\tilde\rho\, \p_{\tilde\rho} u$ and $2\tilde\rho\,\p_{\tilde \rho}w$ can be omitted in the flow equation \eqref{total results of beta functions}.
For constant $w$ one also has $\tilde\xi=0$.
Indeed, the flow equations \eqref{total results of beta functions} have a fixed point with $\tilde\rho$-independent $u$ and $w$, for which $\eta_g=0$, and
\al{
u_*&= \frac{1}{128\pi^2}\left(\tilde{\mathcal N}_U +\frac{20}{3(1-v_*)}\right),\nn[1ex]
w_*&=\frac{1}{192\pi^2}\left(\tilde{\mathcal N}_M +\frac{75}{2(1-v_*)}\right).
\label{fixed point values for u- and w-}
}
This fixed point exists whenever the corresponding relation for $v$,
\al{
v_*=\frac{u_*}{w_*}=\frac{ 3\tilde{\mathcal N}_U (1-v_*)+20}{2\tilde{\mathcal N}_M (1-v_*)+75}\,,
\label{fixed point of v without physical scalar}
}
has a solution for real $v$.
The fixed point corresponds to an acceptable stable theory if 
\al{
&v_*<1,& 
&w_*>0.&
\label{stability condition}
}

Equation\,\eqref{fixed point of v without physical scalar} results in a quadratic equation for $x=1-v_*$,
\al{
2\tilde{\mathcal N}_M x^2+ (3\tilde{\mathcal N}_U -2\tilde{\mathcal N}_M+75)x-55=0.
\label{equation for x}
}
We are interested in solutions with positive $x$ ($v_*<1$) and positive $w_*$.
For $\tilde{\mathcal N}_M>0$ or $\mathcal N_M\geq -43/6$ the condition $w_*>0$ is obeyed for all $x>0$.
For $\tilde{\mathcal N}_M<0$ or $\mathcal N_M<-43/6$ one needs $x$ in the range
\al{
0<x< \frac{75}{|2\tilde{\mathcal N}_M|}.
\label{condition for fixed point}
}
With
\al{
b&=2\tilde{\mathcal N}_M-3\tilde{\mathcal N}_U-75\nn
&=2N_V+4N_F-5N_S-\frac{158}{3},
}
Eq.\,\eqref{equation for x} has the solutions
\al{
x_\pm=\frac{1}{4\tilde{\mathcal N}_M}\left( b\pm \sqrt{b^2 +440\tilde{\mathcal N}_M}\right).
\label{xpm solutions}
}
For $\tilde{\mathcal N}_M>0$ there exists always one solution with $x>0$, whereas the second solution has $x<0$. 
For $\tilde{\mathcal N}_M<0$ and $b>0$ no solution with $x>0$ exists.
For $\tilde{\mathcal N}_M<0$, $b<0$ one finds two solutions with $x>0$, provided
\al{
b^2>440|\tilde{\mathcal N}_M|.
\label{condition when a< 0,b<0}
}
Otherwise no solution exists.
Particularly interesting are the two solutions with both $b$ and $440\tilde{\mathcal N}_M$ negative, e.g. for 
\al{
N_S+N_F>4N_V+\frac{43}{6}.
}
It can be realized for a sufficient large number of fermions and scalars as compared to the number of gauge fields.

\begin{figure}
\includegraphics[width=8.5cm]{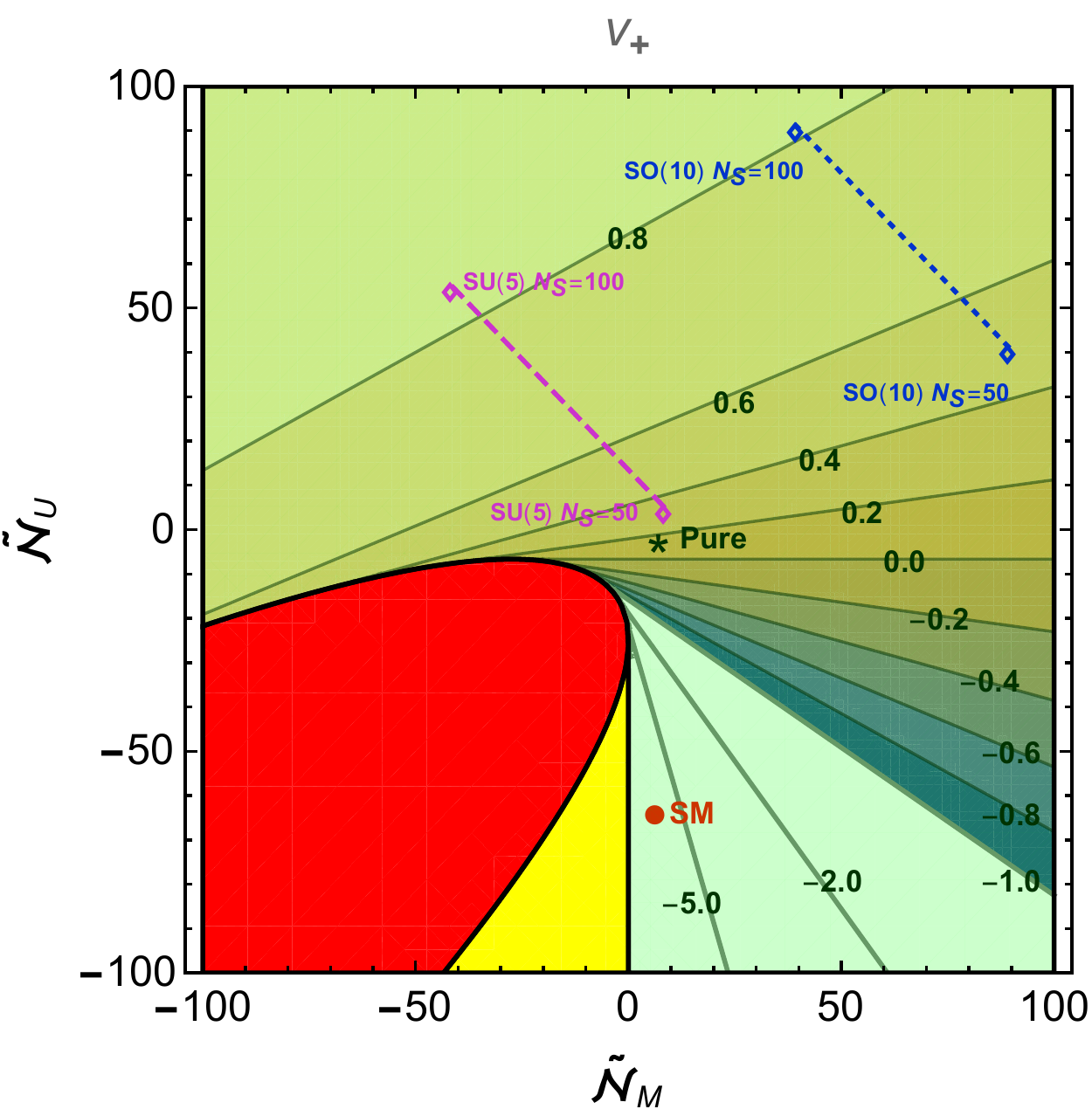}
\caption{
Contour plot of the fixed point value of $v_+$ in the $(\tilde{\mathcal N}_M,\tilde{\mathcal N}_U)$ plane.
In the red region for negative $\tilde{\mathcal N}_U$ and $\tilde{\mathcal N}_M$ no constant scaling solution is found.
In the yellow region the scaling solution is unstable due to $w_+<0$.
The green region admits a stable constant scaling solution.
As one moves toward the region of large positive $\tilde{\mathcal N}_U$ and negative $\tilde{\mathcal N}_M$, corresponding to a large number of scalar fields $N_S$, the fixed point value $v_+$ approaches one.
For $v$ close to one our approximations are no longer reliable.
We indicate the location of pure gravity, the standard model, as well as SU(5) and SO(10) GUT models with a number of scalar fields varying between $N_S=50$ and $100$. 
}
\label{contourfig:FP values of v} 
\end{figure}

\begin{figure}
\includegraphics[width=8.5cm]{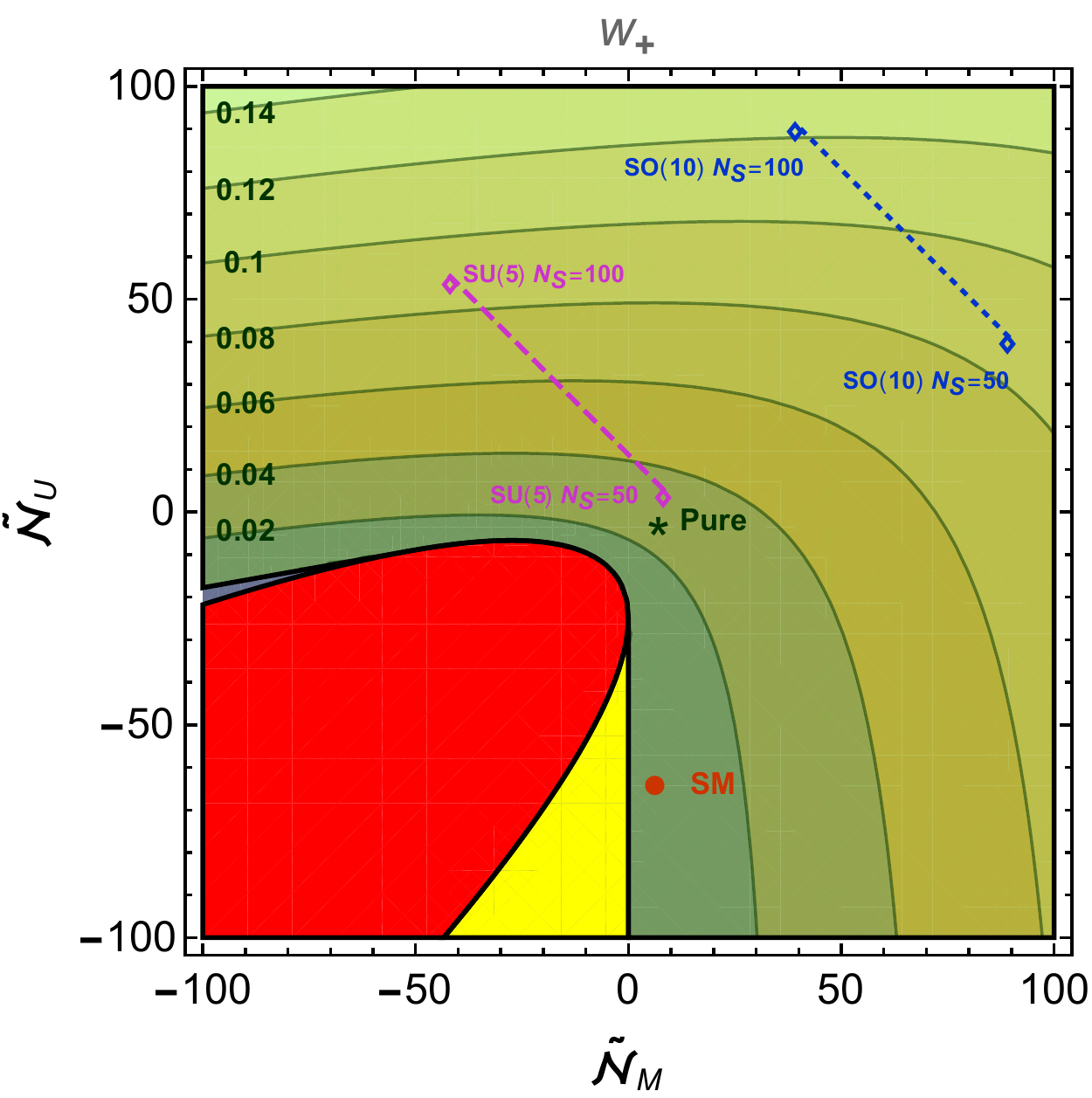}
\caption{
Contour plot of the fixed point value of $w_+$ in the $(\tilde{\mathcal N}_M,\tilde{\mathcal N}_U)$ plane.
The strength of gravity at the fixed point $w_+^{-1}$ increases as one approaches the excluded red and yellow regions for large negative $\tilde{\mathcal N}_U$ and $\tilde{\mathcal N}_M$.
For sufficiently negative $\tilde{\mathcal N}_U$ the fixed point gravity becomes unstable as $\tilde{\mathcal N}_M$ turns negative (red region).
For the small value of $w_+$ near the boundary of the excluded red and yellow regions one may have doubts on the validity of the truncation.
}
\label{contourfig:FP values of w} 
\end{figure}

\begin{figure}
\includegraphics[width=8.5cm]{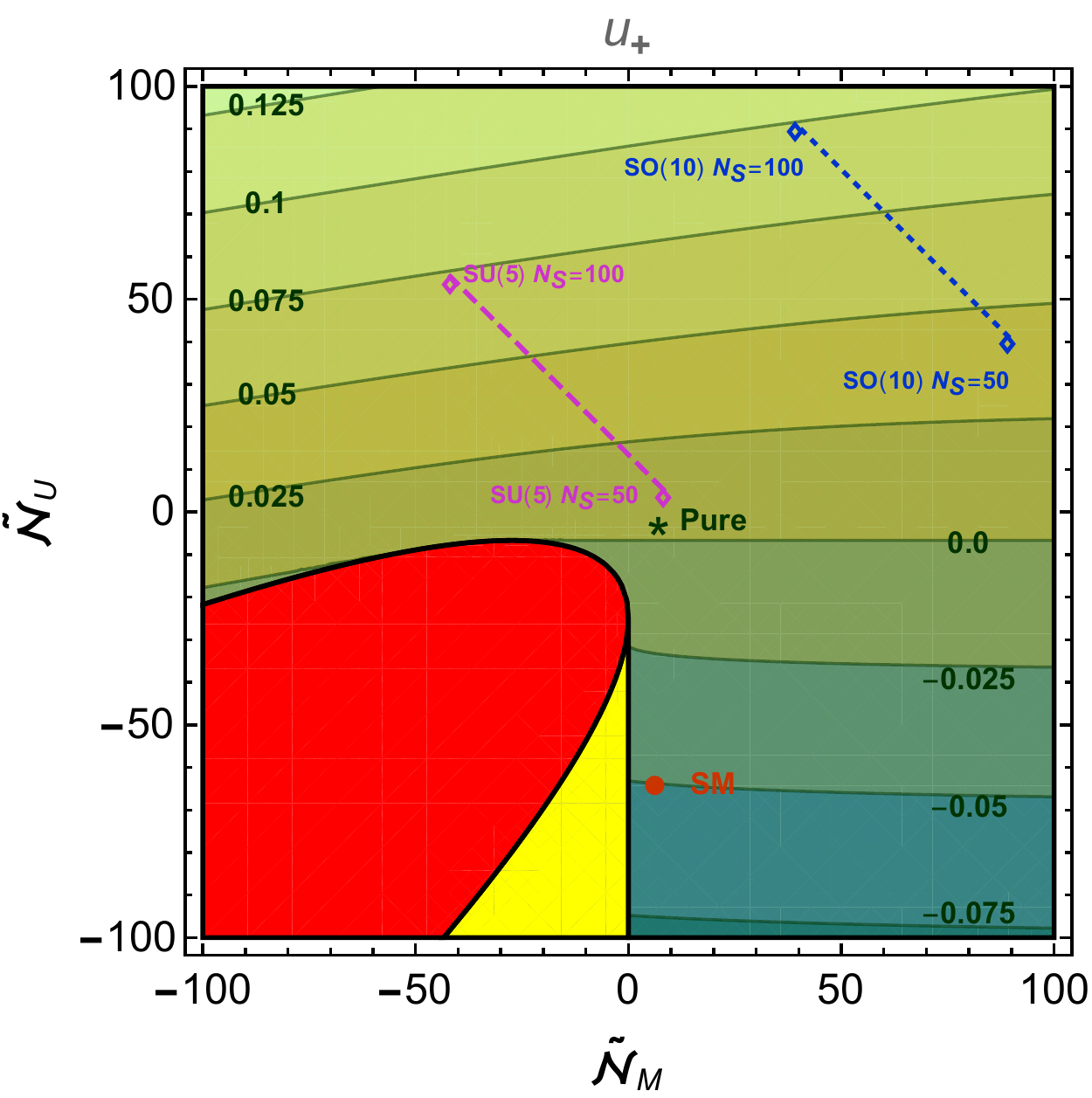}
\caption{Contour plot of the fixed point value of $u_+$ in the $(\tilde{\mathcal N}_M,\tilde{\mathcal N}_U)$ plane.
Roughly the fixed point potential or cosmological constant $u_+$ is positive for positive $\tilde{\mathcal N}_U$ and negative for negative $\tilde{\mathcal N}_U$.
}
\label{contourfig:FP values of u} 
\end{figure}

We can take $\tilde{\mathcal N}_U$ and $\tilde{\mathcal N}_M$ as the two parameters characterizing the particle content of a given model.
In our approximation they specify completely the fixed point for $u$, $w$ and $v$.
Through the dependence on $v_*$, both $u_*$ and $w_*$ depend on both numbers $\tilde{\mathcal N}_U$ and $\tilde{\mathcal N}_M$.
In Fig.\,\ref{contourfig:FP values of v} we present contour plots for $v_*$ in the $(\tilde{\mathcal N}_M,\tilde{\mathcal N}_U)$ plane, and similar for $w_*$ in Fig.\,\ref{contourfig:FP values of w}.
We show $v_+$ and $w_+$, which correspond to $x_+$ with the $+$ sign in Eq.\,\eqref{xpm solutions}. 
Similar plots for $v_-$ and $w_-$, corresponding to $x_-$, are shown in Figs.\,\ref{contourfig:FP values of v_-} and \ref{contourfig:FP values of w_-} in Appendix\,\ref{plots for second scaling solution}.
Allowed regions for stable theories obeying Eq.\,\eqref{stability condition} have different shades of green, while fixed points with instabilities are not acceptable and are indicated with yellow shades.
Regions for which no real solution exists because the argument of the square root in Eq.\,\eqref{xpm solutions} is negative are indicated in red.
Boundaries of these regions are thick lines.
We also present contour plots for $u$ in Fig.\,\ref{contourfig:FP values of u}, with a corresponding plot for $u_-$ in Fig.\,\ref{contourfig:FP values of u_-} in Appendix\,\ref{plots for second scaling solution}.



\subsection{New fixed point}
\label{discussion on New fixed point}
Our investigation shows that for $N_S+N_F>4N_V+43/6$ a new fixed point can emerge.
It is instructive to follow the change of the fixed point values as the parameters $\tilde{\mathcal N}_M$ and $\tilde{\mathcal N}_U$ are changed continuously.
We denote by ``Reuter fixed point'' the one that is connected continuously to the fixed point in pure gravity.

For the pure gravity fixed point one has $N_S=N_F=N_V=0$, and therefore
\al{
&\tilde{\mathcal N}_U=-\frac{8}{3},&
&\tilde{\mathcal N}_M=\frac{43}{6},&
&b=-\frac{158}{3}.&
}
Since $\tilde{\mathcal N}_M>0$ one has $x_-<0$, $v_->1$, which is outside the range of stability.
The Reuter fixed point therefore corresponds to 
\al{
(\text{R}):~~v_*=v_+.
}
For pure gravity one finds 
\al{
&v_*=0.152,&
&u_*=0.00411,&
&w_*=0.0271.&
}

Moving away from the pure gravity fixed point the Reuter fixed point persists for $\tilde{\mathcal N}_M>0$.
Then $x_+$ remains positive for arbitrary $b$.
A change of sign of $b$ is not relevant for the continuation of the Reuter fixed point.
Consider next the limit of small $\tilde{\mathcal N}_M$ and a change of sign of $\tilde{\mathcal N}_M$.
For $b<0$ and $\tilde{\mathcal N}_M$ close to zero one can expand
\al{
&x_+=-\frac{55}{b},&
&x_-=\frac{b}{2\tilde{\mathcal N}_M}+\frac{55}{b}.&
\label{FP for large b value}
} 
For negative $b$ the Reuter fixed point continues to negative $\tilde{\mathcal N}_M$ without any discontinuity.
We can follow the Reuter fixed point on the line $\tilde{\mathcal N}_U=-8/3$ within the green region in Figs.\,\ref{contourfig:FP values of v}--\ref{contourfig:FP values of u}.
This extends to the whole green region on these figures, for which the Reuter fixed point exists and remains associated with stable gravity at the fixed point.

Let us next look at the second solution corresponding in Eq.\,\eqref{xpm solutions} to $x_-$ with a relative minus sign.
As long as $\tilde{\mathcal N}_M$ remains positive, the solution $x_-<0$ remains outside the range of stability.
As soon as $\tilde{\mathcal N}_M<0$, one finds $x_->0$, however.
A new fixed point appears for $\tilde{\mathcal N}_M<0$, 
\al{
(\text{N}):~~v_*=v_-.
}
It starts at small negative $\tilde{\mathcal N}_M$ at $v_-\to -\infty$ .
In this limit the graviton contribution becomes $\sim \tilde{\mathcal N}_M$ and 
\al{
w_- \to \frac{\tilde{\mathcal N}_M}{192\pi^2}\left( 1 +\frac{75}{b}\right)
\nonumber
}
approaches zero.
This limit corresponds to very strong gravity for which our approximations are no longer valid.
The graviton propagator may be dominated by higher order derivative invariants.
Keeping, nevertheless, our truncation, the new fixed point is in the stable range $w_->0$ if 
\al{
-75 < b <0.
}
No second fixed point in the stable range exists for $b<-75$, $\tilde{\mathcal N}_M\to 0$.

As $\tilde{\mathcal N}_M$ becomes more negative, the fixed point (N) may move to more moderate values of $w^{-1}$ for which our approximations are valid again.
The question is whether $w_-$ is positive in this range.
This requires $x_-$ to be sufficiently small such that the inequality \eqref{condition for fixed point} is obeyed.
This condition is obeyed for
\al{
-75+\frac{22}{15}\tilde{\mathcal N}_M < b<0.
\label{condition for NM and b}
}
For a given $b<-75$ the new fixed point $x_-$ may appear in the stable region at nonzero negative $\tilde{\mathcal N}_M$, given by
\al{
\tilde{\mathcal N}_{M,\text{cr}}=\frac{15}{22}(b+75).
}
We conclude that a new fixed point in the stable range exists besides the Reuter fixed point if all three of the following conditions are obeyed
\al{
&b<0,&
&\tilde{\mathcal N}_M<0,&
&\tilde{\mathcal N}_M < \frac{15}{22}(b+75).&
}

\begin{figure}
\includegraphics[width=8.5cm]{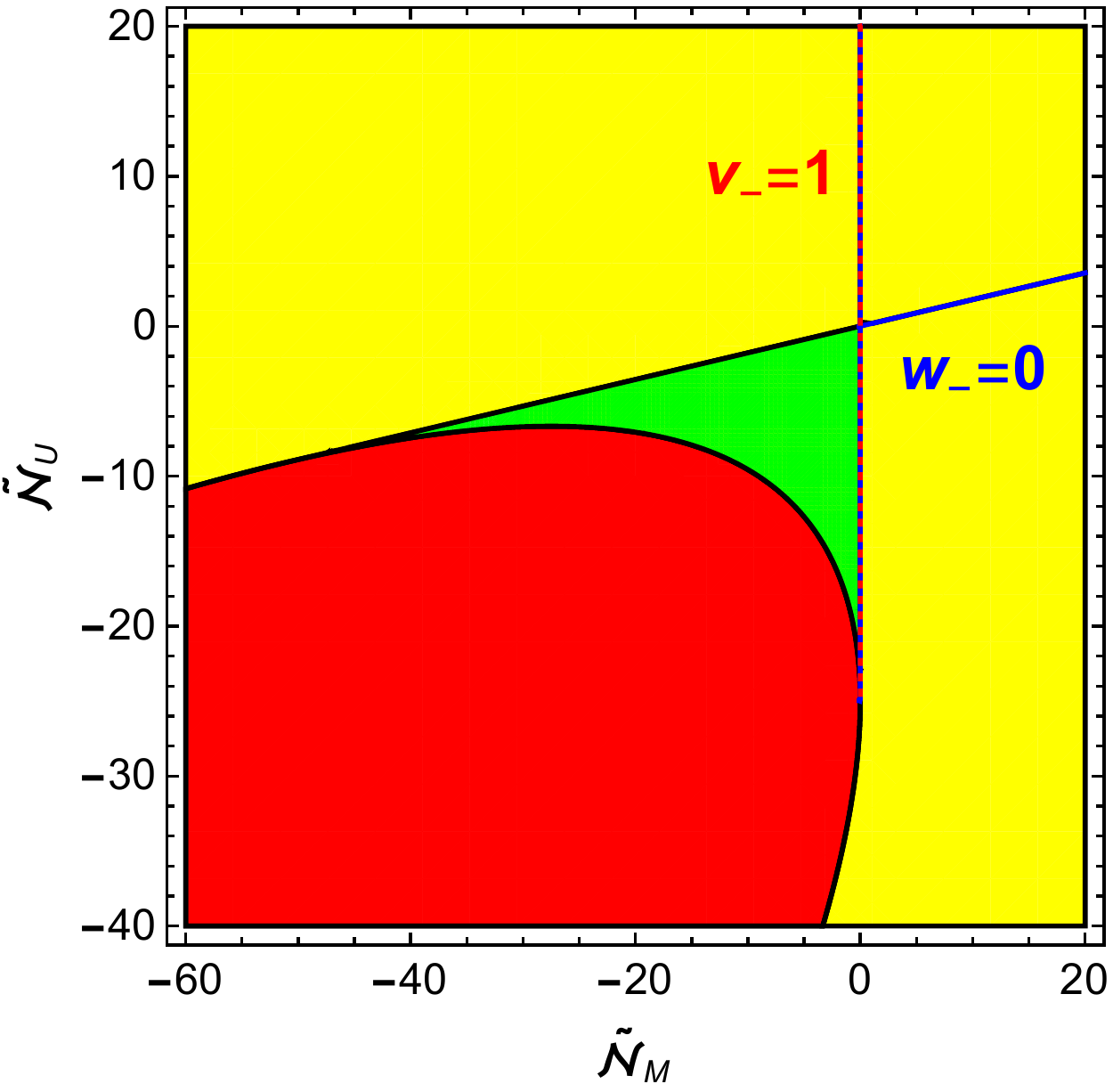}
\caption{Contour plot of the allowed (green) and excluded (red) regions for the new fixed point.
We indicate line plots of $v_-=1$ (red dotted line) and $w_-=0$ (blue solid line) in the $(\tilde{\mathcal N}_M,\tilde{\mathcal N}_U)$ plane.
For the yellow region the fixed point $u_-$, $w_-$ is unstable.
Only for the green region with $\tilde{\mathcal N}_U$ and $\tilde{\mathcal N}_M$ somewhat smaller than zero is the new fixed point stable.
As the unstable yellow region is approached, our truncation is not expected to remain valid.
}
\label{boundary lines} 
\end{figure}

In Fig.\,\ref{boundary lines} we plot in the $(\tilde{\mathcal N}_M,\tilde{\mathcal N}_U)$ plane the lines $w_-=0$ and $v_-=1$, together with the excluded red region.
We also indicate the region where no stable new fixed point exists. 
Only a rather small region of negative $\tilde{\mathcal N}_U$ and $\tilde{\mathcal N}_M$ (green) exists for which the new fixed point is stable. 

Within this region the Reuter fixed point and the new fixed point exist simultaneously.
As a consequence, one expects the existence of crossover trajectories from one fixed point to the other.
The fixed point with a higher degree of stability is attractive for the crossover trajectories.
We will discuss this issue in Sec.\,\ref{particle numbers in Standard Model and Grand Unification}.

The new fixed point typically occurs in a region close to instabilities where our truncation is not very reliable.
Extending the truncation, two outcomes are possible.
The region where the new fixed point exists either shrinks or disappears completely.
Or the region of existence grows larger, making the new fixed point relevant for a larger class of particle physics models.

\subsection{Standard model and grand unification}
\label{particle numbers in Standard Model and Grand Unification}
We next discuss a few particular particle physics models.
For the standard model with $N_S=4$, $N_V=12$ and $N_F=45$, one has
\al{
&\mathcal N_U=-62, \qquad 
\mathcal N_M=-1,&\nn[1ex]
&\tilde{\mathcal N}_U=-\frac{194}{3},\quad
\tilde{\mathcal N}_M=\frac{37}{6},\quad
b=\frac{394}{3}.&
}
Because of the positive value of $\tilde{\mathcal N}_M$ there is only one fixed point solution with $v_*<1$ and positive $w_*$.
One finds
\al{
v_*&=-10.05,&
u_*&=-0.0507,&
w_*&=0.00505.&
}
The graviton contribution is reduced due to the large negative value of $v$.
Also the contribution $\pi_0$ from scalar metric fluctuations will be reduced by a factor of $(1-v/4)^{-1}\approx 0.3$ as compared to the approximation \eqref{cV and cM from scalar metric fluctuations}.
A more accurate estimate would reduce $\tilde{\mathcal N}_U$ by one unit and enhance $\tilde{\mathcal N}_M$ by one unit.
This remains a small effect.
We here comment on the case where the type-I cutoff for gauge fields is employed.
In this case the particle content in the standard model yields $\tilde{\mathcal N}_U=-469/6$ and $\tilde{\mathcal N}_M=-125/6$ for which the fixed point is located in the unstable region (yellow region in Figs.\,\ref{contourfig:FP values of v}--\ref{contourfig:FP values of u}).
While a cutoff of type-I seems to be less well motivated in our view, the change of the standard model fixed point to the yellow region may cast doubts if our truncation is sufficient  for points very close to the boundary.
As discussed in Section\,\ref{sect: Limitation of truncation}, the inclusion of the higher derivative operators in the effective average action could improve this situation.

As another example we may take an SO(10) GUT with $N_V=45$, $N_F=48$ and $\mathcal N_U=N_S-6$, $\mathcal N_M=132-N_S$, or 
\al{
&\tilde{\mathcal N}_U=N_S-\frac{26}{3},\quad
\tilde{\mathcal N}_M=\frac{835}{6}-N_S=139.17-N_S,&\nn[2ex]
&b= \frac{688}{3}-5N_S=229.33-5N_S.&
}
For a large number of scalars $N_S>139$ the combination $\tilde{\mathcal N}_M$ turns negative.
Then $b$ is negative too, such that two solutions could exist provided the constraint \eqref{condition when a< 0,b<0} is obeyed.
This holds indeed for all $N_S$.
The second condition \eqref{condition for NM and b} for positive $w_-$ reads
\al{
\frac{688}{3}-5 N_S> -75 +\frac{22}{15}\left( \frac{835}{6} -N_S\right)
}
or
\al{
N_S <\frac{4510}{159}\approx 28.36.
}
This is not compatible with $N_S>139$.
We conclude that only the Reuter fixed point exists in the stable range for all $N_S$.
No new fixed point is realized for SO(10) GUTs.

\begin{figure*}
\includegraphics[width=8.5cm]{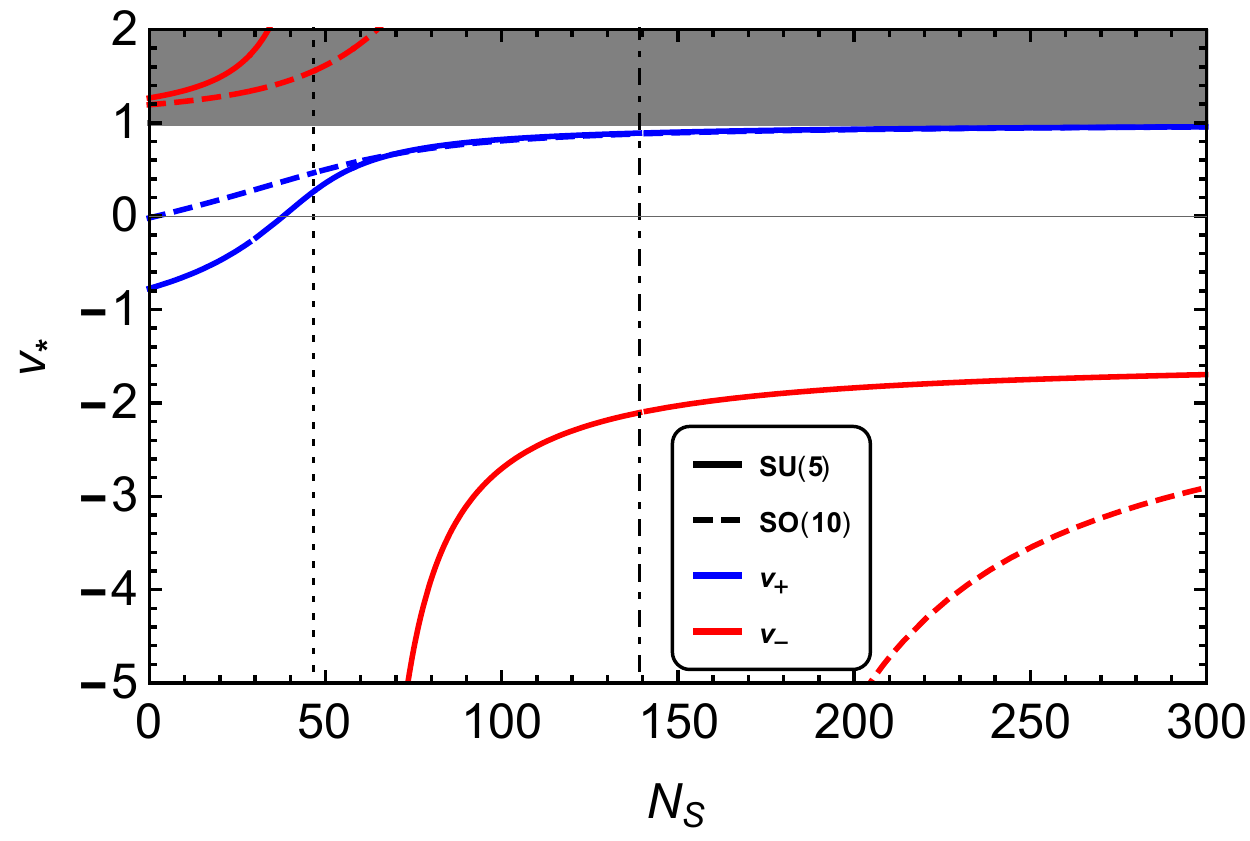}
\setlength\floatsep{10cm}
\includegraphics[width=8.5cm]{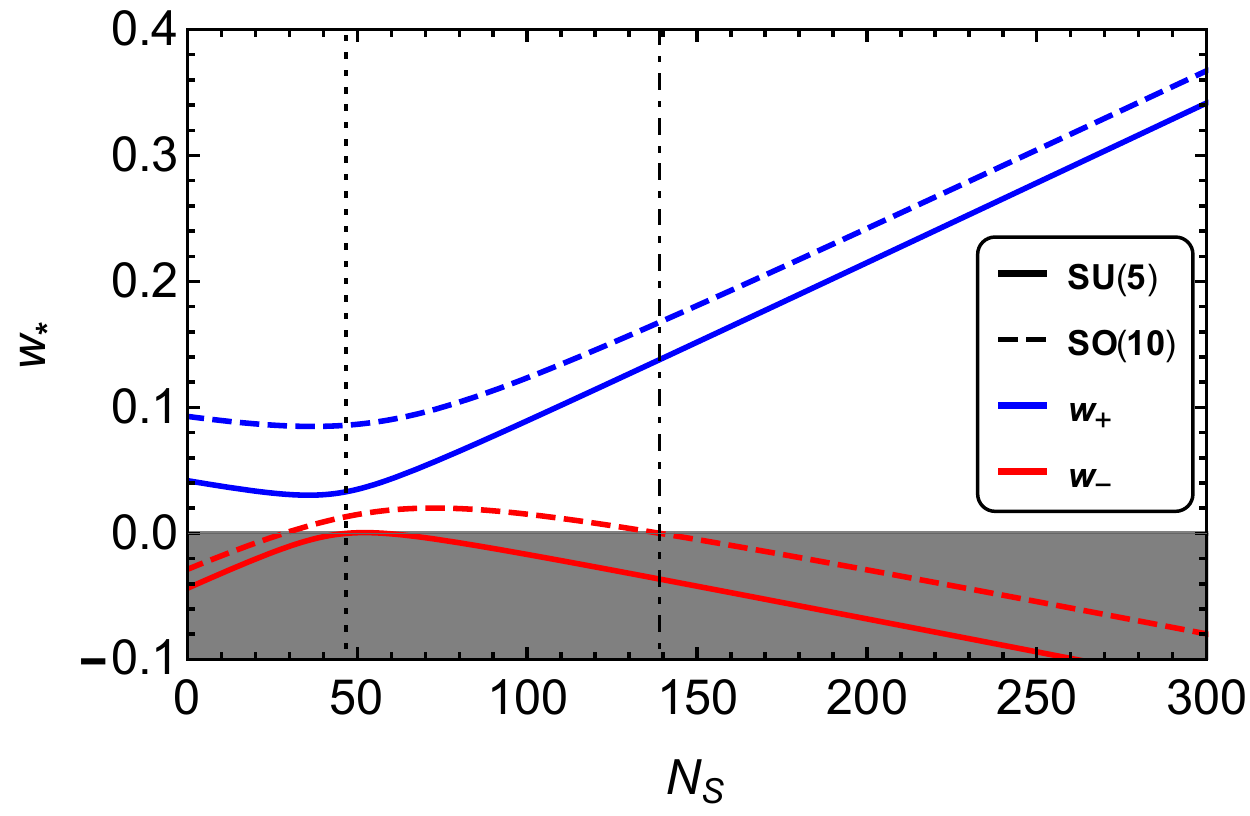}
\includegraphics[width=8.5cm]{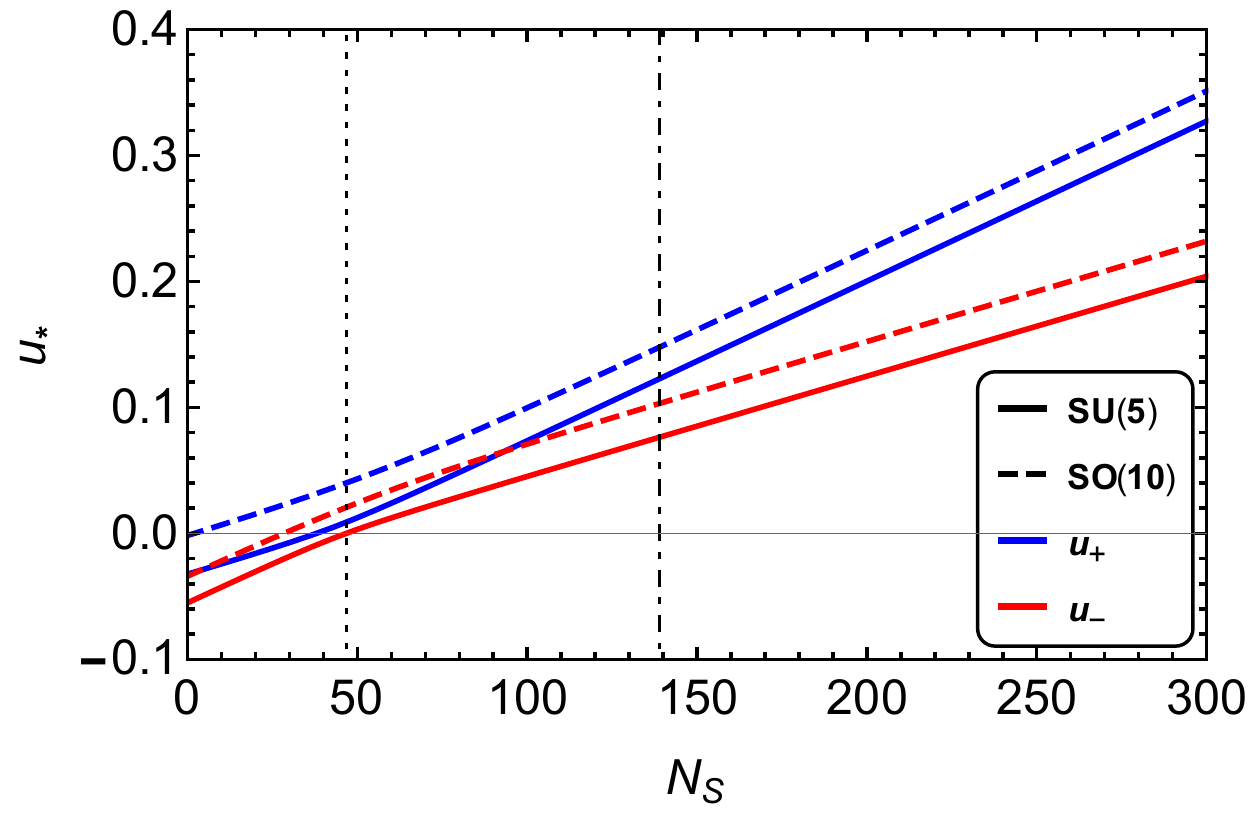}
\caption{Fixed point values of $v_*$, $u_*$ and $w_*$ as functions of $N_S$ for SU(5) (solid lines) and SO(10) (dashed lines) GUT models. 
We display both possible constant scaling solutions $(u_+,w_+,v_+)$ (blue lines in upper parts of the figures) and $(u_-,w_-,v_-)$ (red lines in lower parts of the figures).
For the Reuter fixed point the truncation becomes insufficient as $v_+$ moves close to one for large $N_S$.
For the new fixed values no stable solutions are found for large $N_S$ due to $w_-<0$.
For small $w$ our approximations are doubtful.
The gray regions are not allowed due to the conditions $v_*<1$ and $w_*>0$. The vertical dotted and dot-dashed lines are at $N_S=47$ and $N_S=139$ at which $w_{-*}=0$, respectively.}
\label{fig:FP values} 
\end{figure*}

We plot $u_*$, $w_*$ and $v_*$ as functions of $N_S$ in Fig.\,\ref{fig:FP values}.
Since $\tilde{\mathcal N}_U$ is positive for all realistic $N_S$ one finds positive $w_*$ only for $v_*>0$, and $w_*<0$ for $v_*<0$. 
The Reuter fixed point (blue dashed curves or upper dashed curves) exists for all $N_S$.
For $N_S\gsim 100$ it moves rather close to $v=1$, however, such that our truncation may no longer be reliable.
The new fixed point (red dashed or lower dashed curve) has positive $w_-$ only in a range where $v_->1$, such that it is unstable for all $N_S$.

With
\al{
\mathcal W&=\sqrt{\frac{b^2+440\tilde{\mathcal N}_M}{b^2}}
=\sqrt{1+\frac{440\tilde{\mathcal N}_M}{b^2}}\,,
}
we can write the fixed point solutions as
\al{
v_\pm=1-\frac{15N_S-688}{12N_S-1670}\left( 1\mp \mathcal W \right) .
}
For very large $N_S\to \infty$ one has $\mathcal W \to 0$ such that $v_+$ approaches $1$ and $v_-$ goes to $-3/2$.
Only $v_+$ corresponds to positive $w_*>0$ in this case.
Indeed, for the two solutions $v_+$ and $v_-$ one finds the fixed points for $w$
\al{
w_\pm&=\frac{u_\pm}{v_\pm}\nn
& =\frac{6N_S-835}{1152\pi^2(15N_S-688)}\left( \frac{450}{1\mp \mathcal W}-(15N_S-688) \right) .
}
For large $N_S$ only $w_+$ corresponds to stable gravity, $w_+>0$.
In particular, for $N_S\to \infty$ one has 
\al{
&w_+\to \frac{1}{12672\pi^2}\left( 159N_S-2125\right)\,,\nn[2ex]
&w_-\to \frac{1}{1152\pi^2}\left( -6N_S+925\right)\,.
}

In our truncation we find that a fixed point with constant $u$, $w$ and $v$ exists for arbitrary $N_S$.
The mechanism is a cancellation between negative contributions to $c_M$ from scalar and fermion fluctuations and a large positive contribution from the graviton fluctuations.
As $N_S$ increases, the size of the graviton contribution also has to increase.
This is achieved by moving $v$ close to one, realizing a substantial enhancement of the graviton contribution.
This mechanism implies, however, that for large $N_S$ our approximation becomes questionable since $v$ comes close to one.
Already for $N_S\gsim 50$ one may doubt the validity of our truncation.
This value is too low for a realistic SO(10)-GUT model.
In consequence, we will not be able to make robust statements about SO(10)-GUT models.

For SU(5)-GUT models one has $N_V=24$, $N_F=45$ and therefore
\al{
&\tilde{\mathcal N}_U=N_S-\frac{134}{3}\,,&
&\tilde{\mathcal N}_M=\frac{349}{6}-N_S\,,&
&b=\frac{526}{3}-5N_S\,.&
}
In this case also more moderate numbers of scalar fields are possible.
For a minimal set with a real $24$-plet and a complex $5$-plet one has $N_S=34$ and therefore moderate values of $\tilde{\mathcal N}_U$ and $\tilde{\mathcal N}_M$,
\al{
&\tilde{\mathcal N}_U=-\frac{32}{3}\,,&
&\tilde{\mathcal N}_M=\frac{145}{6}\,.&
}
The corresponding fixed point values are
\al{
&v_*=-0.123\,,&
&u_*=-0.00375\,,&
&w_*=0.0304\,.&
}
They are not far from the values for pure gravity.

\section{Quantum gravity predictions for the Higgs sector}
\label{Higgs and quantum gravity}
In this section we address the question raised in the Introduction, namely whether the quartic coupling $\lambda_H$ of the Higgs sector corresponds to an irrelevant coupling near the UV-fixed point and therefore becomes predictable by quantum gravity.  
For this purpose we have to expand the effective potential $U(\rho)$ in terms of $\rho=h^\dagger h$, where $h$ is the Higgs doublet.
The flow of the first three coefficients of this expansion describes the flow of the cosmological constant, the mass term and the quartic coupling of the Higgs boson.
We will work at fixed values for possible other scalar fields, typically set to zero. 
We also neglect possible small effects from the nonzero gauge and Yukawa couplings of the Higgs doublet.
In this case constant values for $N_F$ and $N_V$ can be taken, and similarly for the number of scalars beyond the Higgs doublet $N_S-4$.
These numbers characterize the short-distance model of particle physics into which the standard model is embedded. 

Besides the expansion of $U(\rho)$ we perform a similar expansion for $F(\rho)$.
We truncate the expansion at second order in the $\rho$ derivatives.
This leaves us with the flow of six couplings describing the deviations from the UV-fixed point or constant scaling solution.
In this space of couplings we compute the stability matrix for small deviations from the fixed point and its eigenvalues, the critical exponents.
For the standard model and GUT models with not too large $N_S$, such that our truncation remains reliable, we find that indeed the quartic Higgs coupling corresponds to an irrelevant parameter at the UV-fixed point.

\subsection{Mass term and couplings}
For the Higgs sector we are interested in $\sqrt{\rho}$ near the Fermi scale $\varphi_0$.
For the range of $k$ of interest here this corresponds to very small values of $\tilde \rho$.
We therefore expand the effective potential $U\fn{\rho}$ around $\rho=0$,
\al{
U\fn{\rho}=V+m^2_H \rho +\frac{\lambda_H}{2}\rho^2\,,
}
and correspondingly for the dimensionless potential $u\fn{\tilde \rho}$,
\al{
u\fn{\tilde \rho}=u_0 +\tilde m_H^2 \tilde \rho +\frac{\tilde \lambda_H}{2}\tilde \rho^2\,.
}
We also expand
\al{
w\fn{\tilde \rho}=w_0 +\frac{\xi_H}{2}\tilde  \rho+\frac{w_2}{2}\tilde \rho^2\,,
\label{w tilde rho}
}
with $\xi_H$ a nonminimal coupling between the Higgs scalar and gravity of the type $-\xi_H h^\dagger h R/2$.
The function $\tilde\xi$ in Eqs.\,\eqref{mass and non-minimal} and \eqref{cV  and cM from gauge scalar} reads 
\al{
\tilde\xi=\xi_H +6w_2 \tilde \rho\,,
}
where $N_\xi=4$ for the standard model and larger suitable $N_\xi$ for larger representations in which the Higgs doublet is embedded in GUT models.

The flow equation for $\tilde m_H^2\fn{\tilde \rho}=\p_{\tilde \rho} u\fn{\tilde \rho}$ is obtained by taking a $\tilde \rho$ derivative of the first equation \eqref{total results of beta functions},
\al{
\p_t \tilde m_H^2= 2\tilde \rho\, \p_{\tilde \rho} \tilde m_H^2  + (A-2) \tilde m_H^2 -\frac{1}{2}A \xi_H v +\frac{1}{32\pi^2}\frac{\p \tilde{\mathcal N}_U}{\p\tilde\rho}\,,
}
where the graviton induced anomalous dimension $A$ reads
\al{
A=\frac{5}{24\pi^2w(1-v)^2}\,.
}
Here we employ $v\fn{\tilde \rho}=u\fn{\tilde \rho}/w\fn{\tilde \rho}$ and
\al{
\p_{\tilde \rho} v=\frac{1}{w}\left( \tilde m_H^2 -\frac{\xi_H v}{2}\right)\,,
\label{flow equation of field-depend v}
}
with $\xi_H\fn{\tilde \rho}=2\p_{\tilde \rho} w\fn{\tilde \rho}$.

Taking a further $\tilde \rho$ derivative of Eq.\,\eqref{flow equation of field-depend v} yields the flow equation for $\tilde \lambda_H$,
\al{
\p_t \tilde \lambda_H &= 2\tilde \rho\, \p_{\tilde \rho}\tilde \lambda_H +A(\tilde \lambda_H  - vw_2) \nn
&+\frac{A}{w} \bigg\{ \frac{2}{1-v} \left( \tilde m_H^2 -\frac{\xi_H v}{2}\right)^2  
-\xi_H \left( \tilde m_H^2 -\frac{\xi_H v}{2} \right) \bigg\}\nn
&\quad
+\frac{1}{32\pi^2}\frac{\p^2 \tilde{\mathcal N}_U}{\p \tilde\rho^2}\,.
}
Similarly, one finds the flow equation for $\xi_H\fn{\tilde \rho}=2 \p_{\tilde \rho}w\fn{\tilde \rho}$ from the $\tilde \rho$ derivative of the second equation \eqref{total results of beta functions},
\al{
\p_t \xi_H= 2\tilde \rho\, \p_{\tilde \rho} \xi_H +\frac{15A}{4} \left( \tilde m_H^2 -\frac{\xi_H v}{2}\right)+\frac{1}{48\pi^2}\frac{\p\tilde{\mathcal N}_M}{\p \tilde\rho} \,.
}
For $w_2\fn{\tilde\rho}=\p^2w/\p \tilde\rho^2=(\p \xi_H/\p\tilde\rho)/2$ one obtains 
\al{
&\p_t w_2 =2\tilde \rho\, \p_{\tilde \rho}w_2 +2w_2 +\frac{15A}{8}\bigg\{ \tilde\lambda_H -w_2 v \nn
&
-\frac{\xi_H}{w}\left( \tilde m_H^2 -\frac{\xi_Hv}{2} \right) +\frac{2}{w(1-v)}\left( \tilde m_H^2 -\frac{\xi_H v}{2}\right)^2 \bigg\}\nn
&\quad
+\frac{1}{96\pi^2}\frac{\p^2\tilde{\mathcal N}_M}{\p \tilde\rho^2}.
}

For $\tilde\rho$-independent $\tilde{\mathcal N}_U$ and $\tilde{\mathcal N}_M$ these flow equations have a simple scaling solution
\al{
&\tilde m_{H*}^2=0,&
&\tilde \lambda_{H*}=0,&
&\xi_{H*}=0,&
&w_{2*}=0,&
\label{Gaussian matter fixed point}
}
which correspond to $\tilde \rho$-independent $u$ and $w$.
The corresponding fixed point values $u_{0*}$ and $w_{0*}$ are given by $u_+$ and $w_+$ as discussed for the constant scaling solutions in Sec.\,\ref{UV fixed point analysis}.
If the gauge and Yukawa couplings are also zero at the fixed point, only the gravitational interactions remain at this fixed point.

Vanishing fixed point values for $\tilde m_H^2$, $\tilde\lambda_H$, $\xi_H$, and $w_2$ follow directly if both $\tilde{\mathcal N}_U$ and $\tilde{\mathcal N}_M$ are constants.
This is only an approximation for small matter couplings.
For the example of a single gauge boson coupling to the Higgs doublet with gauge coupling $e$, the $\rho$ dependence in Eq.\,\eqref{c_M with scalar with gauge boson} generates an additional term for the flow of $\tilde m^2\fn{\tilde \rho}$,
\al{
\Delta \p_t \tilde m_H^2\fn{\tilde \rho}=-\frac{3e^2}{32\pi^2(1+e^2\tilde \rho)^2}\,,
} 
and similar for nonzero Yukawa couplings and quartic scalar couplings.
For $e\neq 0$ the scaling solution is no longer independent of $\tilde\rho$, with $\tilde m_H^2\neq 0$.
For vanishing matter couplings at the fixed point, $e_*^2=0$, these corrections do not change the fixed point.
They modify,  however, the stability matrix for small deviations from the fixed point.

We note at this point that the constant scaling solution \eqref{Gaussian matter fixed point} is not the only possible scaling solution.
For example, one may investigate scaling solutions with $\tilde\rho$-dependent $w$ reaching a form $w\sim \xi_\infty \tilde\rho/2$ for $\tilde \rho\to \infty$.
Such scaling solutions have been discussed in the context of dilaton quantum gravity~\cite{Henz:2013oxa,Henz:2016aoh}.

\subsection{Critical exponents}
For small deviations from this scaling solution we discuss the (truncated) stability matrix $T$ that describes the linear approximation for the vicinity of the fixed point 
\al{
\p_t \tilde g_i=-T_{ij} (\tilde g_j -\tilde g_{j*})\,,
}
with six couplings
\al{
\tilde g_i=(u_0,\, w_0,\,\tilde m_H^2,\, \xi_H,\,\tilde \lambda_H,\,w_2).
}
The stability matrix obtains as
\al{
T_{ij}=-\frac{\p\beta_i}{\p g_j}\bigg|_{g_j=g_{j*}}\,,
}
where
\al{
\p_t g_i =\beta_i\,.
}
The critical exponents are the eigenvalues of the stability matrix.
Eigenvectors with respect to positive critical exponents are relevant couplings, whereas the ones for negative exponents are irrelevant couplings.
The irrelevant couplings are predicted to take their fixed point values.
The six couplings $\tilde g_i=(u_0,\, w_0,\, \tilde m_H^2,\,  \tilde\xi_H,\, \tilde \lambda_H,\,w_2)$ are related to the values of $u\fn{\tilde\rho}$, $w\fn{\tilde \rho}$, $\tilde m_H^2\fn{\tilde \rho}$, etc., at $\tilde \rho=0$.

We first neglect the terms proportional to $\p\tilde{\mathcal N}_U/\p \tilde\rho$ and $\p\tilde{\mathcal N}_M/\p\rho$. 
In this approximation the stability matrix decays into $2\times 2$ blocks.
The first block involves $(\tilde g_1,\, \tilde g_2)=(u_0,\, w_0)$,
\al{
T^{(12)}=\pmat{
4-A &&&& Av\\[3ex]
\displaystyle -\frac{15A}{8} &&&& 2+\displaystyle\frac{15Av}{8}
}.\label{T12}
}
The second block for $(\tilde g_3,\, \tilde g_4)=(\tilde m_H^2,\, \xi_H)$ reads
\al{
T^{(34)}=\pmat{
2-A && &\displaystyle\frac{Av}{2}\\[4ex]
\displaystyle -\frac{15A}{4} &&& \displaystyle\frac{15Av}{8}
},\label{T34}
}
while the third block for $(\tilde g_5,\, \tilde g_6)=(\tilde \lambda_H,\, w_2)$ becomes
\al{
T^{(56)}=\pmat{
-A &&&& Av \\[3ex]
\displaystyle -\frac{15A}{8} &&&& -2 + \displaystyle\frac{15Av}{8}
}.\label{T56}
}
Here $A$ and $v$ are evaluated for $\tilde \rho=0$.
This block structure continues for higher couplings.
It is a result of our simple truncation.
Including for the scalar fluctuation contribution the dependence of $\p_t u$ and $\p_t w$ on $\tilde m_H^2$ and $\tilde \xi$ will mix the different blocks.

Consider first the $(\lambda_H,w_2)$ sector for which the critical exponents are
\al{
\theta_{5,6}&=-\frac{1}{2}\Bigg\{ A+2 -\frac{15Av}{8}\nn
&\quad\pm\sqrt{(A-2)^2-\frac{15A(A+2)v}{4} +\left( \frac{15Av}{8}\right)^2} \Bigg\}\,.
}
For $|v|\ll |(A-2)/(2A)|$ one may expand
\al{
\theta_5&=-A+\frac{15A^2v}{8(A-2)}\,,\nn[2ex]
\theta_6&=-2-\frac{15Av}{4(A-2)}\,.
}
We may also expand for $|Av|\ll (A-2-15Av/8)^2/15$, where
\al{
\theta_5&=-A-\frac{15Av}{8}+\frac{15Av}{4(A-2-15Av/8)^2}\,,\nn[2ex]
\theta_6&=-2-\frac{15Av}{4(A-2-15Av/8)^2}\,,
}
which covers a region of $A$ closer to two.
In the region of validity of these expansions both couplings are irrelevant and therefore predictable.
The eigenvector of the critical exponent $\theta_5$ is mainly $\lambda_H$, while for $\theta_6$ it is $w_2$.
Another expansion for large negative $v$, $-v\gg |(A+2)/(2A)|$, yields
\al{
\theta_5&=\frac{16}{15v}\,,\nn[2ex]
\theta_6&=\frac{15Av}{8}-A-2\,.
\label{critical exponents of 5 and 6 in approximated forms}
}
In this limit one finds again negative critical exponents $\theta_5$ and $\theta_6$.
For small $v$ and negative $v$ both $\lambda_H$ and $w_2$ are irrelevant couplings predicted to be zero at the fixed point.

As $v$ increases from zero toward one, the fixed point approaches the region where our truncation is expected to break down.
For a given $A$ the eigenvalues $\theta_5$ and $\theta_6$ remain real as long as $v<v_\text{cr}$, with 
\al{
v_\text{cr}=\frac{8}{15}\left( 1\mp \sqrt{\frac{2}{A}}\right)^2.
}
The minus sign holds for $v<8/15+16/(15A)$ and the plus sign for $v>8/15+16/(15A)$.
While complex critical exponents for $v>v_\text{cr}$ are no problem, they may nevertheless be artifacts of an insufficient truncation.
For $v=v_\text{cr}$, where the imaginary part starts to set in, one has
\al{
\theta_5=\theta_6=-\sqrt{2A}\,.
}
Both $\lambda_H$ and $w_2$ are irrelevant.

The eigenvalues in the sector $\tilde m_H^2$, $\xi_H$ are shifted by two as compared to $\theta_{5,6}$,
\al{
&\theta_3=\theta_5+2,&
&\theta_4=\theta_6+2.&
}
Similarly, one finds in the sector $\delta u_0$, $\delta w_0$
\al{
&\theta_1=\theta_5+4,&
&\theta_2=\theta_6+4.&
}
Correspondingly, the critical exponents for higher order couplings are shifted to more negative values.
For $-2<\theta_{5,6}<0$ there are four relevant couplings $\delta u_0=u_0-u_{0*}$, $\delta w_0=w_0-w_{0*}$, $\tilde m_H^2$, and $\xi_H$, while all other couplings are irrelevant. 
For more negative $\theta_5$, $\theta_6$ the number of relevant couplings is reduced.

\subsection{Graviton induced anomalous dimension}
\label{anomalous dimension induced by graviton}
A crucial quantity for predictions of the properties of the effective potential for the Higgs scalar is the graviton induced anomalous dimension $A$~\cite{Wetterich:2017ixo,Wetterich:2018poo,Pawlowski:2018ixd,Wetterich:2019qzx}.
For $\eta_g=0$ it is given by
\al{
A=\frac{5}{24\pi^2 w(1-v)^2}&=\frac{5}{24\pi^2 wx^2}
=\frac{80}{x(75+2\tilde{\mathcal N}_Mx)}\,,
\label{graviton induced anomalous dimension}
}
with $x=x_+=1-v_+$ given by the Reuter fixed point \eqref{xpm solutions}.

For pure gravity one finds with $x=0.848$, $\tilde{\mathcal N}_M=43/6$ a value
\al{
A=1.082,
} 
and therefore critical exponents 
\al{
&\theta_5=-1.39+0.491i,&
&\theta_6=-1.39-0.491i.&
}
For the standard model one has $x=11.05$, $\tilde{\mathcal N}_M=37/6$, resulting in 
\al{
A=0.0343,
}
and critical exponents
\al{
&\theta_5=-0.0258.&
&\theta_6=-2.65,&
}

Because of the small value of $A$ the influence of the off-diagonal elements in the matrices \eqref{T12}--\eqref{T56} is small.
In a rough approximation the eigenvectors correspond simply to the couplings $\delta u_0$, $\delta w_0$, $\tilde m_H^2$, $\xi_H$, $\lambda_H$, and $w_2$.
There are three relevant couplings that may be associated with $\delta u_0$, $\delta w_0$, and $\tilde m_H^2$.
The couplings $\xi_H$, $\lambda_H$, and $w_2$ are irrelevant.
The value of $\theta_5$ is rather close to zero, such that the critical exponents for $\delta u_0$, $\tilde m_H^2$, and $\lambda_H$ are not far from the canonical scaling exponents.
The approach of the quartic coupling $\lambda_H$ to its fixed point value $\lambda_{H*}=0$ is rather slow.
As a consequence, the small effects of nonzero gauge and Yukawa couplings in the vicinity of the fixed point have to be taken into account for the flow of $\lambda_H$ away from the fixed point, even for the region of large $k$ where the metric fluctuations are important.

\begin{figure}
\includegraphics[width=8.6cm]{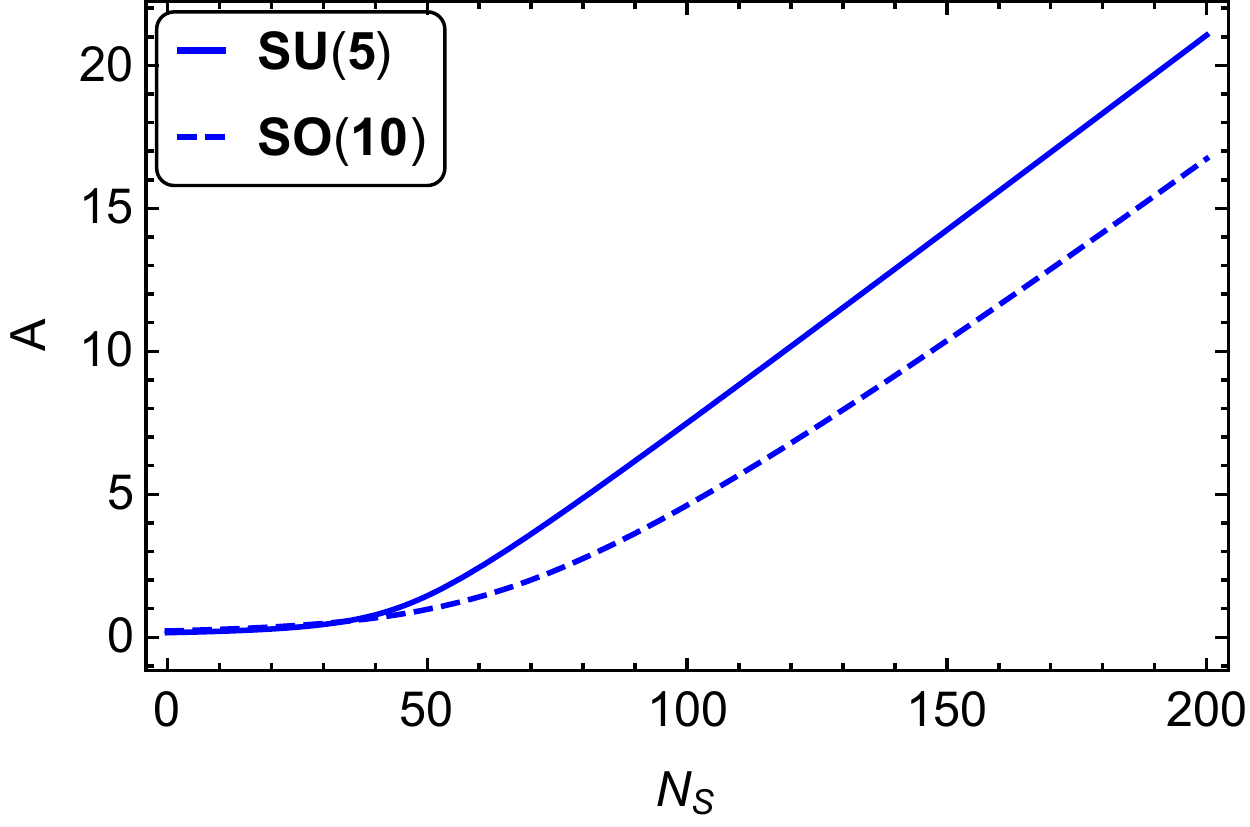}
\caption{Graviton induced anomalous dimension $A$, given by Eq.\,\eqref{graviton induced anomalous dimension}, for SU(5) and SO(10) GUT models as a function of the number of scalars $N_S$.}
\label{fig:Anomalous dimension} 
\end{figure}

To get some intuition on the size of $A$ for GUT models we may consider the limit of large $N_S$.
For $|\tilde{\mathcal N}_M| \ll b^2/440$ one may use Eq.\,\eqref{FP for large b value} for $b<0$, resulting in
\al{
A\approx \frac{16|b|}{825}\left( 1 + \frac{22\tilde{\mathcal N}_M}{15|b|} \right)^{-1}.
}
For large values of $|b|$ the anomalous dimension becomes much larger than one.
For the example of an SO(10) GUT with $N_S=317$ scalars (complex 126, complex 10, and real 45) one has $|b|=1355.7$ and $\tilde{\mathcal N}_M=-177.83$, and one finds $A=32.29$.
For large $N_S$ one observes a linear increase of $A$ with $N_S$, as can be seen from Fig.\,\ref{fig:Anomalous dimension} where we plot $A$ as a function of $N_S$.
The large values of $A$ arise from a fixed point value of $v$ rather close to one.
This is outside the range of validity of our truncation.
The range of very large values of $A$ should therefore not be taken as realistic.

For SU(5) one also has an increase of $A$ for large $N_S$.
For a minimal set with $N_S=34$ one has
\al{
A=1.004.
}
One finds for the critical exponents
\al{
&\theta_5=-0.5075\,.&
&\theta_6=-2.171\,,&
}
Thus $\lambda_H$, $\xi_H$, and $w_2$ are irrelevant couplings, while $\tilde m_H^2$, $\delta u_0$, and $\delta w_0$ are relevant.

\begin{figure*}[!htp]
\includegraphics[width=8cm]{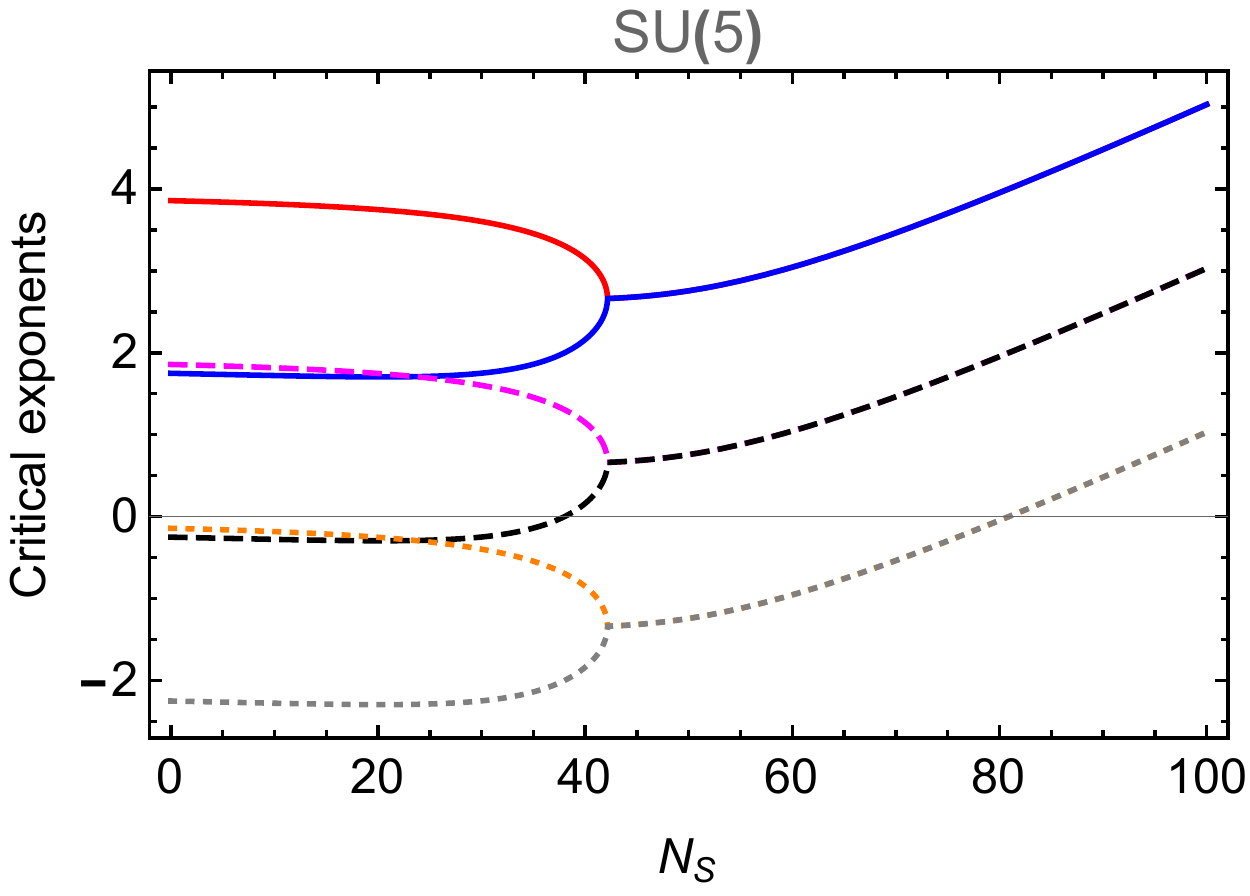}
\setlength\floatsep{10cm}
\includegraphics[width=9.4cm]{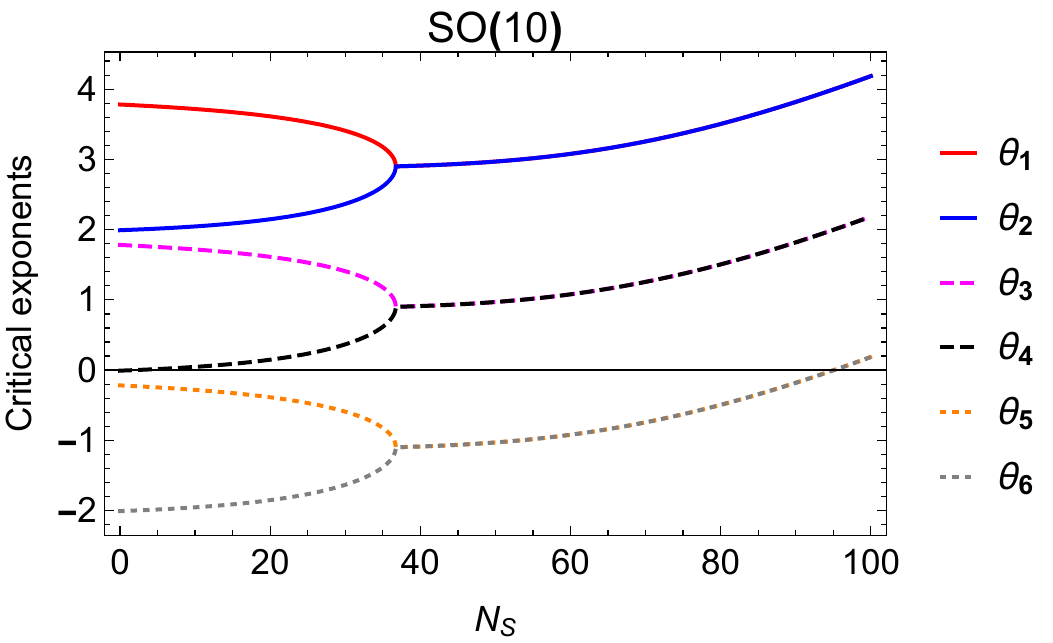}
\caption{Critical exponents (real parts of eigenvalues of $T^{(12)}$, $T^{(34)}$ and $T^{(56)}$) as functions of $N_S$ in SU(5) and SO(10) GUTs.
For small values of $N_S$ the critical exponents are not far from the values given by the canonical mass dimension of the associated parameters.
For the SO(10) model the curves for small $N_S$ from top to bottom correspond approximately to $u_0$, $w_0$, $\tilde m_H^2$, $\xi_H$, $\lambda_H$, and $w_2$.
In particular, the critical exponent for the quartic coupling of the Higgs scalar is negative, making this irrelevant parameter predictable.
As $N_S$ increases, the quartic scalar coupling becomes first more and more irrelevant, with decreasing $\theta_5$.
For $N_S$ close to $40$ the eigenvalues of the stability matrix develop an imaginary part, and the absolute value of the critical exponents starts to increase.
For large $N_S$ our approximations no longer remain valid.
}
\label{fig:CEs of SO10} 
\end{figure*}
In Fig.\,\ref{fig:CEs of SO10} we show the dependence of the critical exponents on the number of scalar fields for SU(5) and SO(10) GUTs.
For SU(10) there is a critical threshold for the number of scalars $N_\text{cr}\approx 37$ such that for $N_S<N_\text{cr}$ all critical exponents are real, while for $N_S>N_\text{cr}$ they develop an imaginary part.
For $N_S<N_\text{cr}$ both $\theta_5$ and $\theta_6$ are negative, while the other critical exponents $\theta_1,..., \theta_4$ are positive.
As $N_S$ approaches $N_\text{cr}$ from below the critical exponent $\theta_5$ decreases, while $\theta_6$ increases. 
Both critical exponents have a common negative value for $N=N_\text{cr}$.
For $N_S>N_\text{cr}$ the real part of the critical exponents $\theta_{5,6}$ increases, even becoming positive as one comes close to  the boundary of the region of validity of our truncation.
This part should not be trusted.
For SU(5) the situation is qualitatively similar, with $N_\text{cr}=43$.
In contrast to SO(10), only three couplings are relevant for $N_S<39$.

\begin{figure*}
\includegraphics[width=8.5cm]{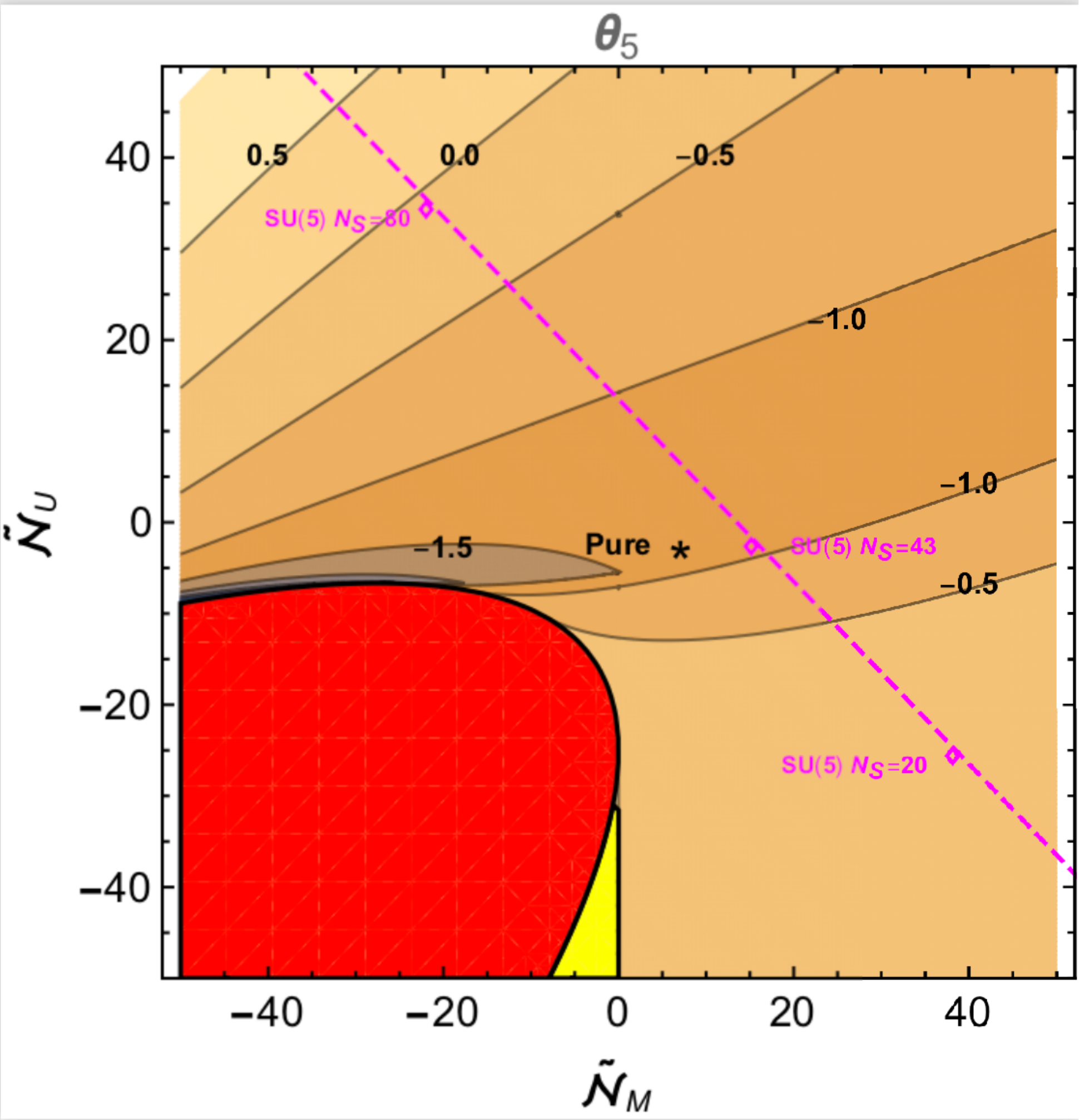}
\setlength\floatsep{20cm}
\includegraphics[width=8.5cm]{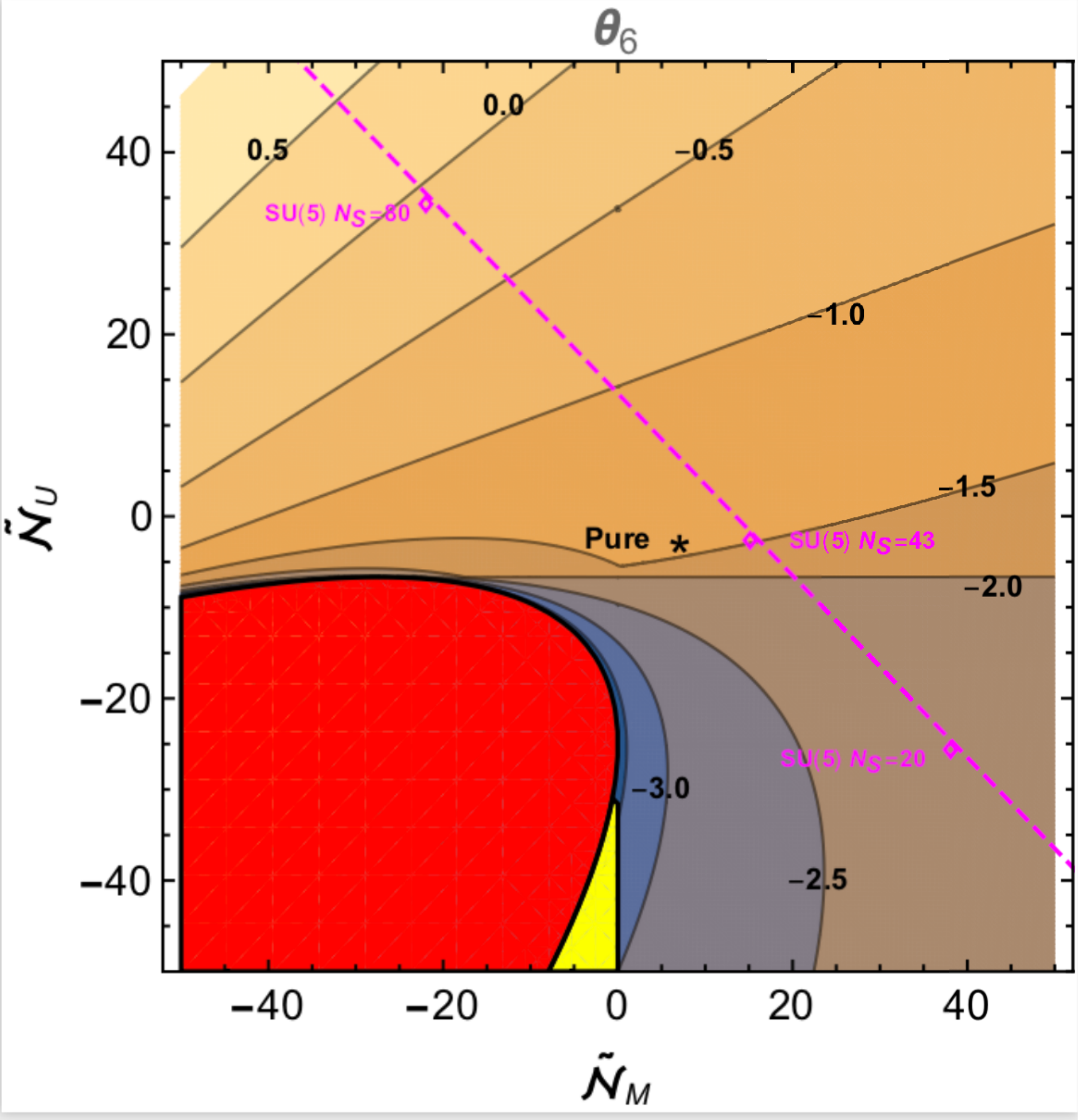}
\caption{
Contour plots for the real parts of $\theta_5$ (left) and $\theta_6$ (right).
The red (no constant scaling solution) and yellow (unstable solution due to $w_+<0$) regions are excluded.
The pink dashed line shows the change of $N_S$ for the SU(5)-GUT model. 
It corresponds to the plots for $\theta_5$ and $\theta_6$ in Fig.\,\ref{fig:CEs of SO10}.
For $N_S\geq 43$ in SU(5) GUT both $\theta_5$ and $\theta_6$ have an imaginary part, so that their real parts are degenerate.
}
\label{fig:critical exponents theta5 and theta6} 
\end{figure*}

In Fig.\,\ref{fig:critical exponents theta5 and theta6} we show contour plots of $\theta_5$ and $\theta_6$ in the ($\tilde{\mathcal N}_U$, $\tilde{\mathcal N}_M$) plane. 
Both exponents are negative except for the upper left corner where our truncation becomes unreliable.

\subsection{Beyond the graviton approximation}
In the graviton approximation $\tilde{\mathcal N}_U$ and $\tilde{\mathcal N}_M$ are treated as constants. 
For this approximation the only $\tilde\rho$ dependence of the flow generators in Eq.\,\eqref{total results of beta functions} (beyond the canonical terms $2\tilde\rho\, \p_{\tilde\rho}u-4u$ and $2\tilde\rho\,\p_{\tilde\rho}w-2w$) arises through the $\tilde\rho$ dependence of $v$.
This results in the block structure of the stability matrix.
The $\tilde\rho$ dependence of the effective particle numbers $\tilde{\mathcal N}_U$ and $\tilde{\mathcal N}_M$ induces a mixing between the different blocks.

Let us start with the $\tilde\rho$ dependence of $\tilde{\mathcal N}_M$ induced by the contribution $-3N_\xi \tilde\xi/2$ [cf. Eq.\,\eqref{particle numbers}] and neglect first other contributions to the $\tilde\rho$ dependence of $\tilde{\mathcal N}_U$ and $\tilde{\mathcal N}_M$.
With the polynomial truncation \eqref{w tilde rho},
\al{
&\tilde\xi=\xi_H +6w_2\tilde\rho\,,&
&\p_{\tilde\rho}\tilde\xi=8w_2\,,&
}
one has 
\al{
&\frac{\p \tilde{\mathcal N}_M}{\p\tilde\rho}=-12N_H w_2\,,&
&\frac{\p^2 \tilde{\mathcal N}_M}{\p\tilde\rho^2}=0\,.&
}
Here $N_\xi=N_H$ is the dimension of the multiplet to which the Higgs doublet belongs.
This results in additional contributions for the flow generators for $w_0$ and $\xi_H$,
\al{
\p_t w_0&=\beta_{w_0}=\cdots-\frac{N_H}{64\pi^2}\xi_H\,,\nn[2ex]
\p_t \xi_H&=\beta_{\xi}=\cdots-\frac{N_H}{4\pi^2} w_2\,.
}

These additional terms do not affect the constant scaling solutions \eqref{fixed point values for u- and w-} and \eqref{Gaussian matter fixed point}.
They influence, however, the stability matrix by mixing the blocks, 
\al{
T_{24}&=-\frac{\p \beta_w}{\p \xi_H}=\frac{N_H}{64\pi^2}\,,\\[2ex]
T_{46}&=-\frac{\p\beta_\xi}{\p w_2}=\frac{N_H}{4\pi^2}\,.
}
This does not change the eigenvalues or critical exponents.
The stability matrix now has a triangle structure
\al{
T=\pmat{
T^{(12)} && \tilde T^{(12)} && 0\\[2ex]
0 && T^{(34)} && \tilde T^{(34)} \\[2ex]
0 && 0 && T^{(56)}
},
}
with $T_{24}$ and $T_{46}$ contributing to the $2\times 2$-matrices ${\tilde T}^{(12)}$ and ${\tilde T}^{(34)}$, respectively. 
In consequence, the eigenvalues remain the eigenvalues of the $2\times 2$-matrices $T^{(12)}$, $T^{(34)}$ and $T^{(56)}$.

The triangular form of the stability matrix remains preserved if we include the dependence of $N_S$ on scalar mass terms.
Let us assume for simplicity that $N_H$ scalars have all mass terms $m_H^2=\tilde m_H^2k^2$, $\tilde m_H^2=\p u/\p\tilde\rho$.
This results in an additional $\tilde\rho$ dependence of $N_S$,
\al{
N_S=N_S^{(0)}+N_H\left( \frac{1}{1+\tilde m_H^2}-1\right).
}
With 
\al{
\frac{\p \beta_u}{\p \tilde m_H^2}&=-\frac{N_H}{32\pi^2(1+\tilde m_H^2)^2}\,,\nn[2ex]
\frac{\p \beta_w}{\p \tilde m_H^2}&=\frac{N_H}{96\pi^2(1+\tilde m_H^2)^2}\,,
}
one finds off-diagonal contributions in the stability matrix
\al{
&T_{13}=\frac{N_H}{32\pi^2}\,,&
&T_{23}=-\frac{N_H}{96\pi^2}\,,&
}
such that
\al{
\tilde T^{(12)}=\frac{N_H}{32\pi^2} \pmat{
1 && 0 \\[2ex]
-1/3 && 1/2
}.
}
Similarly, with 
\al{
\frac{\p \tilde{\mathcal N}_U}{\p \tilde\rho}&=-\frac{N_H \lambda_H}{(1+\tilde m_H^2)^2}\,,\nn[2ex]
\frac{\p \tilde{\mathcal N}_M}{\p \tilde\rho}&=\frac{N_H \lambda_H}{(1+\tilde m_H^2)^2}\,,
} 
one obtains new elements of the stability matrix
\al{
&T_{35}=\frac{N_H}{32\pi^2}\,,&
&T_{45}=-\frac{N_H}{48\pi^2}\,,& 
}
resulting in 
\al{
\tilde T^{(34)}=\frac{N_H}{32\pi^2}\pmat{
1 && 0\\[2ex]
-2/3 && 8
}.
}

We conclude that for the constant scaling solution the stability matrix is very simple.
Its structure remains similar if we take into account the $v$ dependence of the scalar physical metric fluctuation $\pi_0$.
This will be different for possible scaling solutions for which $u$ and $w$ depend on $\tilde\rho$.
In particular, if gauge or Yukawa couplings at the fixed point differ from zero, such a $\tilde\rho$ dependence of the scaling solution will be induced by $\p N_V/\p \tilde\rho$ and $\p N_F/\p \tilde\rho$ that do not vanish for $\tilde\rho\to0$.
This will induce nonzero fixed point values $\tilde m_{H*}^2$, $\lambda_{H*}$, and $w_{2*}$.
We have discussed some details of this case in Sec.\,\ref{gauge and scalar contributions}.

\subsection{Critical exponents for the new fixed point}
We have concentrated our discussion of critical exponents on the Reuter fixed point, since this is the only fixed point for the standard model and the discussed GUT models.
In the small region where two fixed points exist it is interesting to know which one is more stable, e.g. which one has less relevant parameters.
For the new fixed point $v_-$ is negative for the interesting range of $\tilde{\mathcal N}_U$ and $\tilde{\mathcal N}_M$.
For large negative $v$, this has the tendency to make the critical exponent $\theta_6$ more negative according to Eq.\,\eqref{critical exponents of 5 and 6 in approximated forms}.
For the example $\tilde{\mathcal N}_U=\tilde{\mathcal N}_M=-5$ one finds for the new fixed point $v_-=-5.098$ and $w_-=0.0006066$
\al{
(\text{N}):~~
&\theta_5=-0.168,&
&\theta_6=-16.03,&
}
while the critical exponents for the Reuter fixed point are given by
\al{
&(\text{R}):~~
\theta_5=-1.55 + 0.539i,&
&\theta_6=-1.55 - 0.539i.&
}
We conclude that the Reuter fixed point has four relevant parameters, while the new fixed point only has two.
This seems to indicate that the new fixed point is actually the more stable one.

\section{Conclusions}
\label{conclusions}
We have computed the flow equations for the effective potential $U\fn{\rho}$ and the coefficient function of the curvature scalar $F\fn{\rho}$, within quantum gravity coupled to an arbitrary number of scalars, gauge bosons and fermions.
The use of the gauge invariant flow equation constrains the possible form of the effective average action by diffeomorphism symmetry.
Since our setting is formulated in terms of a single macroscopic metric, diffeomorphism symmetry relates, for example, the effective scalar potential and the zero momentum behavior of the graviton propagator.
Furthermore, the gauge invariant flow equation leads to a universal measure contribution to the flow that is a fixed functional of the metric, not involving the matter fields.
This replaces the contribution of ghosts and part of the metric fluctuations in other functional renormalization group investigations of quantum gravity.
For the gauge invariant flow equation the contribution from the physical metric fluctuations decays into the dominant, rather simple graviton contribution (traceless transverse tensor fluctuations) and a physical scalar contribution.
The latter is more involved, also due to mixing with other scalars.
Being subdominant it admits, however, a reasonable approximation that permits us to discuss many aspects analytically.

We concentrate in this paper on the fixed point or scaling solution with field independent $U$ and $F$, and the vicinity of it.We find a scaling solution for all the models we have considered, pure gravity, the standard model and grand unified models based on SO(10) or SU(5) with an arbitrary number of scalar fields $N_S$.
For SO(10) the fixed point is situated, however, outside the range of validity of our truncation, due to the large number of scalar fields needed for a realistic spontaneous symmetry breaking.
For dealing with SO(10) reliably one needs at least to include the effect of the squared Weyl tensor for the graviton propagator. 

For the vicinity of the fixed point we have used a Taylor expansion of $U(\rho)$ and $F\fn{\rho}$ in terms of $\rho$, which is a quadratic invariant formed from scalar fields.
We concentrate on the Higgs doublet $h$ for which $\rho=h^\dagger h$.
We compute the flow of the scalar mass term and quartic scalar coupling, as well as a nonminimal scalar gravity coupling, in the vicinity of the fixed point.
From the corresponding stability matrix and its eigenvalues, the critical exponents, we find that the quartic scalar coupling $\lambda_H$ is an irrelevant parameter for all models considered, restricted to ranges where our truncation does not become invalid. 
This gives support to the prediction of the mass of the Higgs boson in Ref.\,\cite{Shaposhnikov:2009pv}.

On the other hand, for the same range of models the scalar mass term $\tilde m_H^2$ turns out to be a relevant parameter.
Self-adjusted criticality in the Higgs sector, and the associated resurgence mechanism\,\cite{Wetterich:2016uxm}, is not realized for this class of models for the constant scaling solution.
A small ratio between the Fermi scale and the Planck scale is technically natural because of particle scale symmetry~\cite{Wetterich:2019qzx}.
Its value cannot be predicted, however, if the distance from the phase transition is a relevant parameter.
This situation may change for a different scaling solution.
Another interesting possibility for $\tilde m_H^2$ becoming an irrelevant parameter may arise for GUT models with large $N_S$.
The gravity induced anomalous dimension $A$ may grow large for this type of models.
If a more reliable truncation tames the large off-diagonal elements in the stability matrix, this may offer prospects for self-adjusted criticality.

We are aware that an understanding of the effective potential $U$ and the effective squared Planck mass $F$ at the fixed point of quantum gravity, as well as the stability analysis at the fixed point, is only at its beginning.
Nevertheless, already at this stage the gauge invariant flow equation offers many insights, and we hope that a robust picture will arise from extended truncations.

\subsection*{Acknowledgements}
We would like to thank J.\,M.\,Pawlowski, R.\,Percacci, and M.\,Reichert for valuable discussions and comments.
This work is supported by the DFG Collaborative Research Centre ``SFB 1225 (ISOQUANT)."
M.\,Y. is supported by the Alexander von Humboldt Foundation.

\begin{appendix}
\section{Effective average action in the background field formalism and physical gauge fixing}
\label{formulations}
In the main body of this paper we employ the gauge invariant flow equation with a single metric and gauge field.
As we have argued in Sec.\,\ref{description on gauge invariant flow}, this is equivalent to the more standard background formalism with a physical gauge fixing, in the most commonly used truncation where the field dependent inverse propagator is approximated by the second functional derivative of a gauge invariant kernel plus a gauge fixing term.
This equivalence holds if no field derivatives of the flow equation are performed---for the differences concerning field derivatives see Appendix\,\ref{Flow of the graviton propagator appendix}. 
In the background field formalism the IR cutoff depends on a separate background field, not on the macroscopic field as for the gauge invariant flow equation.
In the background formalism one also uses ghost fields with an appropriate IR cutoff.

In the background field formalism the one-loop form of the functional flow equation is exact~\cite{Wetterich:1992yh,Morris:1993qb,Reuter:1993kw,Ellwanger:1993mw}.
It is given as a functional differential equation 
\begin{align}
  \p_t \Gamma_k[\Phi] = \frac{1}{2}\text{Tr} \left[
    \left({\Gamma^{(2)}_k [\Phi] + {\mathcal R}_k } \right)^{-1}\, \p_t {\mathcal R}_k\right],
  \label{eq:wetterich}
\end{align}
where ${\mathcal R}_k$ is an infrared regulator function and $\p_t=k\p_k$ is a dimensionless scale derivative.
The trace in Eq.\,\eqref{eq:wetterich} sums over all internal degrees of freedom of a multifield $\Phi$ that includes ghosts.
It involves a momentum integration and a sum over internal space indices. 
The matrix of second functional derivatives $\Gamma^{(2)}_k$ is the full inverse propagator of
$\Phi$.
For reviews on the functional renormalization group (FRG), one can see Refs.\,\cite{Morris:1998da,Berges:2000ew,Aoki:2000wm,Bagnuls:2000ae,%
  Polonyi:2001se,Pawlowski:2005xe,Gies:2006wv,Delamotte:2007pf,%
  Rosten:2010vm,Braun:2011pp}.

In this appendix we work in the standard background field formalism.
For a physical gauge fixing and a suitable truncation this will produce the same flow equation as in the main text.
This appendix, together with Appendix\,\ref{evaluation of flow equations}, is self-contained. 

\subsection{Setup}
We consider the system of a singlet scalar field $\varphi$ nonminimally coupled to gravity.
Our truncation for the effective action is given by
\begin{align} 
  \Gamma_k&=\Gamma_k^\text{gravity}+\Gamma_k^{S}+\Gamma_k^{V}+\Gamma_k^{F}\,,
  \label{effective action for both}
\end{align}
where
\begin{align}
&  \Gamma_k^\text{gravity}
	 =  -\frac{1}{2}\int_x \sqrt{g}\,F\fn{\rho}R +\Gamma_{\rm gf} +\Gamma_{\rm gh}
  \,,
			\nn[1ex]
&\Gamma_k^{S}
	=	\int_x \sqrt{g}\left\{U\fn{\rho}
		+\frac{Z_\varphi}{2} g^{\mu \nu}\,\p_\mu{\varphi}\,\p_{\nu}\varphi 
		\right\}\,,
				&\label{effective action for scalar}
				\nn[1ex]
&\Gamma_k^{V}
	=	\frac{Z_A}{4e^2}\int_x\sqrt{g}\,F_{\mu\nu}F^{\mu\nu}+\Gamma^{(V)}_{\rm gf} +\Gamma^{(V)}_{\rm gh}\, ,
				\nn[1ex]
&\Gamma_k^{F}
	=	\int_x \sqrt{g}\, \Big\{  iZ_\psi \bar\psi \gamma^\mu D_\mu \psi +y\bar\psi \gamma^5 \psi \varphi \Big\}.
\end{align}
Here $F\fn{\rho}$ is the field dependent Planck mass, $U\fn{\rho}$ is the scalar effective potential, and $Z_\varphi$, $Z_\psi$ and $Z_A$ stand for the field-renormalization factors of $\varphi$, $\psi$ and $A_\mu$, respectively.
The scalar potentials can be expanded as polynomials of $\rho=\varphi^2/2$, namely
\al{
F\fn{\rho}&=M_\text{p}^2+\xi \rho+\cdots,\\[1ex]
U\fn{\rho}&=V+m^2\rho+\frac{1}{2}\lambda \rho^2+\cdots,
}
where $M_\text{p}^2$ is the reduced Planck mass related to the Newton constant $G_N=1/(8\pi^2 M_\text{p}^2)$, $\xi$ is the nonminimal coupling constant that connects between $\rho$ and the Ricci scalar, and $V$ is the cosmological constant.

In our truncation the only violation of the gauge symmetries for the macroscopic metric $g_{\mu\nu}$ and gauge field $A_\mu$ arises from gauge fixing and ghost terms.
In the approximation $\Gamma_k=\bar\Gamma_k+\Gamma_\text{gf}+\Gamma_\text{gh}$ the flow equations will be found to be equivalent to the ones obtained from the gauge invariant flow equation with the same ansatz for the gauge invariant functional $\bar\Gamma_k$.
We emphasize that in the background field formalism the ansatz $\Gamma_k=\bar\Gamma_k+\Gamma_\text{gf}+\Gamma_\text{gh}$ is only an approximation.
The symmetries admit many additional invariants involving the differences $g_{\mu\nu}-\bar{g}_{\mu\nu}$ or $A_\mu-\bar A_\mu$, with $\bar g_{\mu\nu}$ and $\bar A_\mu$ the background fields.
We recall that the equivalence with the gauge invariant flow equation holds only for this truncation and for a physical gauge fixing.

For the gauge bosons the physical gauge fixing and the ghost action are given by
\al{
\Gamma^{(V)}_{\rm gf} &=\frac{1}{2\alpha}\int_x \sqrt{\bar g}\,(\p_\mu A^\mu)^2,\\[1ex]
\Gamma^{(V)}_{\rm gh}&=\int_x \sqrt{\bar g}\,\bar  c \, \p_\mu \p^\mu c\, .
}

To specify the physical gauge fixing for the metric we write the macroscopic metric as
\al{ g_{\mu\nu}={\bar g}_{\mu\nu}+h_{\mu\nu}\,, }
where ${\bar g}_{\mu\nu}$ is a constant background metric and $h_{\mu\nu}$ is a fluctuation field.  
The gauge fixing for diffeomorphism symmetry is given by
\begin{align} 
 \Gamma_{\rm gf} &=
                        \frac{1}{2\alpha}\int_x\sqrt{\bar g}\,
                        {\bar g}^{\mu \nu}\Sigma_\mu\Sigma_{\nu} \,.
                        \label{gauge fixing action}
\end{align}
A class of general gauge fixings reads
\al{
\Sigma_\mu		
	= {\bar D}^\nu h_{\nu \mu}-\frac{\beta +1}{4}{\bar D}_\mu h\,,
	\label{gauge fixing function}
}
where $h={\bar g}^{\mu\nu}h_{\mu\nu}$ is the trace mode.
Bars on operators denote that covariant derivatives are formed with the background metric, and indices of operators are contracted by the background metric as well.
The ghost action associated with the gauge fixing \eqref{gauge fixing function} is given by 
\al{
  \Gamma_{\rm gh}	&=	-\int_x\sqrt{\bar g}\,\bar C_\mu
               \left[ {\bar g}^{\mu\rho}{\bar D}^2+
               \frac{1-\beta}{2}{\bar D}^\mu{\bar D}^{\rho}
               +{\bar R}^{\mu\rho}\right] C_{\rho}\,, \label{ghostaction}
}
where $C$ and $\bar C$ are ghost and antighost fields. 
Equations\,\eqref{gauge fixing action} and \eqref{gauge fixing function} constitute a general family of gauge fixings for diffeomorphism symmetry, specified by two parameters $\alpha$ and $\beta$.
The choice $\beta=-1$ and $\alpha\to 0$ is a ``physical gauge fixing'' that acts only on the gauge modes in $h_{\mu\nu}$.
In this work, we use this gauge choice.
Nevertheless, the next subsection discusses general $\alpha$ and $\beta$.
This will demonstrate explicitly the particular role of the physical gauge fixing.

\subsection{Physical metric decomposition}
A key quantity for the flow equation is the inverse propagator or 1PI two-point function, as given by the matrix of second functional derivatives of $\Gamma_k$. 
To derive the explicit form for the metric two-point function, we split the metric fluctuations into physical and gauge fluctuations~\cite{Wetterich:2016vxu}. 

Accordingly, we decompose the metric fluctuations into
\al{
h_{\mu\nu}=f_{\mu\nu}+a_{\mu\nu}\,,
\label{metric decomposition}
}
where $f_{\mu\nu}$ are the physical metric fluctuations, and $a_{\mu\nu}$ the gauge fluctuations or gauge modes.
The physical metric fluctuations satisfy the transverse constraint $\bar D^\mu f_{\mu\nu}=0$. 
In turn, the physical metric fluctuations can be decomposed into two independent fields as
\al{
f_{\mu\nu}=t_{\mu\nu}+s_{\mu\nu}\,,
\label{physical decomposition}
}
where the graviton $t_{\mu\nu}$ is the transverse and traceless (TT) tensor, i.e., $\bar D^\mu t_{\mu\nu}=\bar g^{\mu\nu}t_{\mu\nu}=0$.
The tensor $s_{\mu\nu}$ is given as a linear function of a scalar field $\sigma$ such that 
\al{
s_{\mu\nu}=\hat S_{\mu\nu} \sigma =\frac{1}{3}P_{\mu\nu} \sigma\,.
}
For a background geometry with constant curvature the projection operator can be found explicitly as
\al{
P_{\mu\nu} =\bigg( \bar g_{\mu\nu}\bar \Delta_S + \bar D_\mu \bar D_\nu - \bar R_{\mu\nu}\bigg) \left(\bar \Delta_S-\frac{\bar R}{3}\right)^{-1},
\label{projection operator Pmunu}
}
with $\bar \Delta_S=-\bar D^2=-\bar D^\mu \bar D_\mu$ the covariant Laplacian acting on scalar fields (spin-0 fields).

Similarly, the gauge modes or unphysical metric fluctuations $a_{\mu\nu}$ are decomposed into a transverse vector mode $\kappa_\mu$, satisfying $\bar D^\mu \kappa_\mu=0$, and a scalar mode $u$. In summary, the metric fluctuations \eqref{metric decomposition} are parametrized by
\al{
f_{\mu\nu}&=t_{\mu\nu}+\frac{1}{3}P_{\mu\nu} \sigma\,, \notag \\[1ex]
a_{\mu\nu}&=\bar D_\mu \kappa_\nu+\bar D_\nu \kappa_\mu - \bar D_\mu  \bar D_\nu  u\,.
\label{parametrized metric}
}
Using the linear combinations
\al{
\sigma&=\frac{3}{4}\left(\frac{\bar \Delta_S-\bar R/3}{\bar \Delta_S-\bar R/4}\right)\left(h+\left(\bar \Delta_S\right)  s\right)\,,\nn[1ex]
u&=\frac{1}{4}\frac{1}{\bar \Delta_S-\bar R/4}\left[ h-3\left(\bar \Delta_S-\frac{\bar R}{3}\right) s\right]\,,
}
we obtain the York decomposition~\cite{York:1973ia} of the fluctuation metric
\begin{align}\nonumber 
h_{\mu\nu}&=t_{\mu\nu}+(\bar D_\mu \kappa_\nu  + \bar D_\nu \kappa_\mu)\\[1ex]
&\quad
+\left(\bar D _\mu \bar D_\nu +\frac{1}{4}\bar g_{\mu\nu}\bar \Delta_S \right) s+\frac{1}{4}\bar g_{\mu\nu}h,
\end{align}
where $h=\bar g^{\mu\nu}h_{\mu\nu}$.
The scalar modes $s$ and $h$ in the York decomposition are given as a mixture of the physical scalar mode $\sigma$ and the gauge mode $u$. 

The ghost fields can be decomposed similarly into vector and scalar fields
\al{
C_\mu&=C_{\mu}^\perp+\bar D_\mu C\,,&
\bar C_\mu&=\bar C_{\mu}^\perp+\bar D_\mu \bar C\,,
}
where $C_{\mu}^\perp$ ($\bar C_{\mu}^\perp$) is the transverse (anti)ghost field and $C$ ($\bar C$) is the scalar (anti)ghost field.

These decompositions yield Jacobians that read
\al{
J_\text{grav1}&=\left[ \det{}_{(1)}'\left(\bar{\mathcal D}_1 \right) \right]^{1/2}, \nn[1ex]
J_\text{grav0}&= \left[ \det{}_{(0)}'\left(\bar{\mathcal D}_0\right)\bar \Delta_S\right]^{1/2},\nn[1ex]
J_\text{gh}&=\left[ \det{}_{(0)}''\left(\bar \Delta_S \right) \right]^{-1}, \label{set of Jacobians}
}
with
\al{
\bar{\mathcal D}_1&=\bar \Delta_V-\frac{\bar R}{4},&
\bar{\mathcal D}_0&=\bar \Delta_S-\frac{\bar R}{4}.&
} 
Here $\bar \Delta_V=-\bar D^2$ is the Laplacian acting on vector fields (spin-1 fields) and a prime denotes a subtraction of the zero eigenmode.
This subtraction, however, does not contribute to the present truncation, so that we hereafter neglect it.

\subsection{Hessians}
The two-point functions (or Hessians) for each degree of freedom in the metric fluctuations defined in Eq.\,\eqref{parametrized metric}, as well as for the scalar field $\varphi$, are obtained by calculating the second order variations of the effective action \eqref{effective action for both} in terms of the fluctuation fields.
For our decomposition of the metric fluctuations, the matrix of the Hessians becomes block diagonal for each degree of freedom or spin.
We neglect scale-independent overall constant factors such as $\sqrt{\bar g}$ and the gauge parameters since they drop out in the flow equations.

For the $t_{\mu\nu}$ mode, we obtain
\al{
\left(\Gamma_{(tt)}^{(2)}\right)^{\mu\nu\rho\sigma}= F\left[ \bar{\mathcal D}_T -\frac{2U}{F} \right]P^{(t)\mu\nu\rho\sigma}\,,
\label{TT mode two point function}
}
where we define the derivative operator
\al{
\bar{\mathcal D}_T=\bar \Delta_T+\frac{2\bar R}{3}\,,
}
with the Laplacian $\bar\Delta_T=- \bar D^2$ acting on transverse traceless tensor fields (spin-2 fields).
The TT-projection operator is given by
\al{
P^{(t)\mu\nu\rho\sigma}=
\frac{1}{2}(P^{\mu\rho}P^{\nu\sigma}+P^{\mu\sigma}P^{\nu\rho})-\frac{1}{3}P^{\mu\nu}P^{\rho\sigma}\,,
}
with $P^{\mu\nu}$ defined by Eq.\,\eqref{projection operator Pmunu}.

The Hessian for the spin-1 gauge mode $\kappa_{\mu}$ is given by
\al{
&\left(\Gamma_{(\kappa \kappa)}^{(2)}\right)^{\mu\nu}=
\bar{\mathcal D}_1\left[  \bar{\mathcal D}_1+\frac{\alpha\bar R}{2} -\alpha U\right] P^{(v)\mu\nu}\,,
}
with $P^{(v)}$ the projection operator on the vector mode, $P^{(v)}{}_\mu{}^\mu=3$.
For $\alpha\to0$ the inverse propagator for the gauge-vector mode becomes independent of $U$
\al{
\lim_{\alpha\to 0}\left( \Gamma_k^{(2)}\right)_{\kappa\kappa}^{\mu\nu}=\left(\bar{\mathcal D}_1\right)^2P^{(v)\mu\nu}.
\label{spin 1 contribution in Landau gauge}
}
\begin{widetext}
We next turn to the Hessian for the scalar modes.
In the $(\sigma,u,\varphi)$-field basis, we obtain
\al{
\Gamma_{(00)}^{(2)}=
\pmat{
\begin{matrix} \left(\Gamma_{(00)}^{(2)}\right)_\text{grav} \end{matrix} & \begin{matrix}  \displaystyle \frac{1}{2}\left[- F' \left(\bar \Delta_S+\frac{\bar R}{4} \right)+ U' \right]\varphi\\[20pt]
\displaystyle \frac{1}{4}\left(-F'\bar R +2U'  \right)\varphi \bar\Delta_{S}\end{matrix} \\[30pt]
 \begin{matrix} \displaystyle \displaystyle \frac{1}{2}\left[ -F' \left(\bar \Delta_S+\frac{\bar R}{4} \right)+ U' \right] \varphi &&&  \displaystyle \frac{1}{4}\left(-F'\bar R +2U' \right)\varphi \bar\Delta_{S} \end{matrix}
&~~~~~~~~
\displaystyle Z_\varphi \bar \Delta_S+m_\varphi^2 -\frac{1}{2} \tilde \xi_\varphi\,\bar R
}\,,
\label{spin 0 matrix}
}
where we define $m_\varphi^2\fn{\rho}=U'+2\rho U''$ the field-dependent scalar mass and $\tilde \xi_\varphi\fn{\rho}=F'+2\rho F''$, and primes denote derivatives with respect to $\rho=\varphi^2/2$.
Here the spin-0 gravitational part is given by the following $2\times 2$ matrix:
\al{
&\left(\Gamma_{(00)}^{(2)}\right)_\text{grav}=
\nn[1ex]&\scriptsize\pmat{
\displaystyle -\frac{F}{6}\left[ \left(\bar \Delta_S -\frac{3}{8}\bar R\right) \bar \Delta_S -\frac{U}{2F}\left( \bar \Delta_S-\frac{\bar R}{2}\right)\right] \left( \bar \Delta_S-\frac{\bar R}{3}\right)^{-1}+\frac{(\beta+1)^2}{16\alpha}\bar \Delta_S
&
\displaystyle \frac{F}{4}\left( \frac{U}{F}-\frac{\bar R}{4}\right)\bar\Delta_{S} + \frac{(\beta +1)(\beta -3)}{16 \alpha} \left(\bar \Delta_S+\frac{\bar R}{\beta -3}\right)\bar\Delta_{S} \\[20pt]
\displaystyle \frac{F}{4}\left( \frac{U}{F}-\frac{\bar R}{4}\right)\bar\Delta_{S} + \frac{(\beta +1)(\beta -3)}{16 \alpha} \left(\bar \Delta_S+\frac{\bar R}{\beta -3}\right)\bar\Delta_{S}
&
\displaystyle \frac{F}{16}\left( \bar \Delta_S-\frac{\bar R}{2}\right)\left( \bar R-\frac{4U}{F}\right) \bar \Delta_S +\frac{(\beta -3)^2 }{16 \alpha } \left( \bar \Delta_S+\frac{\bar R}{\beta -3}\right)^2\bar \Delta_S
}\,.
\label{Hessian spin 0 part}
}

For general $\alpha$ and $\beta$ the Hessian in the scalar sector is rather complicated.
It simplifies considerably for a physical gauge fixing that corresponds to $\beta=-1$.
The $2\times 2$ matrix $\left(\Gamma^{(2)}_{(00)}\right)_\text{grav}$ becomes diagonal.
For $\beta=-1$ the factor $1/\alpha$ remains only in the Hessian for the spin-0 gauge mode $\Gamma^{(0)}_{(uu)}$.
In the limit $\alpha\to 0$ we can further neglect in the matrix \eqref{spin 0 matrix} the elements mixing $u$ with $\sigma$ and $\varphi$.
They do not diverge for $\alpha\to0$ and drop out in the propagator that is the inverse of $\Gamma^{(2)}$.
As a result, for a physical gauge fixing the physical fluctuations and the gauge modes decouple.
The Hessian becomes block diagonal in physical and gauge modes.
For the physical modes one obtains the $2\times2$ matrix
\al{
\scriptsize
&\left( \Gamma_{(00)}^{(2)}\right)_\text{ph}
=\pmat{
\displaystyle -\frac{F}{6}\left[ \left(\bar \Delta_S -\frac{3}{8}\bar R\right) \bar \Delta_S -\frac{U}{2F}\left( \bar \Delta_S -\frac{\bar R}{2}\right)\right] \left( \bar \Delta_S -\frac{\bar R}{3}\right)^{-1}
 &&  \displaystyle \frac{1}{2}\left[ - F' \left(\bar \Delta_S+\frac{\bar R}{4} \right)+ U' \right] \varphi \\[5ex]
\displaystyle \frac{1}{2}\left[- F' \left(\bar \Delta_S+\frac{\bar R}{4} \right)+ U' \right]\varphi && Z_\varphi \bar \Delta_S +m_\varphi^2 -\displaystyle\frac{1}{2} \tilde\xi_\varphi\,\bar R
}\,,\label{physical spin 0 mode two point function}
}
\end{widetext}
while the Hessian for the gauge scalar mode reads
\al{
\left( \Gamma_{(00)}^{(2)}\right)_\text{gauge}=\displaystyle\left(\bar \Delta_S-\frac{\bar R}{4}\right)^2\bar \Delta_S\,.
}
Finally, the Hessians for the ghost field obtains from Eq.\,\eqref{ghostaction} as
\al{\nonumber 
\left(\Gamma_{(\bar C^\perp C^\perp)}^{(2)}\right)^{\mu\nu}&=\bar{\mathcal D}_1\, P^{(v)\mu\nu}\,,\\[1ex] 
\Gamma_{(\bar CC)}^{(2)}&=\left( \bar \Delta_S +\frac{\bar R}{\beta-3}\right)\bar \Delta_S\,.
\label{ghost Hessians}
}

It is an important consequence of the physical gauge fixing $\beta=-1$ and $\alpha\to0$ that the Hessian for the spin-1 and 0 gauge modes and the ghost mode depend on neither $U$ nor $F$.
This also holds for the Jacobians \eqref{set of Jacobians}.
More generally, all these terms do not depend on the form of $\Gamma_k$.
We will see in Appendix\,\ref{evaluation of flow equations} that their contributions to the flow equations can be combined into a universal measure contribution, corresponding to the one employed in the gauge invariant flow equation.
The gravitational measure contribution depends on the metric, but not on the scalar field $\varphi$ or any other matter fields.
All these important simplifications do not hold for a general gauge fixing with arbitrary $\alpha$ and $\beta$.

At this stage we have collected all the ingredients necessary for the flow equation \eqref{eq:wetterich}, up to the cutoff function $R_k$.
The cutoff function will be specified in Appendix\,\ref{evaluation of flow equations}, where we also compute the right-hand side of Eq.\,\eqref{eq:wetterich} in our truncation.
We are interested in the gauge invariant kernel $\bar\Gamma_k$ and therefore evaluate the flow generator for a macroscopic metric equal to the background metric.
The bars on covariant derivatives can therefore be dropped in the following.

\section{Heat kernel evaluation of the flow generator}
\label{heat kernel methods}
The flow generator $\zeta_k$, defined by the right-hand side of the flow equation $\p_t\Gamma_k=\zeta_k$, involves different contributions of the type
\al{
\zeta_k=\sum_i a_i\,\text{tr}\,W\fn{\Delta_i},
}
with $\Delta_i$ appropriate differential operators.
For the contribution of a scalar field one has $\Delta=-D_\mu D^\mu$, $a=1/2$ and $W\fn{\Delta}=\tilde \p_t \ln\,P_k\fn{\Delta}$.
The trace is conveniently evaluated in the heat kernel expansion, that we recall here briefly for convenience.

Assume an operator $\Delta$ with all eigenvalues positive, $\lambda_m>0$.
For the evaluation of $\text{tr}\,W\fn{\Delta}$ we make use of a representation of the $\delta$ distribution,
\al{
\int^{\gamma+i\infty}_{\gamma-i\infty}\frac{\df s}{2\pi i}\, e^{s(z-\lambda_m)}=\delta\fn{z-\lambda_m},
\label{delta distribution function}
}
that holds for real positive finite $\gamma$ provided $\lambda_m>0$.
Insertion of Eq.\,\eqref{delta distribution function} yields
\al{
\text{tr}\,W\fn{\Delta}&=\sum_m W\fn{\lambda_m}= \sum_m \int^\infty_0\df z\,\delta\fn{z-\lambda_m}\,W\fn{z}\nn
&=\int^\infty_0\df z\,W\fn{z}\int^{\gamma+i\infty}_{\gamma-i\infty} \frac{\df s}{2\pi i}e^{sz}\,\text{tr}\,e^{-s\Delta},
}
where we use $\sum_m \exp\fn{-s\lambda_m}=\text{tr}\,\exp\fn{-s\Delta}$. 
One next employs the expansion
\al{
&\text{tr}\,e^{-s\Delta}\nn
&=\frac{1}{16\pi^2}\int_x\sqrt{g} \left\{ c_0\fn{\Delta}s^{-2}+c_2 \fn{\Delta}s^{-1}+c_4\fn{\Delta}+\cdots\right\},
}
for which the coefficients $c_n\fn{\Delta}$ are well known for the operators of interest here.
This yields
\al{
\text{tr}\,W\fn{\Delta}
=\frac{1}{16\pi^2}\sum_{n=0}^\infty Q_{2-n}\int_x\sqrt{g}\,c_{2n}\fn{\Delta},
}
with
\al{
Q_n=\int^\infty_0\df z\,W\fn{z}\int^{\gamma +i\infty}_{\gamma-i\infty}\frac{\df s}{2\pi i} e^{sz}s^{-n}.
\label{Q-function}
}
Evaluating Eq.\,\eqref{Q-function} one finds
\al{
&Q_2=\int^\infty_0\df z\,zW\fn{z},&
&Q_1=\int^\infty_0\df z\,W\fn{z},&\nn[2ex]
&Q_0=W\fn{z=0},& 
}
or, more generally,
\al{
&Q_n=\frac{1}{\Gamma\fn{n}}\int^\infty_0 \df z\,z^{n-1}W\fn{z}, & n\geq 1,&\nn[2ex]
&Q_{-n}=(-1)^n\frac{\p^n W}{\p z^n}\Bigg|_{z=0}. & n\geq 0.&
}
In particular, for $W=\p_t R_k/(z+R_k\fn{z}+m^2)$ the coefficients $Q_n$ are directly related to the threshold functions $\ell_n^d$ that have been widely explored in functional renormalization for different forms of the IR cutoff $R_k$; cf. Eq.\,\eqref{threshold functions}.

The coefficients $c_0$ and $c_2$ are given by 
\al{
&c_0=b_0\,,&
&c_2=b_2 R \,.&
}
We display the value of $b_0$ and $b_2$ for the operators relevant for various fields in Table\,\ref{hkcs}.
The numbers for fermions $\psi$ are given for Majorana spinors.
They have to be doubled for Dirac spinors.

\begin{table*}[htb]
\begin{center}
\caption{Heat kernel coefficients for the individual fields in maximally symmetric four dimensional spacetime. ``T'' and ``TT'' denote ``transverse'' and ``transverse-traceless'', respectively.}
\label{hkcs}
\begin{tabular}{|c|c|c|c|c|c|c|c|}
\hline
    \makebox[1cm]{}  &  \makebox[1.1cm]{tensor} & \makebox[1.5cm]{T-tensor} & \makebox[1.7cm]{TT-tensor} & \makebox[1.1cm]{vector}  & \makebox[1.5cm]{T-vector} & \makebox[1.8cm]{Weyl spinor} & \makebox[1.1cm]{scalar} \\
      & ($h_{\mu\nu}$) & ($f_{\mu\nu}$) & $(t_{\mu\nu})$ & ($A_\mu$) & ($A_\mu^\text{T}$, $\kappa_\mu$, $C^\perp_\mu$) & $(\psi)$ & ($a$, $u$, $\sigma$, $\varphi$, $C$) \\
\hline
& & & & & & & \\
$b_0$ & $10$ & $9$ & $5$ & $4$ & $3$ & $2$ & $1$ \\
& & & & & & & \\
\hline
& & & & & & & \\
$b_2$ & $\displaystyle \frac{5}{3}$ & $\displaystyle \frac{3}{2}$ & $\displaystyle -\frac{5}{6}$ & $\displaystyle \frac{2}{3}$ &$\displaystyle \frac{1}{4}$ & $\displaystyle \frac{1}{3}$ & $\displaystyle \frac{1}{6}$ \\
& & & & & & & \\
\hline
\end{tabular}
\end{center}
\end{table*}

\section{Flow of the graviton propagator}
\label{Flow of the graviton propagator appendix}
In this appendix we compute directly the flow equation for the inverse graviton propagator in flat space.
This will reveal important differences between the gauge invariant flow equation and the background field formalism for the flow of field derivatives of $\Gamma_k$.
Only for the gauge invariant flow do the field derivatives and scale derivatives commute.
We focus on the infrared behavior of the gravitational propagator, i.e., vanishing momentum in flat space, since the main differences are already visible there.

\subsection{Gauge invariant flow equation}
Within the gauge invariant approach with a single metric the inverse graviton propagator is given by the second functional derivative of the gauge invariant effective average action $\bar \Gamma^{(2)}_\text{grav}$,
\al{
\bar \Gamma_k=\frac{1}{2}\int_x t_{\mu\nu}\left( \bar\Gamma_\text{grav}^{(2)} \right)^{\mu\nu\rho\tau} t_{\rho\tau} +\cdots.
}
We compute the flow contribution of a single scalar field.
Our starting point is Eq.\,\eqref{flow generator of scalar},
\al{
\p_t \bar\Gamma_k =\pi_S=\frac{1}{2}\text{tr}\,\tilde\p_t \ln\fn{P_k\fn{z}+m^2}\,,
}
with covariant Laplacian
\al{
z=-D^\mu D_\mu = -g^{\mu\nu} \p_\mu \p_\nu +\Gamma_\mu{}^{\mu\nu}\p_\nu\,.
}
Taking two derivatives one has
\al{
\p_t \left( \bar\Gamma_\text{grav}^{(2)}\right)^{\mu\nu\rho\tau}=\frac{1}{2}\text{tr}\,\tilde\p_{t} \frac{\p^2}{\p t_{\mu\nu} \p t_{\rho\tau}} \ln\fn{P_k\fn{z}+m^2}\,.
\label{Gamma_grav2}
}
For the graviton propagator we need to expand $z=-D^2$ up to second order in $t_{\mu\nu}$
\al{
z&=-\left( \delta^{\mu\nu} -t^{\mu\nu} + t^\mu_{}{\rho} t^{\rho \nu} \right)\p_\mu \p_\nu \nn
&\qquad
-\left( t^{\mu\rho}\p_\mu t^\nu_\rho -\frac{1}{2}\delta^{\mu\nu} \left( t_{\tau\rho}\p_\mu t^{\tau \rho}\right)\right) \p_\nu\,.
}

We consider the graviton propagator in the zero-momentum limit.
This corresponds to $x$-independent $t_{\mu\nu}$.
In this case the operator $z$ becomes diagonal in momentum space
\al{
z\fn{q,q'}=\left( q^2 -t^{\mu\nu} q_\mu q_\nu +t^\mu_{}{\rho} t^{\rho\nu} q_\mu q_\nu \right) \delta\fn{q-q'}\,.
\label{zqqp}
}
Insertion into Eq.\,\eqref{Gamma_grav2} yields, with $\bar P=P_k+m^2$,
\al{
&\p_t \left( \bar\Gamma^{(2)}\right)^{\mu\nu\rho\tau}
=\frac{1}{2} \int_q\tilde\p_t\Bigg\{ \frac{1}{\bar P}\frac{\p \bar P}{\p z}\frac{\p^2 z}{\p t_{\mu\nu} \p t_{\rho\tau}}\nn
&\quad
+\frac{1}{\bar P}\frac{\p^2 \bar P}{\p z^2} \frac{\p z}{\p t_{\mu\nu}}\frac{\p z}{\p t_{\rho\tau}}
-\frac{1}{\bar P^2} \left( \frac{\p \bar P}{\p z}\right)^2 \frac{\p z}{\p t_{\mu\nu}}\frac{\p z}{\p t_{\rho \tau}}\Bigg\}\,,
}
to be evaluated at $t_{\mu\nu}=0$.
Taking into account that $t_{\mu\nu}$ is trace-free one finds, with $z=q^2$,
\al{
&\p_t \left(\bar\Gamma^{(2)}\right)^{\mu\nu\rho\tau}=\frac{1}{2}\int_q \Bigg\{\frac{\p^2}{\p z^2} \left( \frac{\p_t R_k}{P_k+m^2}\right) \nn
&\qquad \times\left( q^\mu q^\nu -\frac{1}{4}q^2 \delta^{\mu\nu} \right)\left( q^\rho q^\tau -\frac{1}{4}q^2 \delta^{\rho\tau} \right)\nn
&\quad
+\frac{1}{2}\frac{\p}{\p z} \left(\frac{\p_t R_k}{P_k+m^2}\right)\bigg( q^\mu q^\rho \delta^{\nu\tau} +q^\mu q^\tau \delta^{\nu \rho} +q^\nu q^\rho \delta^{\mu\tau}\nn
&\quad
+q^\nu q^\tau \delta^{\mu \rho} 
-q^\mu q^\nu \delta^{\rho\tau} -q^\rho q^\tau \delta^{\mu\nu}+\frac{1}{4}q^2 \delta^{\mu\nu} \delta^{\rho \tau}\bigg) \Bigg\}\,.
}
We employ the identities
\al{
&\int_q f\fn{q^2}\,q^\mu q^\nu =\frac{1}{4} \int f\fn{q^2}q^2 \delta^{\mu\nu},\nn[1ex]
&\int_q f\fn{q^2}\,q^\mu q^\nu q^\rho q^\tau\nn
&\quad=\frac{1}{24} \int f\fn{q^2}q^4 \left(\delta^{\mu\nu} \delta^{\rho\tau} +\delta^{\mu\rho} \delta^{\nu\tau}+\delta^{\mu\tau} \delta^{\nu\rho} \right)\,,
}
and
\al{
\int_q=\frac{1}{16\pi^2}\int^\infty_0 \df z \,z \,,
}
in order to obtain 
\al{
\p_t \left( \bar\Gamma_\text{grav}^{(2)}\right)^{\mu\nu\rho\tau} =-\frac{1}{2}\pi_S^{(U)}\hat P^{\mu\nu\rho\tau}\delta\fn{q-q'}\,.
\label{graviton flow equation}
}
Here
\al{
\pi_S^{(U)}=-\frac{1}{32\pi^2}\int^\infty_0 \df z \left( \frac{z^3}{6}\frac{\df^2}{\df z^2}+z^2\frac{\df}{\df z}\right)\frac{\p_t R_k}{P_k+m^2}\,,
\label{pi_SU derivative z}
}
and
\al{
\hat P^{\mu\nu\rho\tau}=\frac{1}{4}\left( 2\delta^{\mu\rho} \delta^{\nu\tau} +2\delta^{\mu\tau}\delta^{\nu \rho}-\delta^{\mu\nu}\delta^{\rho\tau} \right)
}
is a projector that eliminates the trace of $h_{\mu\nu}$.
A flow of the zero-momentum graviton propagator of the form \eqref{graviton flow equation} follows if the part of $\bar \Gamma_k$ not containing derivatives of the metric has the diffeomorphism invariant structure
\al{
\bar\Gamma_k=\int_x \sqrt{g} \,U_g&=-\frac{U_g}{4}\int_x t_{\mu\nu}t^{\mu\nu}\,,\nn[2ex]
\p_t U_g&=\pi_S^{(U)}\,.
} 

For $\p_t R_k$ vanishing fast enough for $z\to \infty$ and $\p_t R_k/(P_k+m^2)$ remaining finite for $z\to 0$ we can perform partial integrations,
\al{
\pi_S^{(U)}&=-\frac{1}{64\pi^2} \int^\infty_0 \df z\,z^2 \frac{\p}{\p z} \frac{\p_t R_k}{P_k+m^2}\nn[1ex]
&=\frac{1}{32\pi^2}\int^\infty_0 \df z\,z \frac{\p_t R_k}{P_k+m^2}\,.
}
Comparison with the part of Eq.\,\eqref{Scalar contribution to F and U} that remains for a vanishing curvature scalar $R=0$ shows that we can identify $U_g$ with the effective scalar potential $U$.
This is what one expects for a diffeomorphism-invariant effective action if a derivative expansion is valid.
On the diagrammatic level several individual diagrams have to combine in a particular way in order to arrive at this simple result.
This combination is dictated by diffeomorphism symmetry.

\subsection{Background field method}
The use of covariant derivatives in the scalar IR-cutoff function, and therefore the dependence of $R_k$ on the macroscopic metric $g_{\mu\nu}$, are crucial for obtaining this simple result.
We can compare this result with the background field method or computations without gauge symmetry for which $R_k$ does not depend on the macroscopic metric.
It may depend on the background metric $\bar g_{\mu\nu}$ that corresponds to flat space in our case.
If we omit the metric dependence of $R_k$, choosing instead of $\sqrt{g} R_k\fn{-D^2}$ a cutoff $R_k\fn{-\p^2}$, $\p^2=\delta^{\mu\nu} \p_\mu \p_\nu$, the derivative $\p/\p z$ in Eq.\,\eqref{pi_SU derivative z} is replaced by $\tilde\p_z$ acting only on the part $\sim z$ in $P_k=z+R_k$, resulting in $\tilde \p_z\bar P=1$, and 
\al{
\tilde \p_z \frac{\p_t R_k}{P_k+m^2}=-\frac{\p_t R_k}{(P_k+m^2)^2}\,.
}
Furthermore the factor $\sqrt{g}$ in the inverse propagator $\sqrt{g}(-D^2+m^2)+R_k\fn{-\p^2}$ is no longer canceled by a similar factor $\sqrt{g}\, \p_t R_k$.
As a consequence, the dependence of $z$ on $t_{\mu\nu}$ is supplemented by
\al{
\delta z=-(\sqrt{g}-1)\delta^{\mu\nu} \p_\mu \p_\nu=\frac{1}{4} t_{\rho\tau}t^{\rho\tau}\delta^{\mu\nu}\p_\mu \p_\nu\,.
}
This adds in momentum space to Eq.\,\eqref{zqqp} a contribution
\al{
\delta z=-\frac{1}{4}q^2\, t_{\mu\nu}t^{\mu\nu}\,\delta\fn{q-q'}\,,
}
resulting in an additional piece 
\al{
\Delta \pi_S^{(U)}=-\frac{1}{32\pi^2}\int^\infty_0 \df z\, z^2\frac{\p_t R_k}{(P_k+m^2)^2}\,.
}

Taking things together, the noncovariant cutoff $R_k\fn{-\p^2}$ replaces $\pi_S^{(U)}$ by $\tilde\pi_S^{(U)}$,
\al{
\tilde \pi_S^{(U)}=-\frac{1}{96\pi^2}\int^\infty_0 \df z\,z^3\frac{\p_t R_k}{(P_k+m^2)^3}\,.
\label{tildepi_SU}
}
In general, one may expect contributions both from a four-point vertex and from three-point vertices, which may by represented graphically as
\al{
\tilde\pi_S^{(U)}\sim
~
\vcenter{\hbox{\includegraphics[width=20mm]{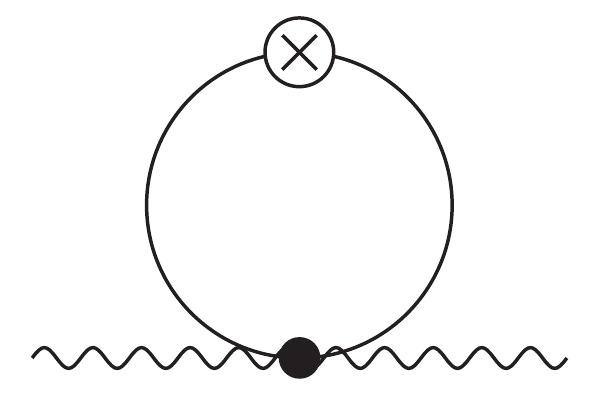}}\vskip1mm}
~~
+
~~
\vcenter{\hbox{\includegraphics[width=30mm]{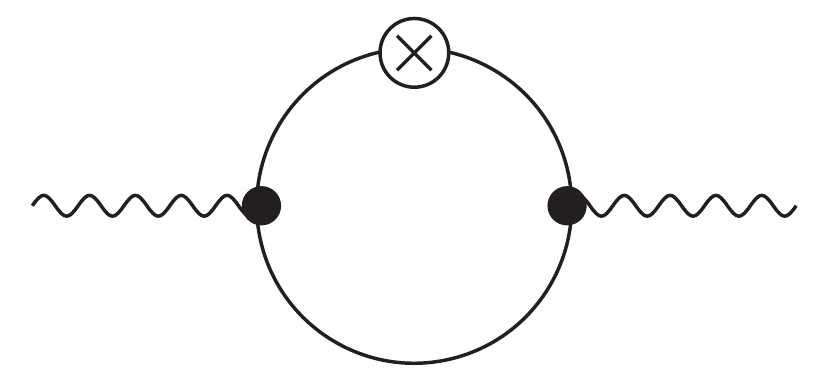}}\vskip0mm}
\,.
}
Here the wiggled lines are gravitons, solid lines are scalar propagators, and a cross denotes the insertion of $\p_t R_k$.
Because of the particular momentum structure the four-point vertex does not contribute, however, and Eq.\,\eqref{tildepi_SU} involves the second graph only.
In contrast, for the gauge invariant flow $\pi_S^{(U)}$ there are further graphs where graviton lines are attached to the cross $\sim \p_t R_k$.

The result \eqref{tildepi_SU} corresponds to a flat-space computation of the flow of the graviton propagator at zero momentum for a cutoff function $R_k$ that violates diffeomorphism symmetry.
This is the type of computation performed in Ref.\,\cite{Meibohm:2015twa}.
As noted there, the sign of $\tilde\pi_S^{(U)}$ is opposite to the sign of $\pi_S^{(U)}$.
For a Litim cutoff one obtains, with $\tilde w=m^2/k^2$,
\al{
\pi_S^{(U)}&=\frac{k^4}{32\pi^2(1+\tilde w)}\,,\nn[1ex]
\tilde \pi_S^{(U)}&=-\frac{k^4}{192\pi^2(1+\tilde w)^3}\,,
}
such that for $\tilde w=0$ one has $\tilde \pi_S^{(U)}=-\pi_S^{(U)}/6$, in agreement with Ref.\,\cite{Meibohm:2015twa}.

The difference between $\pi_S^{(U)}$ and $\tilde \pi_S^{(U)}$ arises only from the different choice of the IR-cutoff function.
Since $\pi_S^{(U)}$ corresponds to a diffeomorphism invariant effective average action, and the difference between $\pi_S^{(U)}$ and $\tilde\pi_S^{(U)}$ is not small, we conclude that the violation of diffeomorphism symmetry in the scalar-induced flow is substantial if $R_k$ is not formulated in terms of covariant derivatives.
If scalar fluctuations play an important role, the truncation where vertices are derived from a gauge invariant $\bar \Gamma_k$ is valid only if $R_k$ involves covariant derivatives.
In other words, if the flow equation is not gauge invariant, additional vertices will play a role.
This generalizes to background gauge fixing if $R_k$ involves a background metric $\bar g_{\mu\nu}$ different from the macroscopic metric $g_{\mu\nu}$.
The additional vertices arise from terms involving the difference $g_{\mu\nu}-\bar g_{\mu\nu}$.

There is, in principle, an ambiguity by which quantity the coefficient of the curvature scalar is defined.
Without gauge invariance the graviton propagator and the graviton vertices are not directly related.
If one wants to use a truncation with a gauge invariant kernel $\bar\Gamma_k$, one has to decide from which quantity one extracts the flow of $F(\rho)$.
As we have argued, the effects of gauge symmetry breaking by a noncovariant cutoff function can be substantial.
They seem to be reduced if one employs the graviton vertices, with a certain degree of encouraging universality between the three- and four-point vertices~\cite{Eichhorn:2018akn}.

\subsection{Comparison}
In principle, one is free to choose a cutoff function provided some general properties for the behavior at large and small (covariant) momenta are obeyed.
Different choices of $R_k$ correspond to different flow trajectories on which the quantum effective action is reached for $k\to 0$.
We want to choose initial conditions for the flow at large $k$ such that for $k=0$, where $R_k$ vanishes, the effective action is diffeomorphism invariant. 
For $k=0$ it should involve only a single metric once the dependence on the gauge modes is removed by a partial solution of the field equations, taking the leading contribution from the physical gauge fixing term.

For the gauge invariant flow equation this is achieved whenever we start at large $k$ with a gauge invariant $\bar \Gamma_k$.
In contrast, for cutoff functions not involving the macroscopic metric one needs instead to take for the initial value at large $k$ an effective action that features terms violating diffeomorphism symmetry.
In this context the gauge invariant flow equation seems to offer more control of the truncation.
The issue whether one can find a choice of macroscopic fields and a precise definition of $\bar\Gamma_k$ such that the gauge invariant flow equation becomes exact is not yet settled.
Despite this shortcoming, it is our opinion that truncations of the gauge invariant flow are more reliable than a flow that violates diffeomorphism symmetry.

One may ask which is the correct graviton propagator that describes the propagation of gravitational waves and encodes the information about the primordial cosmic tensor fluctuations according to the formalism of Ref.\,\cite{Wetterich:2016vxu}.
As discussed in a similar investigation for Yang-Mills theories, this question concerns essentially the coupling to physical sources~\cite{Wetterich:2017aoy}.
For gravity, the physical sources are given by a conserved energy momentum tensor.

By construction, the field equations derived from the gauge invariant effective action $\bar\Gamma$ involve a conserved energy momentum tensor.
One actually defines in this setting the physical energy momentum tensor by the field equations derived from $\bar\Gamma$ in the presence of matter fields, suitably averaged in the case of inhomogeneous matter distributions. 
The propagator $G_P$, defined by the inverse of $\bar\Gamma^{(2)}$ on the subspace of physical fluctuations by Eq.\,\eqref{propagator and Gamma2}, describes indeed the propagation of metric perturbations induced by physical sources.
It obeys all the necessary criteria for the physical graviton propagator.
For the formulation of the gauge invariant effective average action $\bar\Gamma_k$ leading to the gauge invariant flow equations, these properties extend to arbitrary $k$.

For formulations where the gauge fixing, the Faddeev-Popov determinant and a possible IR cutoff do not involve the macroscopic metric, but rather an independent fixed background metric (which may be a flat space metric), the issue is more complicated.
The field equations derived from the effective action no longer involve a conserved energy momentum tensor.
Indeed, we have argued that for fixed background fields the effective average action $\Gamma_k$ is not gauge invariant, even if a physical gauge fixing is used.
For $k=0$ one may recover the physical properties of diffeomorphism symmetry by the use of Slavnov-Taylor identities and Becchi-Rouet-Stora-Tyutin (BRST) symmetry.
This can be extended to those flow trajectories for $k\neq 0$ that recover for $k=0$ the physical properties of diffeomorphism symmetry.
This procedure involves modified Slavnov-Taylor identities~\cite{Ellwanger:1994iz} or associated ``background field identities"~\cite{Reuter:1993kw}, or a $k$-dependent version of BRST symmetry~\cite{Asnafi:2018pre}.
These identities control, in principle, the diffeomorphism violating terms in the effective action.
They are rather different to handle in practice, however.
The gauge invariant effective average action $\bar\Gamma_k$ and the use of gauge invariant flow equations circumvents all these complications, treating directly with objects of interest.

We conclude that the inclusion of graphs from the field dependence of $R_k$ is crucial for a simple justification of gauge invariant truncations.
A similar situation was previously discussed for the graviton contribution to the flow of the graviton propagator at zero momentum~\cite{Wetterich:2018poo}.
Of course, one may sometimes encounter situations where the omission of graphs from the field dependence of $R_k$ has only modest consequences for the flow equations.

\section{Flow generator}
\label{evaluation of flow equations}
In this appendix we compute the flow equation for the effective potential $U$ and the coefficient of the curvature tensor $F$ in the background field formalism with physical gauge fixing and a truncation $\Gamma_k=\bar\Gamma_k+\Gamma_\text{gf}$, with gauge invariant $\bar\Gamma_k$ and $\Gamma_\text{gf}$ the gauge fixing term.
In this approximation we find the same flow equations as for the gauge invariant setting.
In particular, the more detailed result in the sector of physical scalar fluctuations can be taken over directly to the gauge invariant flow, extending the approximative result for $\pi_0$ in Sec.\,\ref{discussion on scalar metric fluctuation}.  
As discussed in Appendix\,\ref{Flow of the graviton propagator appendix}, this simple correspondence does not hold for quantities involving field derivatives, as propagators or vertices.
The contribution of Appendixes\,\ref{formulations} and \ref{evaluation of flow equations} is self-contained and may be considered as an independent computation of the flow equation.

Within the standard background field approach~\cite{Reuter:1993kw} we derive the flow generator that is given as
\al{
\p_t \Gamma_k=\zeta_k=\pi_2+\pi_0+\eta_1+\eta_0+\tau\,.
}
Here $\pi_2$ and $\pi_0$ are the contributions from the spin-2 graviton ($t_{\mu\nu}$) and the spin-0 scalar fields ($\sigma$ and $\varphi$), respectively, while $\eta_1$ and $\eta_0$ are the spin-1 ($\kappa_\mu$, $C_\mu^\perp$) and spin-0 ($u$, $C$) measure contributions.
We also have to regularize the Jacobians.
The corresponding contribution to the flow equation is part of the measure contribution $\eta_1+\eta_0$.
Contributions of additional matter fields are denoted by $\tau$.
We concentrate on gravity coupled to a single scalar field, for which the main ingredients for the flow equation are given in Appendix\,\ref{formulations}.

\subsection{Infrared cutoff function}
The infrared cutoff function ${\mathcal R}_k$ has to obey several criteria: 
(i) It should regulate the propagator in the infrared such that the momentum integration in the flow equation \eqref{eq:wetterich} remains finite for small $q^2$. 
(ii) The derivative $\p_t {\mathcal R}_k$ should decay fast for high momenta $q^2 \gg k^2$ such that for fixed fields the momentum integral is also UV finite. 
(iii) The typical scale should be set by $k$, with $\mathcal R_k$ vanishing for $k\to0$.
With these requirements only a finite momentum interval of $q^2$ near $k^2$ contributes effectively to the flow equation.
(iv) In the background field formalism the effective average action should be invariant under combined gauge transformations of the macroscopic and background metrics.
Then ${\mathcal R}_k$ should be formulated in terms of covariant derivatives involving the background metric.
(For the gauge invariant flow equation one uses instead covariant derivatives involving the macroscopic metric.)

These criteria limit the choice of ${\mathcal R}_k$, but many different forms remain possible.
We require here two additional properties:
(v) The physical part of the cutoff function should be a sufficiently smooth function of the covariant momenta.
In particular, it should not contain explicit projectors on particular modes that would induce additional strong nonlocalities.
A natural choice for the metric is a cutoff ${\mathcal R}_k(\hat{\mathcal D}_f)$ with $\hat{\mathcal D}_f$ given by Eq.\,\eqref{derivative hat D} and related directly to the second variation of the curvature scalar.
Similarly, for the scalars we will employ ${\mathcal R}_k({\mathcal D}_S)$ with ${\mathcal D}_S$ the covariant scalar Laplacian.
In the ghost sector $R_k$ should be a function of the differential operator in Eq.\,\eqref{ghostaction}.
While the use of explicit projectors is avoided by these definitions of ${\mathcal R}_k$, an effective projection on the different modes will take place for maximally symmetric geometries.
In this case the relevant operators as $\hat{\mathcal D}_f$ become block diagonal, and the same happens for ${\mathcal R}_k(\hat{\mathcal D}_f)$.

The next criterion (vi) requires a separate cutoff for the gauge modes that diverges $\sim 1/\alpha$ for $\alpha\to0$.
This is needed for an effective cutoff in this sector, since otherwise the gauge fixing term $\sim1/\alpha$ would not be regularized.
The cutoff for the gauge modes should act only on the gauge fluctuations, not on the physical fluctuations.
This avoids mixing between the physical and gauge modes also for $k>0$.
Finally, the Jacobians \eqref{set of Jacobians} need a regularization by ${\mathcal R}_k$ as well.
Otherwise, they would induce strong nonlocalities.
Different choices and normalizations of modes yield different Hessians and different Jacobians.
The regularization for the Jacobians should be of a type that makes the regularization independent of the precise definition of fields.

Finally, we include in ${\mathcal R}_k$ prefactors as $F$ such that ${\mathcal R}_k$ has a similar structure as the second variation of kinetic terms such as $FR_k$ or $Z_\varphi \p^\mu \varphi \p_\mu \varphi$. 
With these prescriptions the addition of the IR cutoff ${\mathcal R}_k$ replaces kinetic operators as $F\hat{\mathcal D}_f$ by $FP_k(\hat{\mathcal D}_f)$.
Here $P_k(\mathcal D)$ is of the form $P_k(\mathcal D)=\mathcal D+R_k(\mathcal D)$.
For the flow equations we have to add the IR-cutoff to the Hessian $\Gamma^{(2)}_k$.
This replaces in Eq.\,\eqref{TT mode two point function} $\bar{\mathcal D}_T$ by $P_k(\bar{\mathcal D}_T)$, or in the $\varphi$--$\varphi$ element of the matrix \eqref{spin 0 matrix} $Z_\varphi \bar\Delta_S\to Z_\varphi P_k(\bar\Delta_S)$.

\subsection{Physical metric fluctuations}
We first evaluate the contributions from the physical fluctuations.
There are the TT mode ($t_{\mu\nu}$) and the physical spin-0 modes ($\varphi$ and $\sigma$), whose forms of the flow generators are given by
\al{
&\pi_2=\frac{1}{2}\text{Tr}_{(2)}\frac{\p_t \mathcal R_k}{\Gamma_k^{(2)}+\mathcal R_k}\Bigg|_{tt},\nn[2ex]
&\pi_0=\frac{1}{2}\text{Tr}_{(0)}\frac{\p_t \mathcal R_k}{\Gamma_{k}^{(2)}+\mathcal R_k}\Bigg|_\text{ph},
\label{physical mode generators}
}
respectively.
The two-point functions $\Gamma^{(2)}_k$ for the TT mode and the physical spin-0 modes are shown in \eqref{TT mode two point function} and \eqref{physical spin 0 mode two point function}, respectively.

\subsubsection{Spin-2 TT mode}
Using the heat kernel method summarized in Appendix\,\ref{heat kernel methods} we can evaluate the contributions \eqref{physical mode generators}.
Our criteria for the IR cutoff function correspond for the TT mode to a type-II cutoff function in the naming of Ref.\,\cite{Codello:2008vh}, with
\al{
W_2\fn{z}=\frac{\p_t (FR_k\fn{z})}{F\left(P_k\fn{z} -vk^2\right)}\,.
}
This is precisely the formula of Sec.\,\ref{evaluation of graviton contributions} and results in
\al{
\pi_2&=\frac{1}{16\pi^2}\int_x\sqrt{\bar g}\bigg[ \frac{20}{3} k^4\ell_0^4\fn{-v} \left( 1- \frac{\eta_g}{8} \right) \nn
&\qquad \qquad \qquad \qquad
- \frac{25}{4} k^2\ell_0^2\fn{-v} \left(  1- \frac{\eta_g}{6} \right) \bar R \bigg],
}
with $\eta_g=-\p_t \ln w$.
The threshold functions for the Litim cutoff are evaluated as
\al{
&\ell_0^4\fn{\tilde w}=\frac{1}{2(1+\tilde w)},&
&\ell_0^2\fn{\tilde w}=\frac{1}{1+\tilde w}.&
}
Using the formulas in Appendix\,\ref{IR cutoff scheme}, one can find for comparison the case where a type-I cutoff is employed.

\subsubsection{Spin-0 modes}
Let us calculate contributions from the spin-0 physical scalar fields.
For our cutoff function only the diagonal part of the Hessian is replaced with $P_k\fn{z}=z+R_k\fn{z}$, with $z=\bar\Delta_S$.
The projection of the cutoff on the $\sigma$ mode yields
\al{
{\mathcal R}_k^\text{ph}\fn{z}=\pmat{
\Gamma_k^{\sigma\sigma}\fn{P_k}-\Gamma_k^{\sigma\sigma}\fn{z}  && 0\\[3ex]
0 &&   Z_\varphi \,R_k\fn{z}
},
}
where
\al{
&\Gamma_k^{\sigma\sigma}\fn{z}= -\frac{F}{6}\left[ \frac{ z -\frac{3\bar R}{8}}{ z -\frac{\bar R}{3}} z -\frac{U}{2F}\frac{ z -\frac{\bar R}{2}}{ z -\frac{\bar R}{3}}\right].
}
This apparently somewhat complicated form is only a result of the projection, the original cutoff $R_k(\hat{\mathcal D}_f)$ being much simpler.

We again extract the flow generator $\pi_0$ from the heat kernel method.
The physical spin-0 mode contributions are given as
\al{
\pi_0&=\frac{1}{2} \text{Tr}_{(0)}\tilde \p_t \ln\fn{\Gamma_{k}^{(2)}+\mathcal R_k}\Big|_\text{ph}
=\frac{1}{2}\text{Tr}_{(0)}\,W_0\fn{z}\nn
&=\frac{1}{2(4\pi)^2} \int_x\sqrt{\bar g}\left\{ b_0^{(0)}Q_2[W_0]+ b_2^{(0)}Q_1[W_0] {\bar R} \right\},
\label{pi 0 contributions}
}
where $\tilde \p_t$ is the $t$ derivative acting on only the scale dependence in the regulator.
The heat kernel function for the spin 0 modes involves the eigenvalues of the regulated inverse propagator matrix \eqref{physical spin 0 mode two point function}, with $\bar\Delta_S$ replaced by $P_k(\bar\Delta_S)$.
It reads 
\al{
W_0\fn{z}&=\tilde \p_t \ln\Bigg\{ 3\rho \left[ -F'\left( z + \frac{\bar R}{4}\right)+U'\right]^2\nn
&\quad
+\frac{F}{18}\left[ \frac{P_k-\frac{3\bar R }{8}}{ P_k -\frac{\bar R}{3}}P_k-\frac{U}{2F}\frac{ P_k-\frac{\bar R}{2}}{ P_k -\frac{\bar R}{3}}\right]\nn
&\qquad\times
\left(Z_\varphi P_k +m_\varphi^2 -\displaystyle\frac{\tilde\xi_\varphi}{2} \bar R \right)\Bigg\},
\label{heat kernel of spin 0 physical contribution}
}
where the first term in the logarithm corresponds to the mixing contribution between $\sigma$ and $\varphi$.

To extract the explicit form of the beta functions, we define the dimensionless quantities
\al{
&\tilde \rho=\frac{Z_\varphi\varphi^2}{2k^2}=Z_\varphi\rho\,,\qquad
u\fn{\tilde\rho}=\frac{U\fn{\rho}}{k^4}\,,\nn[2ex]
&w\fn{\tilde \rho}=\frac{F\fn{\rho}}{2k^2}\,,\qquad
v\fn{\tilde \rho}= \frac{2U\fn{\rho}}{F\fn{\rho}k^2}=\frac{u\fn{\tilde\rho}}{w\fn{\tilde\rho}}\, ,\nn[2ex]
&\tilde n_\varphi\fn{\tilde \rho}=\frac{\tilde \xi_\varphi}{Z_\varphi }=\frac{F'+2\rho F''}{Z_\varphi}=2w'+4\tilde \rho w''\,,\nn[2ex]
&\tilde m_\varphi^2\fn{\tilde \rho}=\frac{m_\varphi^2}{Z_\varphi k^2}=\frac{U'+2\rho U''}{Z_\varphi k^2}=u'+2\tilde \rho u''\,.
}

\begin{widetext}
Evaluating Eq.\,\eqref{pi 0 contributions} with Eq.\,\eqref{heat kernel of spin 0 physical contribution} one finds
\al{
\pi_0&=\frac{k^4}{32\pi^2} \left[  \Upsilon^{(\sigma\sigma)}_{1,0,0,0,0,1} + \Upsilon^{(\varphi\varphi)}_{0,1,0,0,0,1} \right] \int_x\sqrt{\bar g} 
+ \frac{k^2}{32\pi^2} \Bigg[\frac{1}{6} \left\{ \Upsilon^{(\sigma\sigma)}_{1,0,0,0,0,0} + \Upsilon^{(\varphi\varphi)}_{0,1,0,0,0,0} \right\} 
-\frac{1}{2}\left\{ \Upsilon^{(\sigma\sigma)}_{1,0,0,1,1,0} + \Upsilon^{(\varphi\varphi)}_{0,1,0,1,1,0} \right\} \nn
&\qquad
-\frac{1}{24}\left\{ \Upsilon^{(\sigma\sigma)}_{2,0,1,0,1,1}+\Upsilon^{(\varphi\varphi)}_{0,0,1,1,1,1} -\Phi\fn{v} \right\} 
-\tilde n_\varphi \left\{ \Upsilon^{(\sigma\sigma)}_{0,0,0,1,1,1} +\Upsilon^{(\varphi\varphi)}_{0,2,0,0,1,1}  \right\} 
  \Bigg]\int_x\sqrt{\bar g}{\bar R}.
}
Here the threshold functions are defined as
\al{
 \Upsilon^{(\sigma\sigma)}_{i,j,k,l,m,n}&=\int^{\infty}_0\df x\,   \left( x^n  f_g\fn{x}\right)\frac{(p\fn{x}+m_\varphi^2)^{i}\left(p\fn{x}-v/4\right)^{j}\left(1-v/p\fn{x}\right)^{k}\left(\frac{3\tilde \rho (-2w'x+u')^2}{2w}\right)^{l}}{ \left[ \left(p\fn{x} -v/4\right) ( p\fn{x}+m_\varphi^2 )+3\tilde \rho (-2w' x+u' )^2/2w \right]^{m+1} }\,,\nn[2ex]
 \Upsilon^{(\varphi\varphi)}_{i,j,k,l,m,n}&=\int^{\infty}_0\df x\,   \left( x^n  f_\varphi\fn{x}\right)\frac{(p\fn{x}+m_\varphi^2)^{i}\left(p\fn{x}-v/4\right)^{j}\left(1-v/p\fn{x}\right)^{k}\left(\frac{3\tilde \rho (-2w'x+u')^2}{2w}\right)^{l}}{ \left[ \left(p\fn{x} -v/4\right) ( p\fn{x}+m_\varphi^2 )+3\tilde \rho (-2w' x+u' )^2/2w \right]^{m+1} }\,\nn[2ex]
\Phi\fn{v}&= \int^\infty_0 \df x\, f_\Phi\fn{x}  \frac{ (v/p\fn{x})(p\fn{x}+\tilde m_\varphi^2) }{\left(p\fn{x} -v/4\right) ( p\fn{x}+m_\varphi^2 )+3\tilde \rho (-2w'x+u' )^2/2w}\,.
}
\end{widetext}
They involve the dimensionless combinations
\al{
f_\varphi\fn{x}&=\frac{\p_t (Z_\varphi R_k\fn{z})}{Z_\varphi k^2}=-\eta_\varphi r\fn{x} +\frac{\p_t R_k\fn{z}}{k^2}\, ,\nn[5pt]
f_g\fn{x}&=\frac{\p_t(F R_k\fn{z})}{Fk^2}=(2-\eta_g) r\fn{x} +\frac{\p_t R_k\fn{z}}{k^2}, \nn[5pt]
f_\Phi\fn{x}&=   \left(4+ \p_t \ln v - \eta_g \right) r\fn{x}+\frac{x}{p\fn{x}}\frac{\p_t R_k\fn{z}}{k^2}, \nn[5pt]
r\fn{x}&=\frac{R_k\fn{z}}{k^2},\quad
p\fn{x}=\frac{P_k\fn{z}}{k^2},
\quad
x=\frac{z}{k^2},
}
and the anomalous dimensions
\al{
&\eta_g=-\frac{\p_t  w}{w},&
&\eta_\varphi=-\frac{\p_t Z_\varphi}{Z_\varphi}.&
}

The result is somewhat lengthy, partly because of the mixing between the scalar modes.
It simplifies considerably if this mixing can be neglected.
Assume that in the denominator of the threshold functions, one has
\al{
\left(p\fn{x} -v/4\right) ( p\fn{x}+m_\varphi^2 )\gg 3\tilde \rho (2w'x+u' )^2/2w\,.
}
In this case the mixing term is suppressed and can be neglected.
In particular, the mixing is absent if $\tilde\rho=0$.

Neglecting the mixing, $\pi_0$ can be evaluated as $\pi_0=\pi_k^{(S)}+\pi_k^{(\sigma)}$, where $\pi_k^{(S)}$ is the contribution from the $\varphi$ fluctuation given in Eq.\,\eqref{pi_S} and
\al{
&\pi_0^{(\sigma)}= \frac{k^4}{32\pi^2} \Upsilon^{(\sigma\sigma)}_{1,0,0,0,0,1}\int _x\sqrt{\bar g} \nn
& 
+\frac{k^2}{32\pi^2}\bigg[ \frac{1}{6}  \Upsilon^{(\sigma\sigma)}_{1,0,0,0,0,0} 
-\frac{1}{24}\left( \Upsilon^{(\sigma\sigma)}_{2,0,1,0,1,1}-\Phi \right) \bigg]\int _x\sqrt{\bar g}\bar R\,.
}
The threshold functions simplify in the absence of mixing, and we obtain for the Litim cutoff
\al{
\pi_0^{(\sigma)}&=\frac{1}{32\pi^2}\int_x\sqrt{\bar g}\bigg[\frac{4}{3}\left( 1-\frac{\eta_g}{8}\right)k^4\nn
&\quad
 +\bigg\{ \frac{1}{2(1-v/4)}\left( 1-\frac{\eta_g}{6}\right)\nn 
 &\quad -\frac{1-v}{18(1-v/4)^2}\left( 1-\frac{\eta_g}{8}\right) 
+\frac{1}{24}\Phi\fn{v} \bigg\}k^2 \bar R \bigg],
\label{pi0 sigma contribution without mixing term}
}
where 
\al{
\Phi\fn{v}&= v\int_0^\infty \df x\frac{\left(4+ \p_t \ln v - \eta_g \right) r\fn{x}+\frac{x}{p\fn{x}}\frac{\p_t R_k\fn{z}}{k^2}}{p\fn{x}(p\fn{x}-v/4)}\nn
&=\frac{3v}{1-v/4}\left( 1 +\frac{\p_t \ln v - \eta_g}{6} \right).
\label{Phi threshold function}
}

The result \eqref{pi0 sigma contribution without mixing term} and \eqref{Phi threshold function} has already a simple structure from which the contributions to $c_V^{(\sigma)}$ and $c_M^{(\sigma)}$ are easily extracted.
We observe the appearance of various factors $(1-v/4)$, where the part $\sim v$ arises from the mass term in the $\sigma$ propagator.
For analytic discussions it is often sufficient to set $v=0$, since the $c_V^{(\sigma)}$ and $c_M^{(\sigma)}$ are subdominant.
We have discussed in the main text that this is a valid approximation for our purposes.
Setting $v=0$, Eq.\,\eqref{pi0 sigma contribution without mixing term} results in Eq.\,\eqref{approximated pi0}.

\subsection{Measure contribution}
\label{Measure contributions}
We next evaluate the measure contribution $\eta_k$.
As argued in Ref.\,\cite{Wetterich:2016ewc}, this contribution takes a simple form
\al{
\eta_k=-\frac{1}{2}\text{Tr}_{(1)}\frac{\p_t P_k\fn{\bar{\mathcal D}_1}}{P_k\fn{\bar{\mathcal D}_1}}-\frac{1}{2}\text{Tr}_{(0)}\frac{\p_t P_k\fn{\bar{\mathcal D}_0}}{P_k\fn{\bar{\mathcal D}_0}}\,,
\label{measure contributions}
}
with 
\al{
&\bar{\mathcal D}_1=\bar \Delta_V-\frac{\bar R}{4},&
&\bar{\mathcal D}_0=\bar \Delta_S-\frac{\bar R}{4}.&
}
It is shown by explicit calculations in \cite{Pawlowski:2018ixd} that actually the measure contribution is given indeed by Eq.\,\eqref{measure contributions}, with $\mathcal D_1=\mathcal D_0=q^2$ in the case of flat background $\bar g_{\mu\nu}=\delta_{\mu\nu}$.
We generalize here the result to arbitrary background metrics.

The measure contribution arises from the gauge fluctuations, the ghost fluctuations, and the regularization of the Jacobian.
Only the combination of all contributions results in the simple expression \eqref{measure contributions}.
For the regularized Jacobians we again replace the relevant differential operator $\bar{\mathcal D}$ by $P_k(\bar{\mathcal D})$.
This is necessary since otherwise the Jacobians would induce strong nonlocalities.
 The flow contributions from the regularized Jacobians read
\al{
&J_{\mathrm{grav1}}= \frac{1}{2}\Tr_{(1)} \frac{\p_t P_k\fn{\bar{\mathcal D}_1}}{ P_k\fn{\bar{\mathcal D}_1}}\, , \nn[2ex]
&J_{\mathrm{grav0}}=- \frac{1}{2}\Tr_{(0)} \frac{\p_t P_k\fn{\bar{\mathcal D}_0}}{ P_k\fn{\bar{\mathcal D}_0}} 
-\frac{1}{2}\Tr_{(0)} \frac{\p_t P_k\fn{\bar \Delta_S}}{ P_k\fn{\bar\Delta_S}} \,,
\nn[2ex]
&J_\text{gh}= \Tr_{(1)} \frac{\p_t P_k\fn{\bar\Delta_S}}{ P_k\fn{\bar\Delta_S}}\, .\label{aux3}
} 
 
By a separate computation of all other individual contributions we will show explicitly that the combined measure contribution is given by the simple form \eqref{measure contributions}.

The spin-1 measure contribution is given by
\al{
\eta_1=\delta^{(1)}_k-\epsilon^{(1)}_k,
\label{eta_1k}
}
with the spin-1 vector gauge mode \eqref{spin 1 contribution in Landau gauge} and the Jacobians \eqref{aux3} for the spin-1 gauge field 
\al{
\delta^{(1)}_k &= \lim_{\alpha\to 0}\frac{1}{2}\text{Tr}_{(1)}\frac{\p_t \mathcal R_k}{\Gamma_k^{(2)}+\mathcal R_k}\Bigg|_{\kappa\kappa} +J_\text{grav1}\nn[1ex]
&=\text{Tr}_{(1)}\frac{\p_t  P_k\fn{\bar{\mathcal D}_1}}{P_k\fn{\bar{\mathcal D}_1}} -\frac{1}{2}\text{Tr}_{(1)}\frac{\p_t  P_k\fn{\bar{\mathcal D}_1}}{P_k\fn{\bar{\mathcal D}_1}}\nn[1ex]
&=\frac{1}{2}\text{Tr}_{(1)}\frac{\p_t  P_k\fn{\bar{\mathcal D}_1}}{P_k\fn{\bar{\mathcal D}_1}},
}
and the spin-1 ghost mode \eqref{ghost Hessians}
\al{
\epsilon^{(1)}_k &=\text{Tr}_{(1)}\frac{\p_t \mathcal R_k}{\Gamma_k^{(2)}+\mathcal R_k}\Bigg|_{\bar C^\perp C^\perp}
=\text{Tr}_{(1)}\frac{\p_t  P_k\fn{\bar{\mathcal D}_1}}{P_k\fn{\bar{\mathcal D}_1}}.
}
This sums up to the measure contribution from the spin-1 modes
\al{
\eta_1=-\frac{1}{2}\text{Tr}_{(1)}\frac{\p_t  P_k\fn{\bar{\mathcal D}_1}}{P_k\fn{\bar{\mathcal D}_1}}.
}

Next we discuss the spin-0 measure contribution.
The spin-0 gauge mode coming from the metric fluctuation \eqref{physical spin 0 mode two point function} and the corresponding Jacobian \eqref{aux3}  is
\al{
\delta^{(0)}_k &= \lim_{\alpha\to 0}\frac{1}{2}\text{Tr}_{(0)}\frac{\p_t \mathcal R_k}{\Gamma_k^{(2)}+\mathcal R_k}\Bigg|_\text{gauge} + J_\text{grav0} \nn
&=\left( \text{Tr}_{(0)}\frac{\p_t  P_k\fn{\bar{\mathcal D}_0}}{P_k\fn{\bar{\mathcal D}_0}} 
+\frac{1}{2}\text{Tr}_{(0)}\frac{\p_t  P_k\fn{\bar{\Delta}_S}}{P_k\fn{\bar{\Delta}_S}} \right) \nn[1ex]
&\quad -\frac{1}{2}\left( \text{Tr}_{(0)}\frac{\p_t  P_k\fn{\bar{\mathcal D}_0}}{P_k\fn{\bar{\mathcal D}_0}}
 + \text{Tr}_{(0)}\frac{\p_t  P_k\fn{\bar{\Delta}_S}}{P_k\fn{\bar{\Delta}_S}}\right) \nn[1ex]
&=\frac{1}{2}\text{Tr}_{(0)}\frac{\p_t  P_k\fn{\bar{\mathcal D}_0}}{P_k\fn{\bar{\mathcal D}_0}}\,,
}
while the spin-0 ghost mode \eqref{ghost Hessians} is
\al{
-\epsilon^{(0)}_k &=-\text{Tr}_{(0)}\frac{\p_t \mathcal R_k}{\Gamma_k^{(2)}+\mathcal R_k}\Bigg|_{\bar C C}
+J_\text{gh}\nn[1ex]
&=-\left(\text{Tr}_{(0)}\frac{\p_t P_k\fn{\bar{\mathcal D}_0}}{P_k\fn{\bar{\mathcal D}_0}} + \text{Tr}_{(0)}\frac{\p_t P_k\fn{\bar\Delta_S}}{P_k\fn{\bar \Delta_S}}\right) \nn
&\qquad+ \text{Tr}_{(0)}\frac{\p_t P_k\fn{\bar \Delta_S}}{P_k\fn{\bar\Delta_S}}\nn[1ex]
&=-\text{Tr}_{(0)}\frac{\p_t P_k\fn{\bar{\mathcal D}_0}}{P_k\fn{\bar{\mathcal D}_0}}\,.
}
In consequence, the measure contribution of the spin-0 modes becomes
\al{
\eta_0&=\delta^{(0)}_k-\epsilon^{(0)}_k=-\frac{1}{2}\text{Tr}_{(0)}\frac{\p_t  P_k\fn{\bar{\mathcal D}_0}}{P_k\fn{\bar{\mathcal D}_0}}.
}

This concludes the proof that  the total measure contribution $\eta_k=\eta_1+\eta_0$ is given by Eq.\,\eqref{measure contributions}. 
For both the spin-1 and 0 measure contributions, the simple relation $\delta_k=2\epsilon_k$ holds.
We have employed here the type-II cutoff scheme; namely the regulator $R_k$ is employed to replace $\bar{\mathcal D}_1$ and $\bar{\mathcal D}_0$ with $P_k$.
Even if one uses the type-I cutoff function such that $\bar\Delta_V$ and $\bar\Delta_S$ are replaced by $P_k$, these relations hold.

The evaluation of Eq.\,\eqref{measure contributions} is done in Sec.\,\ref{evaluation of measure contribution} and yields
\al{
\eta_k=-\frac{1}{16\pi^2}\int_x \sqrt{\bar g}\left(4 k^4 \ell_0^4\fn{0} +\frac{17}{12}k^2 \ell_0^2\fn{0} \bar R \right).
}
The use of Eq.\,\eqref{Type I and Type II transformation} in Appendix\,\ref{IR cutoff scheme} allows us to obtain the case of the type-I cutoff functions.

\subsection{Contributions from other free particles}
We summarize contributions from massless free particles in the background field formalism.
For $N_S$ scalars, $N_F$ Weyl fermions, and $N_V$ gauge bosons, we have contributions,
\al{
\tau&=\pi_k^{(S)}+\pi_k^{(F)} +\zeta_k^{(V)} \nn[1ex]
&=\frac{N_S}{2}\text{Tr}_{(0)}\frac{\p_t P_k\fn{\Delta_S}}{P_k\fn{\Delta_S}}
-\frac{N_F}{2}\text{Tr}_{(\frac{1}{2})}\frac{\p_t P_k(-\Slash{\mathcal D}^2)}{P_k(-\Slash{\mathcal D}^2)}\nn
&\quad+\frac{N_V}{2}\left(\text{Tr}_{(1)}\frac{\p_t P_k\fn{\mathcal D_T}}{P_k\fn{\mathcal D_T}} -\text{Tr}_{(0)}\frac{\p_t P_k\fn{\Delta_S}}{P_k\fn{\Delta_S}} \right)\nn[1ex]
&=\frac{k^4}{16\pi^2}(N_S-2N_F+2N_V)\ell_0^4\fn{0} \int_x\sqrt{\bar g}\nn
&\quad
+\frac{k^2}{96\pi^2}\left( N_S+N_F-4N_V\right) \ell_0^2\fn{0} \int_x\sqrt{\bar g}\,\bar R.
}
The choice of the cutoff follows the criteria developed at the beginning of this appendix.
In the language of ref.\,\cite{Codello:2008vh} this corresponds to a type-I cutoff for bosons, while for fermions and gauge bosons the type-II cutoff is used.
As advocated, this yields (for $\tilde\xi=0$) the same flow equations \eqref{cV and cM general form}.

\section{Different IR-cutoff scheme}
\label{IR cutoff scheme}
For a comparison of different implementations of IR cutoffs we consider 
\al{
\Gamma_k^{(2)}=\Delta+cR+\tilde w k^2=\mathcal D+\tilde w k^2,
}
with $\Delta$ the negative covariant Laplacian in an appropriate sector.
For a cutoff function $R_k\fn{\mathcal D}$ one obtains
\al{
\zeta_1&=\frac{1}{2}\text{tr}\,\tilde \p_t\ln\fn{\mathcal D+R_k\fn{\mathcal D} +\tilde w k^2}\nn[1ex]
&=\frac{1}{16\pi^2}\int_x\sqrt{g}\, \Big\{ b_0 \ell_0^4\fn{\tilde w}k^4+\left( b_2-cb_0\right)\ell_0^2\fn{\tilde w}k^2  R\Big\}.
\label{general discussion of cutoff II}
}
This cutoff is of type II in the classification of Ref.\,\cite{Codello:2008vh}.
If we choose instead a type-I cutoff function $R_k\fn{\Delta}$ the result is
\al{
\zeta_2&=\frac{1}{2}\text{tr}\,\tilde \p_t\ln\fn{\mathcal D+R_k\fn{\Delta} +\tilde w k^2}\nn[1ex]
&=\frac{1}{16\pi^2}\int_x\sqrt{g}\, \bigg\{ b_0 \ell_0^4\fn{\tilde w+\frac{cR}{k^2}}k^4\nn
&\qquad \qquad \qquad \qquad +b_2 \ell_0^2\fn{\tilde w+\frac{cR}{k^2}}k^2 R\bigg\}.
}

Expanding in a power of $R$ yields
\al{
\zeta_2&=\frac{1}{16\pi^2}\int_x\sqrt{g}\,\Big\{ b_0\ell_0^4\fn{\tilde w}k^4  \nn
&\qquad\qquad +\left[ b_2  \ell_0^2\fn{\tilde w}-c\,b_0k^2 \ell_1^4\fn{\tilde w} \right] k^2 R\Big\}\,,
}
with 
\al{
\ell^d_1\fn{\tilde w}=-\frac{\p}{\p \tilde w}\ell_0^d\fn{\tilde w}.
}
Comparison with $\zeta_1$ replaces in Eq.\,\eqref{general discussion of cutoff II} for the coefficient of $R$
\al{
(b_2-cb_0)\ell_0^2\fn{\tilde w}\to b_2 \ell_0^2\fn{\tilde w}-cb_0 \ell_1^4\fn{\tilde w}
\label{Type I and Type II transformation}
}
such that for the term $\sim c$ the threshold function $\ell_0^2\fn{\tilde w}$ is replaced by a different threshold function $\ell_1^4\fn{\tilde w}$.
Unless there are particular cancellations of the type $cb_0\approx b_2$ or $cb_0 \ell_1^4\approx b_2 \ell_0^2$, the two types of cutoff give qualitatively similar results.
Quantitative differences fall into the general class of cutoff differences.
We recall, however, that for fermions a chirally invariant cutoff in terms of the Dirac operator does not allow for an arbitrary choice of $R_k$

\section{Constant scaling solution for the new fixed point}\label{plots for second scaling solution}
\begin{figure}
\includegraphics[width=9cm]{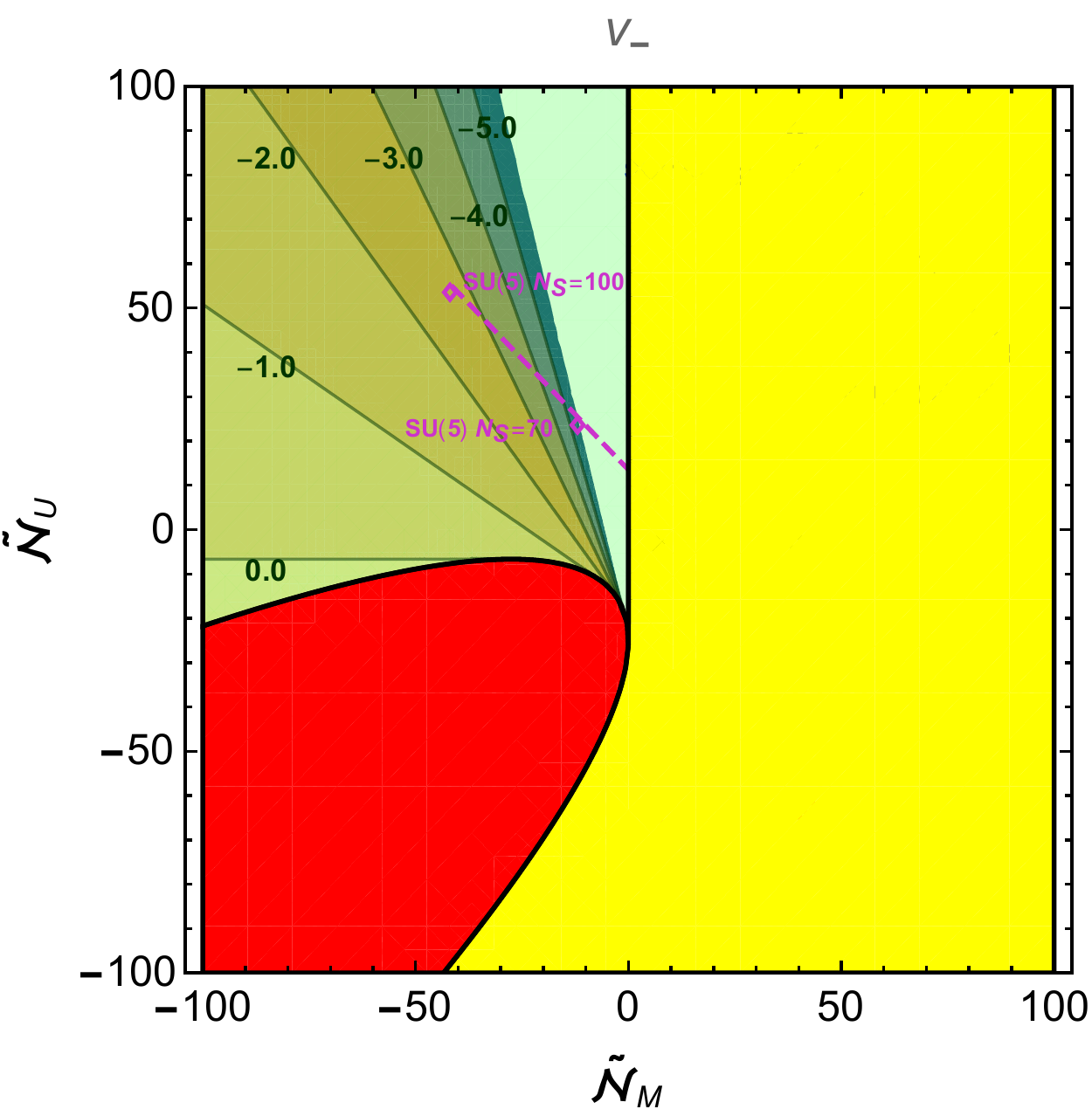}
\caption{
Contour plot of the fixed point value of $v_-$ in the $(\tilde{\mathcal N}_M,\tilde{\mathcal N}_U)$-plane.
For the yellow region on the right of the figure there is no stable solution due to $v_->1$.
For the red region no constant scaling solution is found.
}
\label{contourfig:FP values of v_-} 
\end{figure}

\begin{figure}
\includegraphics[width=9cm]{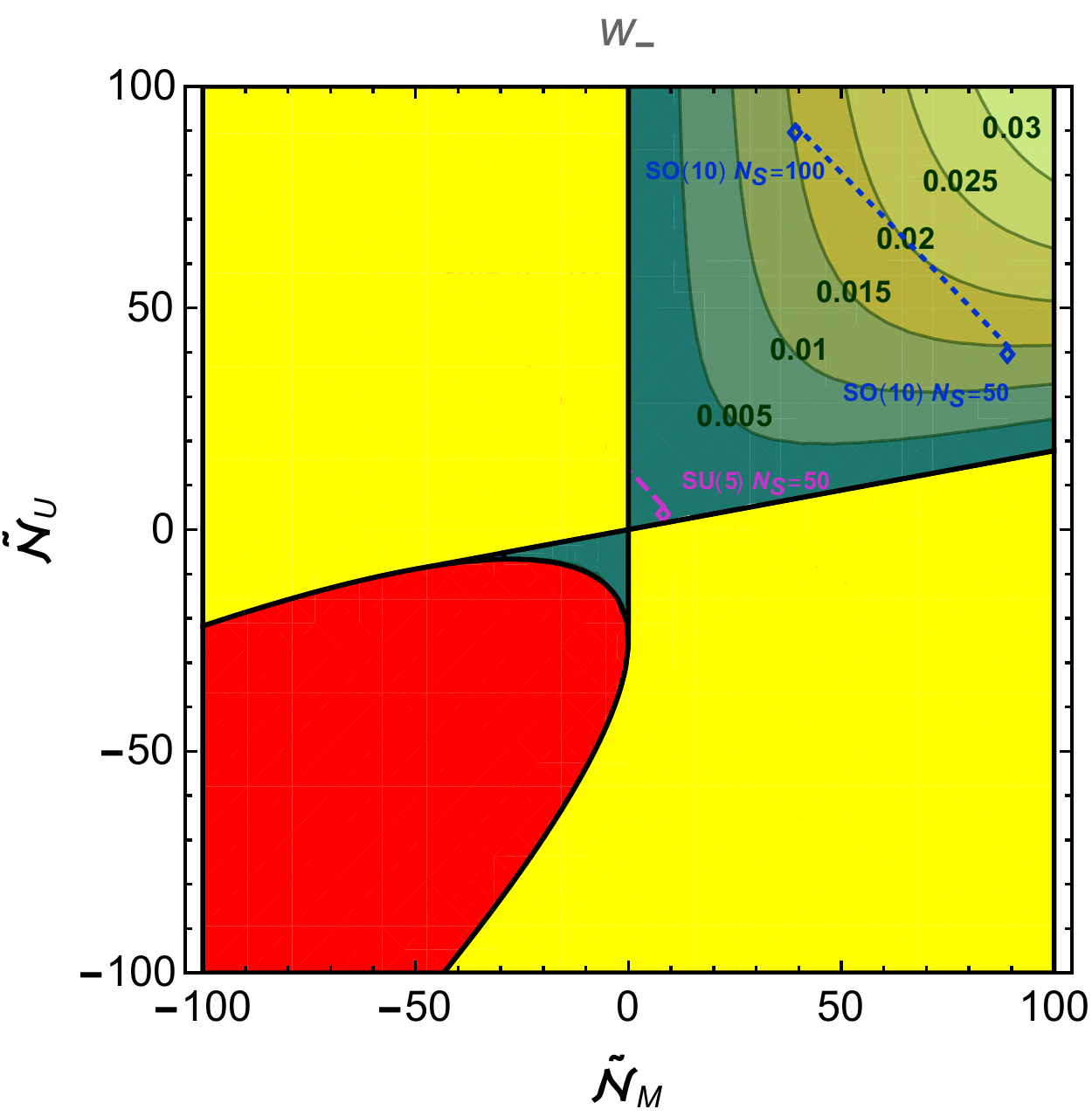}
\caption{
Contour plot of the fixed point value of $w_-$ in the $(\tilde{\mathcal N}_M,\tilde{\mathcal N}_U)$ plane.
For the yellow region the new fixed point has unstable gravity due to $w_-<0$.
The combination of the yellow and red regions in Figs.\,\ref{contourfig:FP values of v_-} and \ref{contourfig:FP values of w_-} leads to the excluded red and yellow regions in Fig.\,\ref{boundary lines}.
}
\label{contourfig:FP values of w_-} 
\end{figure}

\begin{figure}
\includegraphics[width=9cm]{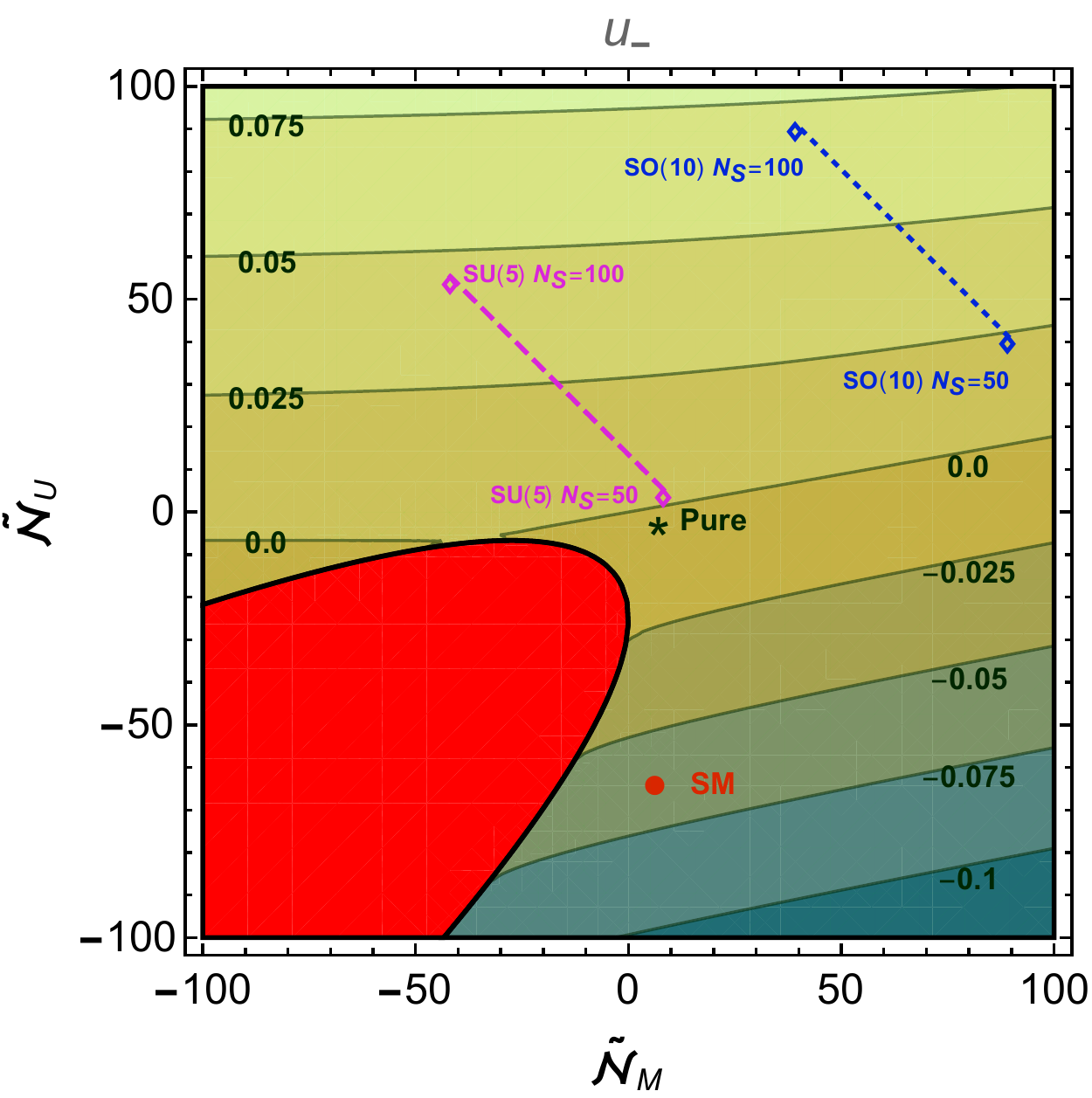}
\caption{Contour plot of the fixed point value of $u_-$ in the $(\tilde{\mathcal N}_M,\tilde{\mathcal N}_U)$ plane.}
\label{contourfig:FP values of u_-} 
\end{figure}

We display in this appendix the contour plots for the second scaling solution $v_-$, $w_-$, and $u_-$ in Figs.\,\ref{contourfig:FP values of v_-}, \ref{contourfig:FP values of w_-}, and \ref{contourfig:FP values of u_-}, respectively.
In the red regions there is no stable fixed point.
The conditions, $w_->0$ and $v_-<1$, are not satisfied in the yellow regions.
As discussed in Sec.\,\ref{discussion on New fixed point}, the second scaling solution can be allowed only for the green region in Fig.\,\ref{boundary lines}, which is the intersection of the green regions in Figs.\,\ref{contourfig:FP values of v_-} and \ref{contourfig:FP values of w_-}.

\end{appendix}
\newpage
\bibliography{refs}
\end{document}